# Improving the Estimation of Ship Length via ISAR

March 6, 2026

John R. Bennett
Bennett Radar Software
1945 Briargate Place
Escondido, California 92029
858-922-9732
mulebennett5@gmail.com

Published Software in:
github.com/mulebennett5-source/IAT_Published_Tybee

# Improving the Estimation of Ship Length via ISAR

John R. Bennett[1]

***Abstract* - A method for estimating the aspect angle of ships at sea from an ISAR is developed. The ISAR AutoTrack (IAT) algorithm uses the information from the adaptive motion compensation velocity to improve the tracker estimation of the ship aspect angle and thus to improve the estimation of ship length. The IAT is based on classical methods of autofocus for synthetic aperture radar. The average mocomp velocity yields the error in the in-range component of the ship velocity; the linear time trend of the velocity determines the cross-range component of the ship velocity. The IAT has two methods for implementing the algorithm, the Search and Analytical methods. Both methods benefit from an intelligent smoothing process that removes system errors, random noise, and ocean waves. The goal of the IAT is to measure ship length to within 10 percent over all azimuth angles and ranges relative to the aircraft and for (unsigned) aspect angles from 5 to 85 degrees. Using the IAT allows a major reduction in the radar resources dedicated to tracking; and since the IAT creates its estimates during the ISAR time window it is unaffected by ship maneuvers. Recommendations for further development and testing of the IAT are presented.**

## I. INTRODUCTION

This paper describes progress in maritime surveillance radar driven by one major technical trend and one key processing idea.

The major technical trend is the increasing use of GPS for the validation and development of maritime radars. There are two variations of this. First, standard and differential GPS sensors with recording systems have been fielded on test ships so that the data can be analyzed along with the radar data (Section IX). Second, nearly all large ships carry Automatic Identification Systems (AIS) [16] that transmit their GPS measurements and other information to a wide area. This is mainly used for collision avoidance, but the data can easily be received and recorded by an aircraft operating an imaging radar to give time records of the ship's geographical position, course, and speed. The AIS also transmits the ship's MMSI number, so that the analyst can use the internet to find its current cargo and port schedule. With AIS recording a radar aircraft can execute an R&D or operator training program that captures data from many ships at long ranges during a single flight. However, in routine operations the resulting system cannot count on AIS data since many of the targets of this surveillance could be Dark Ships, ships that turn off their AIS, or encrypt the transmissions, or use it to spoof a fake course and position. The opportunity to collect lots of data makes algorithm development much easier. For example, in the data analyzed in Sections V-VIII there were enough ship cases to determine that the usual technique of using a tracker to estimate the aspect angle of the ship is not accurate enough to measure ship length adequately. This led to the key processing idea.

The key processing idea is not due to a major scientific breakthrough or an engineering insight. Instead, it is an example of memory. I realized that an old autofocus algorithm for synthetic aperture radar could be repurposed for the analysis of radar data of ships at sea to improve the estimation of ship length, which is a critical parameter for ship identification. This algorithm is called Map-Drift Autofocus and the full algorithm for exploiting it to estimate ship length is the ISAR AutoTrack (IAT) algorithm. The only rub is that this process requires that the full motion compensation process for the system needs to be known to the processing software. Once that is done there are a few simple rules that can make the algorithm achieve maximum accuracy for real systems operating in a wide variety of ocean environments. The algorithm developed from the Map-Drift principle may use tracker information as a first guess; but this is not essential – it can run without this.

In the paper I have tried to fill out many of the necessary details for implementing the algorithm. As usual in science the concepts here are easily understood, but the practical details for implementing them in a flying system require a lot of thought and testing. And there are important issues on how to integrate AIS/GPS and improved algorithms into R&D efforts, training programs, and operations. This paper is a contribution to this discussion.

A critical parameter for identifying ships at sea is the length of the ship (Length Overall, or LOA). This is a fundamental discriminator for various ship types. How well can an ISAR measure ship length?

The approach here is to investigate this question under the assumption of a radar system like existing operational radars. Thus, I assume that the radar has a single antenna used for both transmitting and receiving and has a range resolution sufficient to have up to a few hundred range cells over the LOA. This definition excludes some possible future systems that use

Bennett Radar Software, 1945 Briargate Place,
Escondido, California, 92029, (email: mulebennett5@gmail.com).

multiple apertures to produce interferometric measurements of the three-dimensional structure of the ships. Such systems are being investigated now but it is not clear whether existing aircraft can support them at the required ranges of hundreds of kilometers. In any case, the existing operational ISAR systems for maritime surveillance are still in active production and given the very long lead times for this development, it is safe to conclude that the single-aperture assumption will apply to a large proportion of existing radars for many years to come.

The basic method for estimating LOA by radar is to first estimate the Range Extent (RE) of the ship – the difference in range between the farthest detected scatterer and the nearest detected scatterer. The RE must be combined with an estimate of the aspect angle of the ship to the radar to form the Thin Ship Approximation of the LOA:

$$\text{LOA} = \text{RE}/\text{Cosine}\ (\text{AspectU}) = \text{RE}*\text{Secant}\ (\text{AspectU}) \qquad \text{Eq. 1}$$

The aspect angle is defined as the difference between the bearing (the direction from the aircraft to the target) and the ship heading angle. But sometimes this is shifted by 180 degrees to make the zero angle indicate motion of the ship towards the radar. The variable AspectU is the unsigned aspect, the magnitude of the aspect angle shifted to the range of 0-90 degrees. This is useful for setting requirements since the sign of the aspect does not affect the length estimate and it also does not matter whether the ship is moving toward or away from the radar. An exception for this rule is that sometimes the direction can matter for a physical interpretation of the results. For example, some ships with large mirror-like deckhouses have very different signatures for approaching and receding aspects. Another angle that needs to be considered for analysis of this algorithm is the azimuth angle, the angle to the target relative to the nose of the aircraft. This affects the problem for mathematical reasons that will be covered in the analytical theory of Section III.

Most ISAR systems use a pair of radar modes to estimate the RE and aspect angle:

(1) Aspect Angle: A large-scale search real-aperture mode with coarse range resolution and a large revisit time (perhaps seconds). During this mode, the system tracks multiple ships for many minutes to estimate the ships' locations, courses, and speeds.

(2) Range Extent: A higher resolution coherent ISAR mode pointed at a selected ship, with a high pulse repetition frequency, robust adaptive motion compensation and advanced image focus algorithms. The ISAR mode receives a cue from the search mode that gives estimates of the initial position of the ship, the ship speed, and the aspect angle – which is derived from the ship course, the bearing to the ship and the aircraft motion.

Practical algorithms for this may add other details to the estimation process to correct for multipath scattering, ship width, and other effects; but this is a reasonable place to start. For a given value of RE the dependence on the cosine of the aspect angle implies that the fractional error in LOA should be the product of the tangent of the aspect angle (the fractional derivative of the secant) and the expected error of aspect angle in radians. For example, if we assume an aspect error of 3 degrees that is independent of aspect, the fractional error in LOA would be the red curve in Figure 1. At an aspect of 45 degrees the tangent is 1 and the error is about 5 percent. However, this model predicts that the tangent term yields very large LOA errors at higher aspect angles.

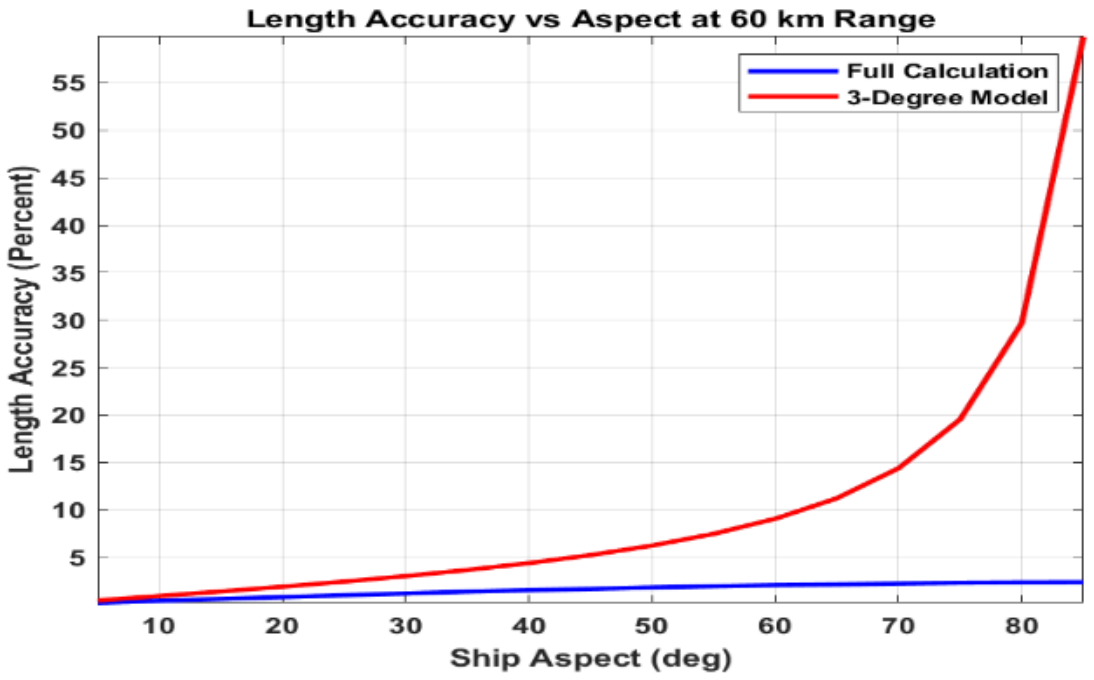

***Figure 1: Theoretical ship length accuracy for a circular scan radar***

It is important to stress that the assumption that the error in aspect is independent of aspect has no support in analysis or data – it is simply a common hypothesis. And when we do a more detailed calculation, allowing for the effects of cross-range position errors due to the radar beamwidth, the result is the blue curve in Figure 1. This shows that a better simulation yields a much smaller LOA error and a much-reduced trend with higher aspect angles. This same simulation also directly evaluates the accuracy of the aspect angle – Figure 2 shows that, in contrast to the constant error hypothesis, the accuracy of aspect improves strongly with aspect, cancelling a large part of the LOA trend for the constant error assumption. The main physical reason for this improvement is that at higher aspect angles the cross-range component of the ship velocity is easier to estimate.

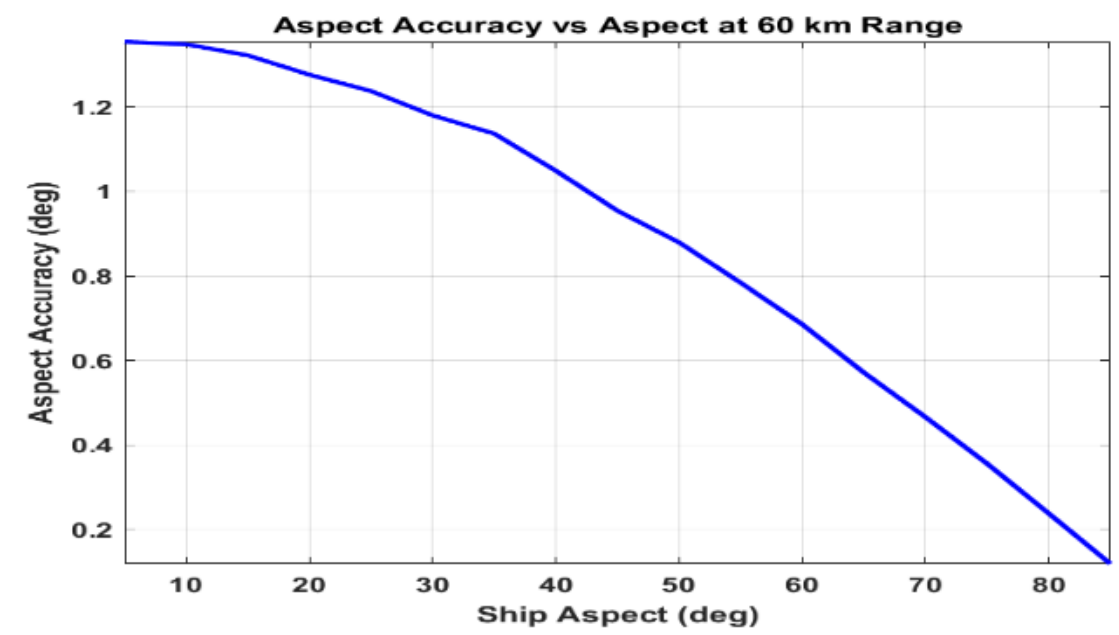

***Figure 2: Aspect Angle Accuracy vs Aspect Angle for a circular scan radar***

This simulation is based on an ideal model of ship tracking and should not be considered correct in detail. Real trackers have other major sources of error. First, the ship is not a single target but multiple targets at different ranges and times. Second, the tracker may confuse the detections from different ships – it only requires a few mistakes in associating radar detections with the ships in a scene to create large errors in the estimates of the ships' courses, speeds, and Lat/Lon positions. Third, achieving useful aspect angle estimates may require over 100 independent detections over 15 minutes or so. During this time the ship rotates relative to the radar up to 90 degrees or more just due to the aircraft motion. Thus, one may start a track when the ship is bow-on to the radar and then end with a broadside flash, which makes it very difficult to estimate the ship's location. Finally, a ship can change course during the tracking window.

Another important issue in evaluating the accuracy of ship length is that the low aspect angle range – where the constant aspect error hypothesis predicts the lowest errors – is also the region where important errors in RE occur. This is true especially at low grazing angle, where most maritime ISAR systems operate. For example, many ships have a large superstructure that can shadow one end of the ship, causing an underestimation of LOA. And my experience is that multipath scattering can also be important at low aspect, often causing overestimation of LOA. But as the aspect angle leaves the low range the ISAR processor has an easier time seeing the two ends of the ship and thus measuring RE better.

These simulations suggest that ISAR processors and ATR algorithms that put strong limits on the allowed aspect angles are not wise. If, for example, a processor eliminates all detections over 45 degrees then it is only applicable to half the expected range of aspects. We assume here that the useful range of aspect angle goes from 5 degrees to 85 degrees, about 88 percent of the expected aspects. The algorithm presented here, called ISAR AutoTrack (IAT), is a method to accomplish this goal. The tracker and ISAR modes yield complementary information that can be used for estimating the ship motion and aspect angle. The purpose of IAT is to combine the information from the two modes to produce more accurate aspect angles and better LOA estimates than either mode alone.

An earlier paper, [1] gives a detailed analysis of the potential for 3D-ISAR**.** This paper is available for download from arXiv but the main results relevant to the present discussion are summarized here. The algorithm in the 3D Paper estimates LOA based on the RE calculated using a robust algorithm that tries to remove the effects of multipath scattering and that corrects for the effect of the ship width on the length estimate. The input of the 3D LOA algorithm is the ISAR imagery plus the list of target reports generated by the ISAR processor to focus the imagery.

Figure 3 is a scatter plot of the LOA error for 43 MSR2 cases (see Section V for this definition), including both broadside and forward looks. The aspect angles for these cases are computed from the AIS ship heading reports and the known radar pointing, in contrast to the IAT results reported later. The mean ship length here is 250 meters and the RMS error is 20 meters for a relative error of 8 percent. There is a slight (3%) bias low since there isn't always a detected scatterer at each end of the ship – this could be corrected empirically.

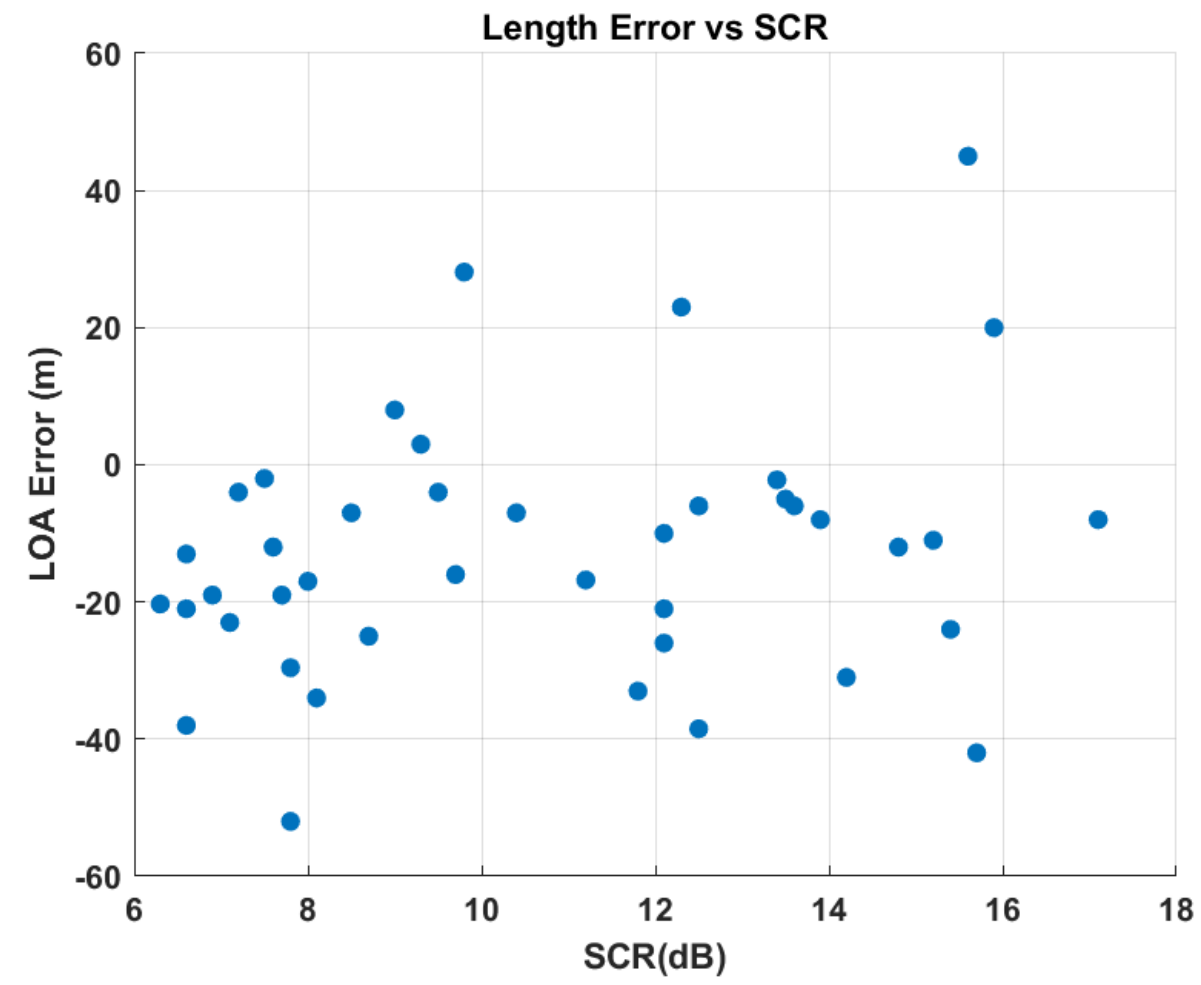

***Figure 3: Ship Length error versus Signal-to-Clutter Ratio for 43 MSR2 cases***

The 3D algorithm also estimates the deviation over time from the mean aspect angle [Figure 4] but cannot reliably estimate the mean aspect angle itself because this value is not used in the main ISAR processor, which just produces focused range-Doppler images. The algorithm also needs the mean tilt angle; but this is readily calculated from the aircraft altitude and the range, with the normal corrections for earth curvature and refraction. But once the mean aspect and tilt angles are known then the 3D algorithm can estimate the rotation rates in the horizontal and vertical planes and then use this information to convert the range-Doppler coordinate system to meters in the three dimensions.

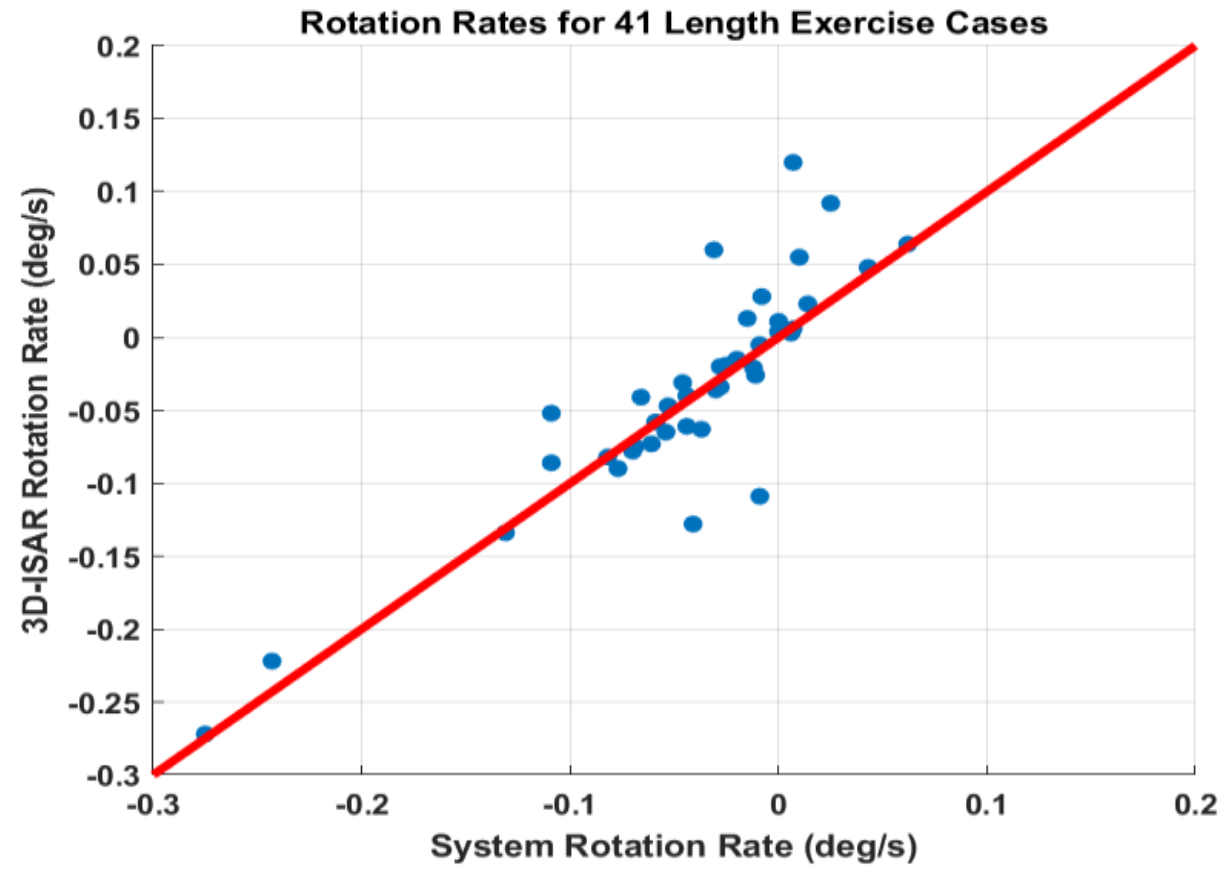

***Figure 4: Apparent rotation rate - Aux vs 3D ISAR***

A critical result of the 3D mathematical formulation in the 3D ISAR paper is based on the fundamental relation between the rotation rates about the two axes of rotation. A thin-ship version of the full equation in the 3D paper is:

$$\frac{\langle r * \dot{r} \rangle}{\langle r^2 \rangle} = -\dot{\varphi} * \tan\varphi - \dot{\theta} * \tan\theta \qquad \text{Eq. 2}$$

Here $\varphi$ is the aspect angle of the ship and $\theta$ is the tilt angle. From a point of view of a fixed ship position and a variable radar position $\varphi$ corresponds to a horizontal variation of the radar and $\theta$ corresponds to a vertical variation of the radar. The time derivatives are the effective scale factors. This equation is for a particular statistical moment. But, as discussed in the 3D ISAR paper, any method that uses linear regression can be parsed into relations between the moments. The left-hand side of this equation is information on the range and range-rate offsets that can be derived from the ISAR images or from the set of target reports used in the motion compensation and autofocus algorithms. It is one of the fundamental moments used in the algorithm, the range-Doppler covariance function. This is essentially the regression slope of the ship in the range and Doppler coordinates, and it has a dimension of 1/time, the same dimension as the rotation rate. The two terms on the right-hand side are the primary terms that determine the scale factors that relate the Doppler to the two 'cross-range' coordinates in meters. Even assuming one knows which angle term is relevant for a given ship and time, there is a problem with the relationship here. The relation between the left-hand (observable) moment, and the rotation rate scale factors also involves the tangents of the two angles for the aspect and the tilt of the ship relative to the radar beam.

The 3D ISAR algorithm does not use the above form of Eq. 2. Instead, it uses a higher order version of the Thin Ship Approximation. And to resolve the need for estimation of the two angles and angle-rates it makes use of additional moments that arise naturally from the two-dimensional regression used in the ISAR focus algorithm. But in all these relations the algorithm does not get any simpler. There is no approximation that would allow one to reduce the problem to the estimation of a single scale factor.

The bottom line here is that the scale factors for the two coordinates cannot be determined from just the ISAR data. It needs both an estimate of the plane of rotation for the time of the image and it needs the center or the mean values of the two angles. In other words, the aspect angle is not needed just for the estimation of the ship length but for any algorithm that attempts to convert the ISAR estimate of Doppler to cross-range in meters.

The main lesson from Figures 3 and 4 is that the mean aspect angle critically determines both the LOA and the angle rates. And in the 3D algorithm the angle rates help identify the image frames that show vertical structure (Profile Views) and those that show horizontal structure (Plan Views) – these are the frames that can allow ATR to identify of the ship. Thus, an improvement in aspect angle can be expected to improve nearly all measurements for a 3D ISAR ATR algorithm.

## II. PREVIOUS WORK

The 3D-ISAR paper discussed 3D-ISAR logic, which works from the output of the ISAR processor. A major output of this method is the identification of the ISAR frames that represent either Profile or Plan views of the ship. The purpose of this identification is to allow an ATR algorithm to compare features in the image to known physical features of the ships [2,3,4].

The 3D-ISAR Paper is an extension of an earlier method for focusing ship images and estimating length [5]. This paper was developed as part of SAIC's contribution to the ONR program on ship ATR called the Small Craft ATR (SCATR) program. During this program the ship length estimates were passed to another SCATR participant, Scott Musman.

Musman [6] developed an ATR algorithm prior to SCATR in collaboration with the Naval Research Laboratory. Then in [7] he led a team that applied it to the SCATR data, using the SAIC ship length estimates for over 2000 frames. He concluded that they are accurate within +/- 10.5 ft for a set of ships: a 47-ft dive boat, the Lois Ann, and Coast Guard cutters of 25 and 33.5 meters long (Point Stuart and Tybee). But they show that accuracy can be improved using a frame selection procedure based on a neural network, achieving an accuracy of +/- 2.3 ft. These errors are smaller than those found in the present analysis – probably because the flat dive boat and the cutters are not as susceptible to shadowing and multipath scattering.

But both the SAIC and the Musman teams assumed that the aspect angle could be measured independently of the ISAR system. The SCATR data was collected from a shore site at the USN NIWC Pacific San Diego and the ships' courses and speeds were measured with portable GPS receivers on the ships. Thus, this data is less relevant to aspect angle accuracy compared to radars without this information but can be used for insight into the algorithms discussed here.

A recent paper [8] presents a method for using 3D ISAR processing to both estimate the aspect angle and identify ship types. In this technique the target is represented by a 3D point cloud. The paper compares two approaches to classification using point clouds of maritime targets. The first approach makes use of features extracted from the point cloud of the target from different perspectives (i.e. side, top and front views) to form three-point density images (PDI). These are then fed into a convolutional neural network (CNN) to classify the targets. The second approach uses an offline target database comprising size and a coarse target silhouette and classifies using a few simple rules. In contrast to the method in the 3D ISAR paper, which assumes only a single aperture, the calculations in [8] develop 3D information via output from three apertures. The present study is concerned mainly with algorithms that can readily be applied to currently deployed systems. Most such systems cannot support the large interferometric baselines required for a direct approach to 3-D based on multiple apertures.

There are also papers on methods of using two receivers, which can be implemented easier than the 3-aperture system above [9,10,11,12]. Like nearly all the length algorithms in the literature, these systems would clearly benefit from improvement in aspect angle estimation by the algorithm presented here. But so far, I have not seen a full engineering discussion of the ranges and operating conditions where an interferometric system could produce 3D ISAR imagery better than a single-aperture radar.

The Maritime Classification Aid (MCA) [13] is a commercial software product that attempts to apply ATR principles to ISAR data and to estimate ship length. The input specs for the MCA give insight into the industry experience on ship ID. First, the MCA requires an estimate of the ship course better than 2 degrees. This is much better than can be obtained from a tracker due to the errors summarized in the Introduction. Second, the MCA requires the aspect angle to be less than 45 degrees, consistent with the elementary theory of the red line in Figure 1. These specs mean that the MCA can expect large errors in ship length and that it will only work on at most half the expected cases.

There are also papers that assume that the aspect angle can be estimated from the tracker [2, 3] and therefore concentrate on the estimation of the RE. The classical paper [2] is a complete ATR method that does not use ship length explicitly. But the algorithm is only tested with ships either heading away from or towards the radar so by construction the absolute value of the cosine of the aspect is one.

A subject closely related to 3D ISAR and aspect angle estimation is the study of cross-range scaling of ISAR data. In fact, since the scaling factor is simply the rotation rate of the target, it is reasonable to consider this subject to be only a synonym for 3D ISAR. See the discussion of Eq. 2 in the introduction for more background on the two rotational degrees of freedom for ship ISAR. For example, [14] is a study of the scaling factor that is very similar to the analysis of [5], a precursor study of the 3D ISAR paper discussed in the introduction. In both papers the rotation rate is calculated as the square root of the derivative of the acceleration with the range coordinate and this derivative is estimated by linear regression in range of the estimates of acceleration from individual scattering centers. But papers [5] and [14] are outdated now since the 3D ISAR paper does a much more detailed analysis of the problem, modeling the rotational motions as both aspect and tilt variations (rotation axes both vertical and horizontal). And it models all relevant terms of the covariance function that appear in the general regression equation for the derivatives of acceleration with respect to both range and Doppler, produces Plan and Profile images, and does consistency checks to evaluate the quality of the answers.

## III. THEORETICAL BASIS OF IAT

This paper describes an algorithm called **'ISAR AutoTrack'** (IAT) that uses the adaptive motion compensation data estimated during the coherent ISAR mode to improve the estimates of the ship aspect angle and thus improve the ship length accuracy.

The basic idea for IAT is that one should examine the full processing stream to estimate the ship's course and speed, not just use either the tracker or the ISAR imagery in isolation.

The IAT algorithm should be a natural and obvious method to one who has studied synthetic aperture radar processing. It depends on two fundamental principles. First, the mean value of the adaptive motion compensation velocity is a measure of the error in the in-range component of the velocity vector of the ship. Second, the apparent mean rate of change of the motion compensation velocity is a measure of the error in the quadratic focus function, which is proportional to the square of the cross-range component of the velocity vector. This is the basis of the classical SAR autofocus map-drift algorithm [15].

But there are important differences between ship ISAR, and the SAR autofocus problems. For one thing, the ship ISAR problem must deal with the effects of ocean waves causing motion compensation signatures that have the potential to swamp the signal used by the IAT to estimate the errors in the ship's course and speed. Also, the method usually used for estimating the rate of change of the adaptive motion compensation velocity (the Doppler acceleration) is based on SAR image formation done twice, each image using one half of the normal integration time. The apparent shift in the two images, the map drift, is a measure of the Doppler shift between the two image halves. This method is not useful for ISAR since an ISAR process produces many complex image frames; and one can simply estimate Doppler acceleration from the time derivative of the adaptive motion compensation velocity. The fundamental rule used to convert the Doppler acceleration, however, is similar between the map-drift algorithm and the IAT. Both algorithms are based on the estimated error in the Doppler acceleration (i.e., the adaptive motion compensation):

$$A * R = -2 * V * u + u^2 \qquad \text{Eq. 3}$$

Where $A$ is the time derivative of the adaptive motion compensation velocity, $V$ is the mean difference between the cross-range component of the aircraft velocity and the cross-range component of the tracker velocity, $R$ is the mean range to the target ship, and $u$ is the error in the cross-range component of the ship. Note that the term that completes the square on the right-hand side is the square of $V$, which is the deterministic motion compensation. This would yield the full formula for the standard SAR quadratic focus function. If this were a map-drift algorithm for a SAR image the negative sign in this expression would be a plus sign – an error in the aircraft velocity has the oppositive sign as an error in the target motion. But this does not change the fundamental mathematics of the problem.

This formula has two fundamental assumptions. First, the displacement of the target over the ISAR dwell time is small compared to the mean range. This is reasonable since, in contrast to the tracker's estimate over many minutes, the ISAR dwell is only perhaps a few tens of seconds. Second, the division of the in-range and cross-range directions from the tracker information is approximately correct. In other words, the radar azimuth pointing is not far off – it may have an error of a degree or two. If it were much larger than this, the radar would not likely be able to get a return. But aside from the platform Doppler effect discussed in Section VIII this has a negligible effect on the IAT solution.

The IAT estimates $u$ from the Doppler acceleration by solving this equation using the quadratic formula. The code chooses the +/- choice in that formula by selecting the plus sign when $V$ is negative and minus sign when $V$ is positive. This choice selects the value that is consistent with the linear approximation that one gets from dropping the quadratic term and avoids the root that yields a value of $u$ that is approximately the same as the aircraft velocity. Note that the IAT uses the full quadratic formula solution because the linear approximation is only valid for cases where the error is small compared to the cross-range component of the aircraft velocity – and this is not true near the azimuth directions close to the bow or tail views where this value is zero.

## IV. DETAILS OF THE IAT ALGORITHM

The IAT algorithm parses the system motion estimates for the platform and the target and uses the ISAR adaptive motion compensation data to improve them.

The fundamental relation in the IAT algorithm is the formula for the deterministic motion compensation velocity, **D**. The formula calculates the radial velocity as a function of time based on a set of parameters given by the tracker or modified by the algorithm.

$$\mathbf{D(t;C,S,B,R,U(t),V(t))}$$

The time variable covers the coherent ISAR window. The parameters are:

**C: The ship's course, assumed independent of time**

**S: The ship's speed, assumed independent of time**

**B: The Initial bearing to the ship**

**R: The initial range to the ship**

**U(t): The eastward component of the platform velocity versus time**

**V(t): The northward component of the platform velocity versus time**

Figure 5 outlines a hypothetical full 3-D ATR algorithm for ship ID using ISAR. The blue arrows denote data flows, understood to be cumulative in the sense that a downstream algorithm can use any of the upstream outputs. For example, the tracker produces estimates of the ship speed and heading and its Lat/Lon position at the start of the ISAR dwell. However, this information would normally be used only in the ISAR Processor and ISAR AutoTrack boxes where it would be replaced by improved versions of these parameters.

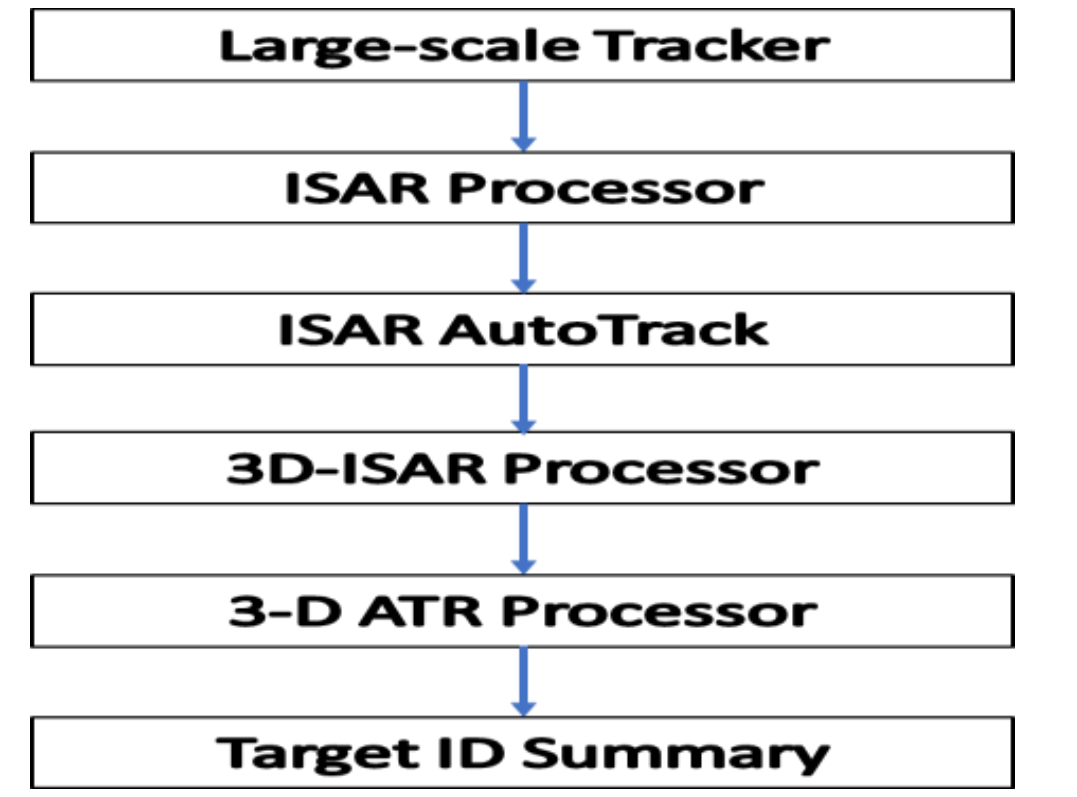

***Figure 5: Processing Stream for a 3-D ATR Processor***

The inputs and outputs of these 6 boxes are:

Large-scale Tracker: Radar data from a non-coherent scan mode that surveys a large area of the ocean at a coarse time sampling (perhaps 5 to 10 seconds). Output of estimates of the ship's speed, heading and Lat/Lon at the start of the ISAR dwell. All of these parameters are subject to correction in downstream algorithms.

ISAR Processor: A series of complex or intensity images that can be displayed in an ISAR movie plus information on the time, range, Doppler and acceleration of 'bright points' in the movie that are selected by the autofocus algorithms. For systems that do not normally output autofocus details, the 3D-ISAR block can derive equivalent information from a stream of intensity images. The output of this stage is the image array plus the series of target reports and the adaptive motion compensation used to maintain the target centered in Doppler.

ISAR AutoTrack: The IAT algorithm uses the above formula defining the deterministic motion compensation process in two ways. First, it calculates $D_0$, the deterministic motion compensation based on the tracker estimates of course and speed, the system values of the initial bearing and range, and the time measurements of the platform velocity. Then it executes a search process for the best estimates of the course and speed. The criterion for the search is a minimum of the difference between D and the total mocomp velocity, defined as the sum $D_0$+AdaptiveMocomp. Here the adaptive mocomp should theoretically include all sources of mocomp applied in the ISAR processor – estimates from an initial acquisition mode, time-dependent mocomp applied to keep the image centered in Doppler, and possibly corrections applied in the Phase Gradient Autofocus (PGA) method. In its present configuration the IAT ignores the PGA contribution, but it contains an option for outputting this information as a diagnostic. The output of the IAT consists of improved estimates of the ship speed, heading and aspect angle and its Lat/Lon position at the start of the ISAR mode.

3D-ISAR Processor: As written in my software the 3-D ISAR Processor has two processing stages. First, it reads the adjusted mean aspect angle from the IAT and the set of target reports from the ISAR Processor and uses these to estimate the ship length. This stage is the culmination of the processes studied in this paper. The second stage of the processing involves algorithms that are described in [1] to do a full 3D ISAR analysis of the ISAR data, including the estimation of the time series of the tilt and aspect angle over time and the construction of composite images for Profile, Plan and 3D image frames.

The last two processors have not been designed or written yet but are included here for completeness. Anyone who wants to implement a fully 3D ATR processor would need to decide whether to accept the improved LOA estimates of the first stage of the 3D-ISAR Processor or to go further. For example, some engineers may decide to just harvest the 'low hanging fruit' of improved ship length without committing to a more difficult software effort necessary to do the full 3D-ISAR Processor or to create new algorithms to exploit the 3-D data in a 3-D ATR Processor.

The main implication of the use of the deterministic formula in both the initial motion compensation and the IAT is that to apply the IAT to an operational system requires knowledge of both the motion estimates and detailed knowledge of the system algorithms for D. Imperfect knowledge of the actual formula for D used in the system will cause errors in the IAT estimate of the aspect angle and the LOA of the ship. For example, a given system algorithm may use either Cartesian or spherical coordinates to estimate the range to the target from the Lat/Lon positions of the aircraft and the ship. And each system may have different values for the speed of light and different methods of accounting for diffraction. The IAT is a sensitive coherent algorithm that needs to know all such details to work properly.

The basic assumptions of the IAT are that the aircraft path and the initial target location from the tracker are correct. The IAT processor searches for or analytically calculates only two values, the course and speed of the ship.

a. The system data provided to the IAT is the aircraft track, an estimate of the initial location of the ship, and the tracker estimate of the ship's course and speed. These parameters yield the **deterministic** motion compensation estimate.

b. During the ISAR processing the system also **adaptively** corrects the deterministic motion estimate with coherent estimates of the ship's range and range-rate versus time using the high-resolution ISAR data.

c. The IAT assumes that the **adaptive** motion compensation data represents an error term to the **deterministic** estimate.

d. The IAT estimate of ship course and speed is formed by a search process over deviations from the tracker course and speed. The program used here currently uses a heading grid of 1441 values spanning the range [-90 to +90 degrees] and a speed grid of 81 values spanning the range of [0.25 to 1.75] times the tracker speed. The search strategy is simple – the motion compensation velocity residual is calculated for every value in the grid. The criterion for convergence is the minimum variance of the residual.

e. The goal of this search is to find a deterministic solution that is as close as possible to the sum of the original deterministic solution and the adaptive solution.

Note that the IAT is fully consistent with SAR principles. The primary principle is that the mean range-rate from the adaptive motion compensation measurement is due to the range motion of the ship. The second principle is based on classical SAR autofocus methods (e.g., the SAR Map-Drift algorithm [13]) – the time derivative of the range-rate (the Doppler acceleration) is a measure of the error in the cross-range component of motion. In strip-map SAR this is assumed to be an error in the platform motion; but here it applies to the unknown target motion since today we can assume that the platform motion is well-measured by its Global Positioning and Inertial Navigation systems.

In addition to the Search method described above, the code for the IAT algorithm includes an Analytical method that explicitly estimates the motion following the formulation described in Section III.

Both methods are preceded by a Data Preparation algorithm that tries to minimize the effects of radar interference and ocean waves. The outputs of this stage are the mean and the time derivative of the adaptive motion compensation velocity calculated by linear regression after the time series is passed through a smoothing process. The smoothing includes a median filter to remove rare events such as pulses corrupted by radar interference and then a normal three-point smoother to reduce random noise. To avoid the errors due to ocean waves the smoother also removes an adaptive narrow-band fluctuation using a chapeau function routine to estimate the phase of the oscillation over time. In this method a first estimate of the center frequency of the oscillation is created and then the phase of the oscillation is modified with time by comparing the local phase relative to the carrier using a window function to isolate a portion of the time series. The final answer is the center tone multiplied by the time varying phase. However, I have only tested this with a couple of wave signatures so I cannot argue that this is the only way to do the problem – I fully expect that anyone implementing the IAT would experiment with different narrow-band filters.

A confounding problem for both the Search and Analytical methods is that the tracker produces errors in course and speed that they can readily estimate; but it also has errors in the initial ship geographical position. This error also has in-range and cross-range components. The in-range component causes a mean range shift that can be corrected in the ISAR acquisition process. However, the error in cross-range position yields a mean motion compensation velocity equal to the product of the angle displacement and the cross-range component of the platform velocity. Since the platform speed is normally much larger than the ship speed, this can cause an error in the mean motion compensation velocity that competes with the use of the mean to estimate the in-range velocity error. The error due to position is much less for azimuth angles near the forward and backward look directions since the cross-range value of the platform velocity is smaller there.

Fortunately, the error in cross-range position can be reduced for all azimuth angles by using an advanced radar data collection process to measure this value directly and correct for it. There are two basic methods for this – (1) varying the antenna azimuth during the data collection to find the direction of maximum returned power and (2) using multiple apertures to improve the estimate of the bearing to the ship. These are standard methods used in the ISAR community – for example, 30 years ago I participated in a design study that used the first one to create an adaptive pointing algorithm to keep the target in the fat part of the beam. None of the engineers involved thought that this was difficult. Although the IAT algorithm can still greatly improve the accuracy of aspect angle and ship length without either of these methods, the gains in performance for improved bearing measurement are considerable.

It is also worth noting that there is a simple operational method that can ensure that the map-drift equation is put into the linear mode. If there is a very important target ship dead ahead of the radar, one could just turn the aircraft 30 degrees or so to the right or left during the ISAR mode, generating a cross-range velocity of half the aircraft velocity.

## V. TEST OF THE IAT WITH MSR2 DATA

The Inverse-SAR Processor used here was initially developed in 1994. Since then, the algorithm has been applied to multiple ISAR systems. In the last 30 years I have collected or been sent many test results. I have permission to use these data sets for scientific work and algorithm development but not to publicize the systems or the data owners. The data used here is from flight tests of a relatively modern X-band test radar that I call Marine Surveillance Radar #2 (MSR2). All this data is from commercial ships – mostly container ships but a few tankers, bulk carriers and car carriers. The data has a range resolution of 1 meter and a PRF of 512 Hz. Range compression is done via stretch processing. All ISAR images and calculations presented here were done with my own Fortran and Matlab software from the raw data.

The MSR2 data had limited information on the system power and other variables needed for the radar equation and thus I could not estimate the radar cross sections of the ships. Fortunately, the data had a lot of information in auxiliary files, and I was also given software to sort the raw AIS files by MMSI number. All other information had to be derived from the data. I selected a group of 39 cases that had both tracker and AIS data. For general quality I used on the Signal-to-Clutter Ratio, defined as C/(1-C) where C is the pulse-pair correlation coefficient. My threshold for SCR was set at 0 dB. This eliminated some smaller targets and some cases where the tracker position error was so large that the beam missed the target. I also removed one case where the ISAR imagery clearly showed a large container ship but the MMSI number for the case was for a 30-foot sailing vessel. And I eliminated 3 cases for U. S. combatant ships since these might be sensitive. For all 22 remaining cases I viewed the ISAR movies and examined the number of and the SNR values of the scattering centers identified by my Spatially Variant Autofocus and Phase Gradient Autofocus algorithms. The ISAR output showed that there were many scattering centers for each case that met my normal SNR threshold. I concluded that this ISAR data was of high quality. I concluded that the adaptive mocomp data in the auxiliary files was reasonable because my own ISAR processor identified only small corrections to it. I concluded that the absence of the radar equation parameters did not harm my ability to evaluate the quality of the data.

However, the deterministic mocomp in the auxiliary files was not consistent with the variables used to create it; and I needed both the data result and the formulas for computing it. Thus, I wrote my own algorithm for this from the tracker output and the aircraft velocities using an earth tangent plane and Cartesian coordinates. This may have caused errors in aspect angle due to the inconsistency error caused by the difference between my estimate and that from the real-time software. But this inconsistency error cannot be too large since when I replaced the position at the start of the ISAR mode from the tracker Lat/Lon estimate to the true Lat/Lon from the ship AIS the RMS aspect error dropped from 8 to 4.6 degrees. In this process I also found that the differences between the two measurements had an RMS value of about 1 km, which seems to confirm that the system had correctly identified the targets.

For these 22 cases the RMS tracker error in course or aspect is 26.6 degrees and the RMS error in speed is 4.0 m/s (7.8 knots).

Figures 6-9 illustrate the basic motion compensation data that drives the IAT for 4 of the 22 cases selected to cover the variety of ships and results in the full data set. The smoothed value of the motion compensation velocity is the default for the algorithm. As expected, the plots show errors in the mean motion compensation velocity, indicating error in the in-range component of the ship velocity; and errors in the mean slope of the curves, indicating error in the cross-range component.

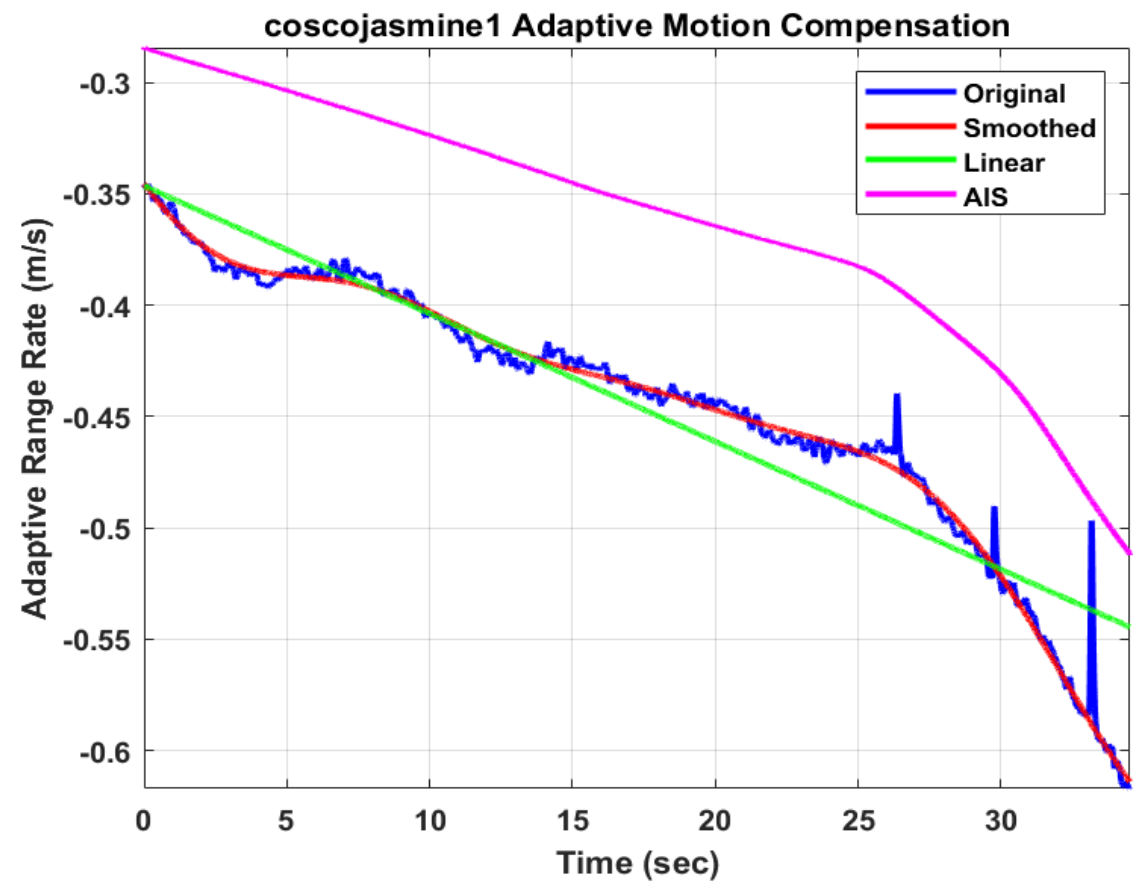

***Figure 6: Analysis of the adaptive motion compensation data for the ship case coscojasmine1***

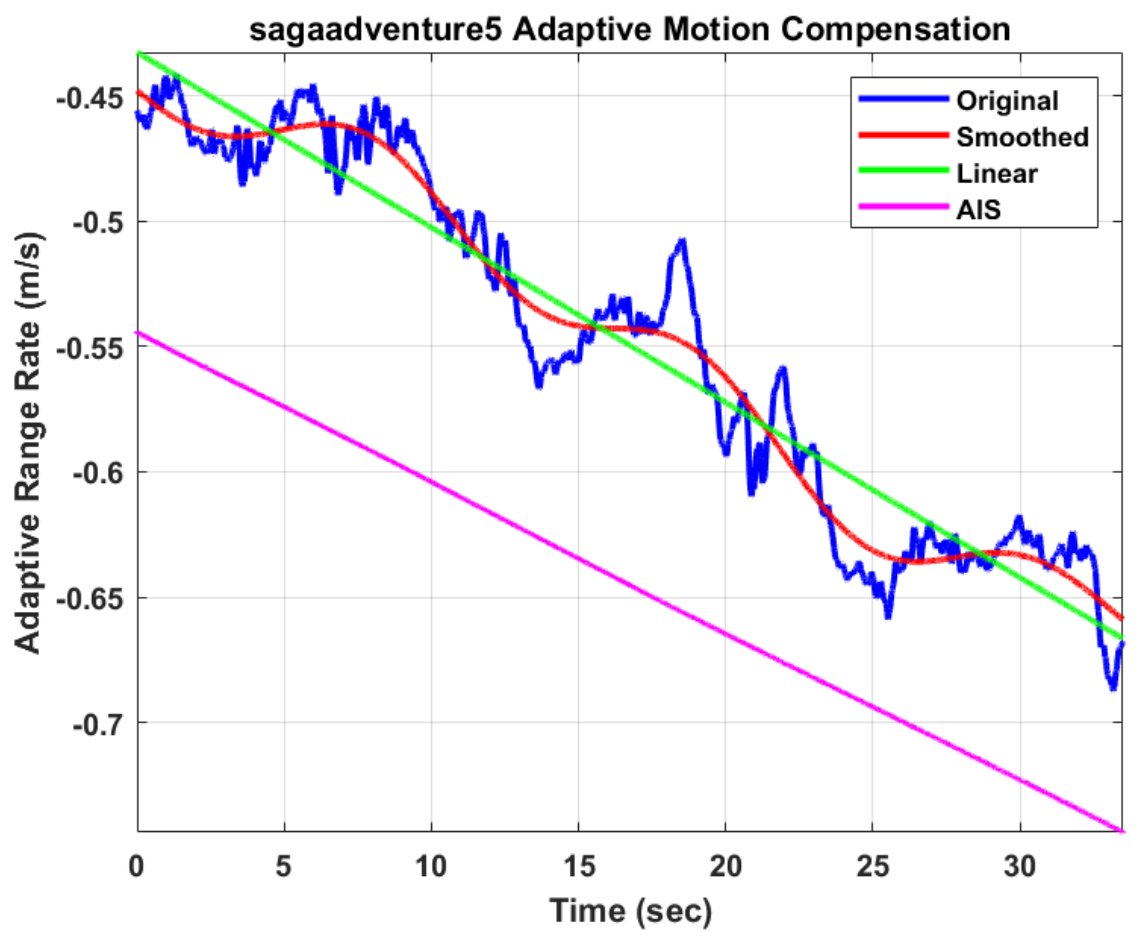

***Figure 7: Analysis of the adaptive motion compensation data for the case sagaadventure5***

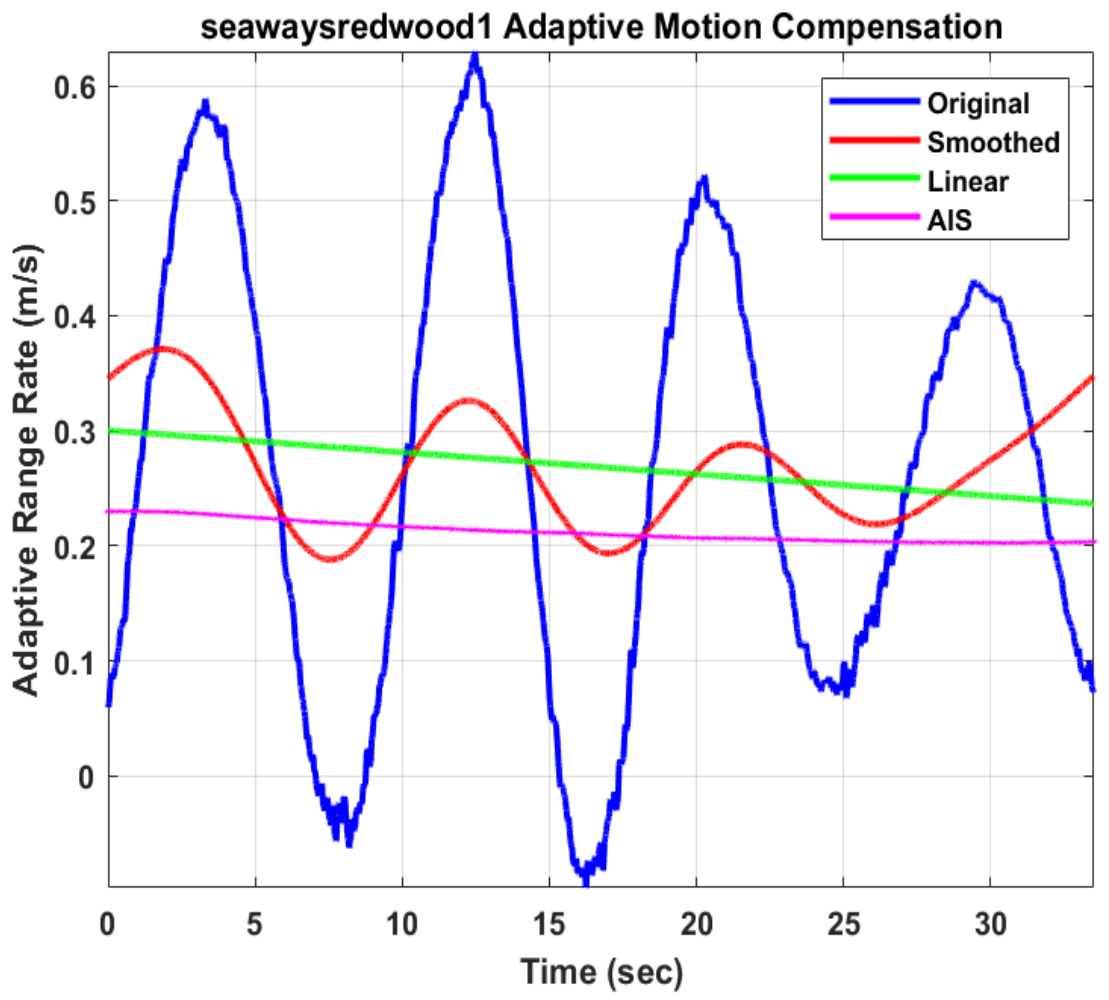

***Figure 8: Analysis of the adaptive motion compensation data for the case seawaysredwood1***

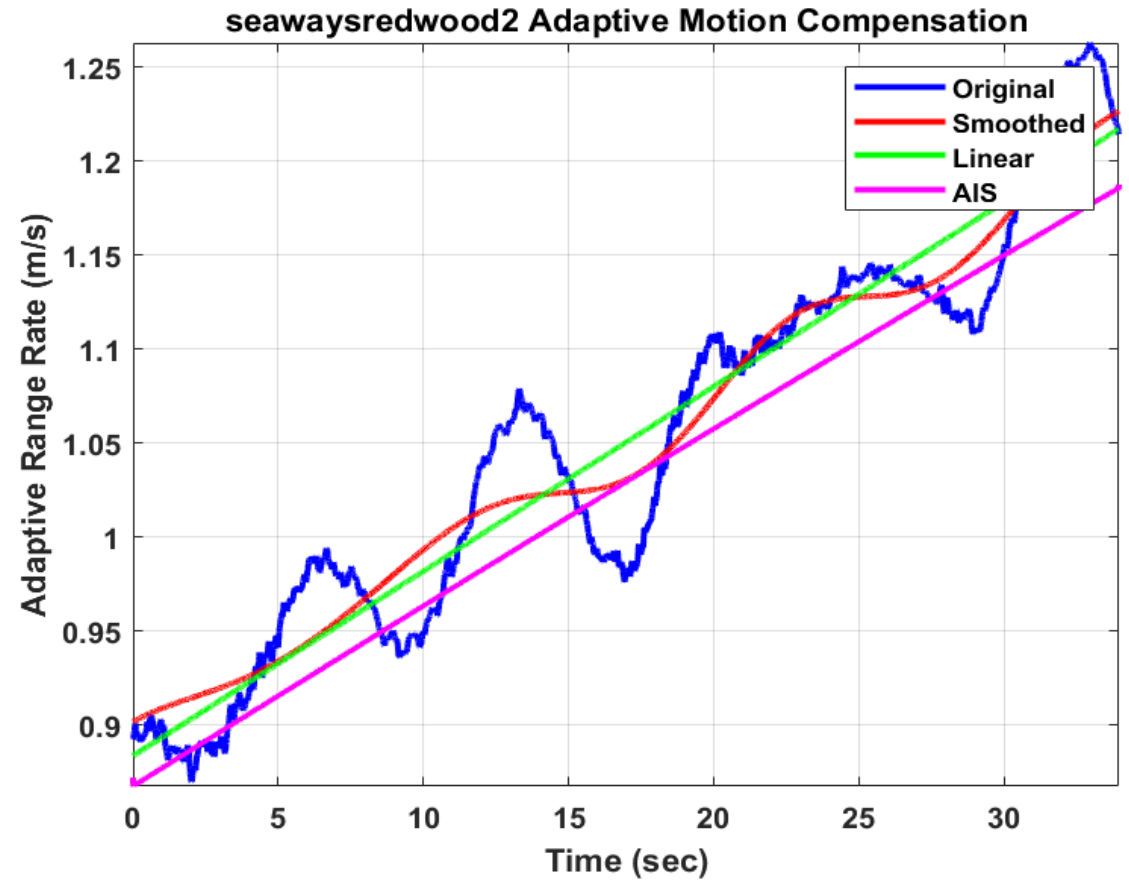

***Figure 9: Analysis of the adaptive motion compensation data for the case seawaysredwood2***

Figures 10-13 plot the error score versus the heading offset for these 4 cases. These plots of error versus heading offset are done at the computed optimum value of ship speed.

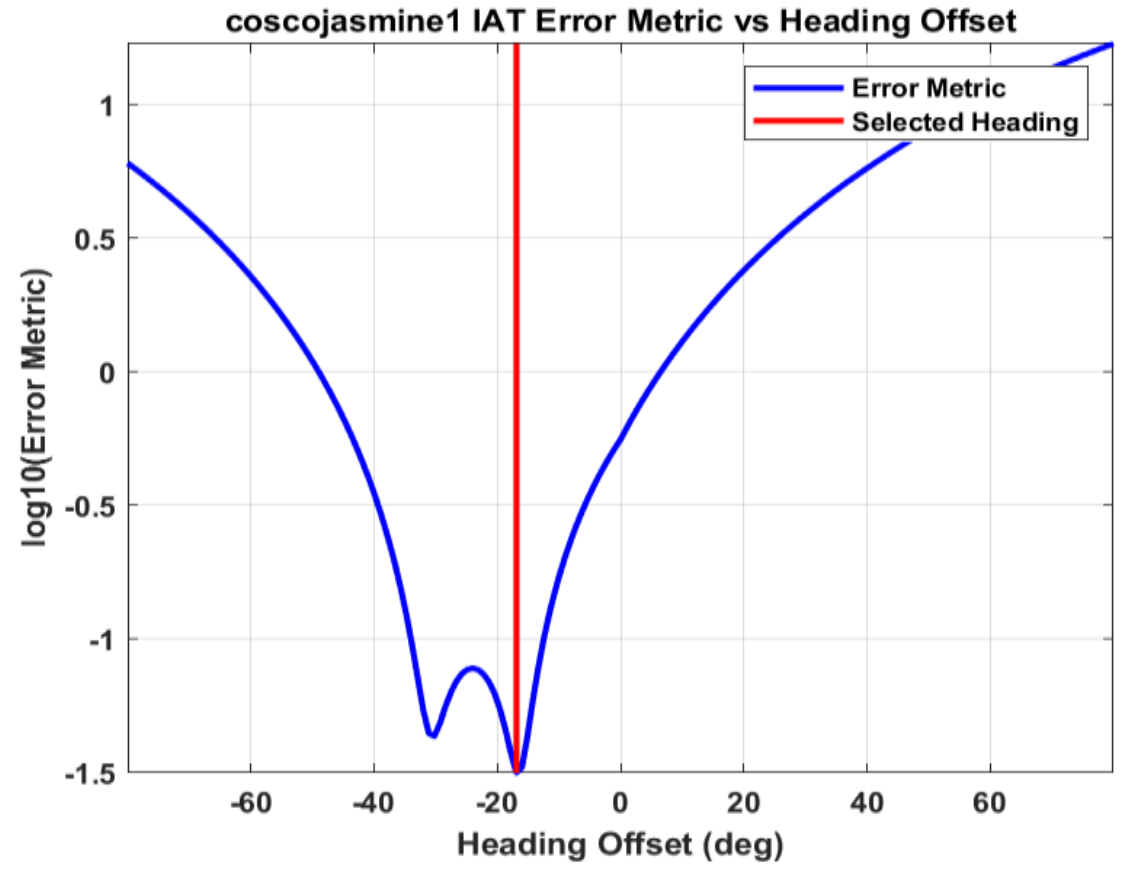

***Figure 10: Error metric versus heading offset for coscojasmine1***

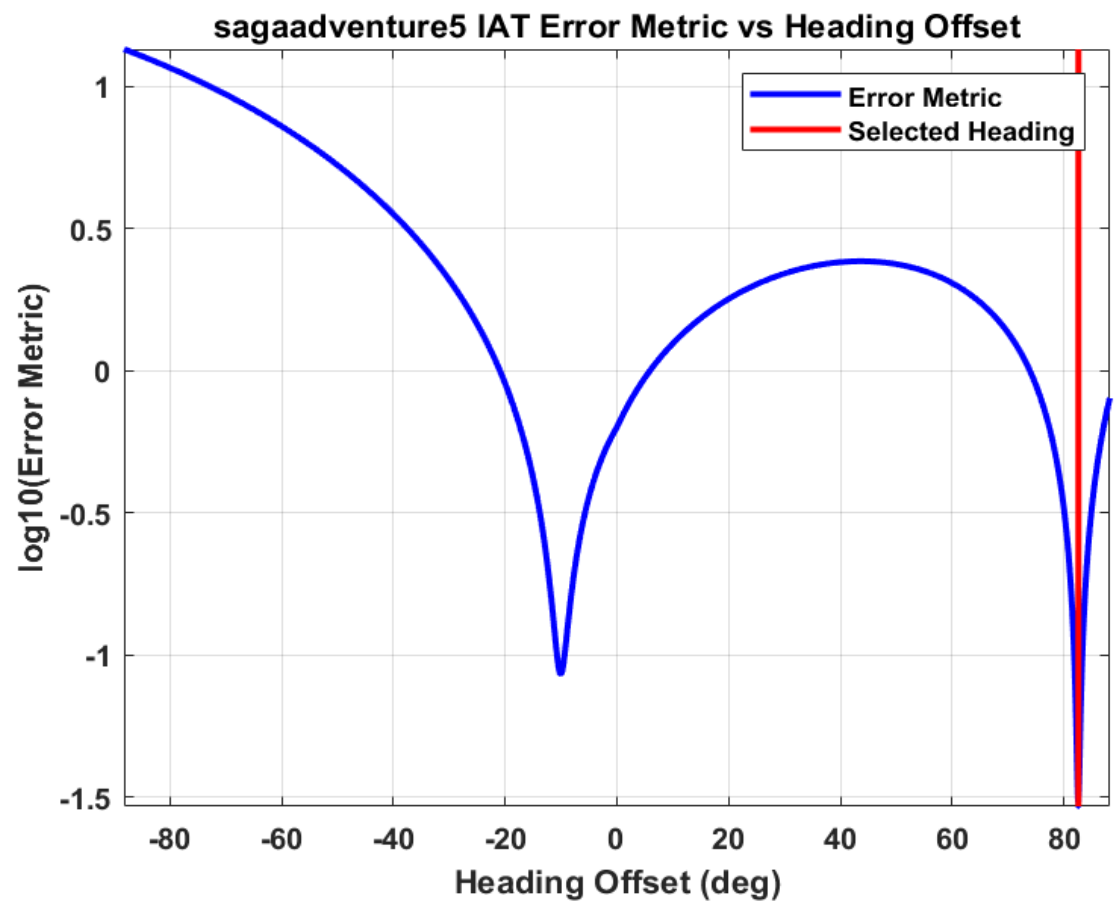

***Figure 11: Error metric versus heading offset for sagaadventure5***

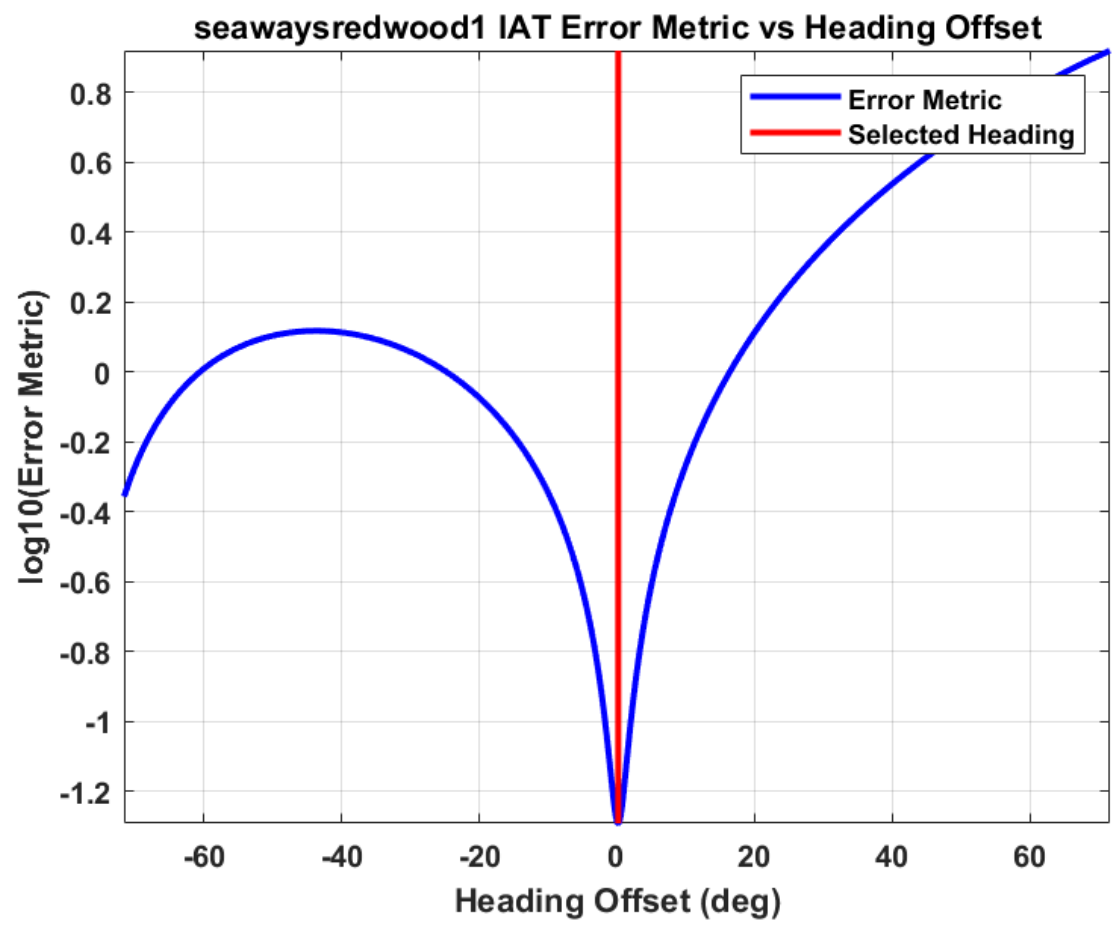

***Figure 12: Error metric versus heading offset for seawaysredwood1***

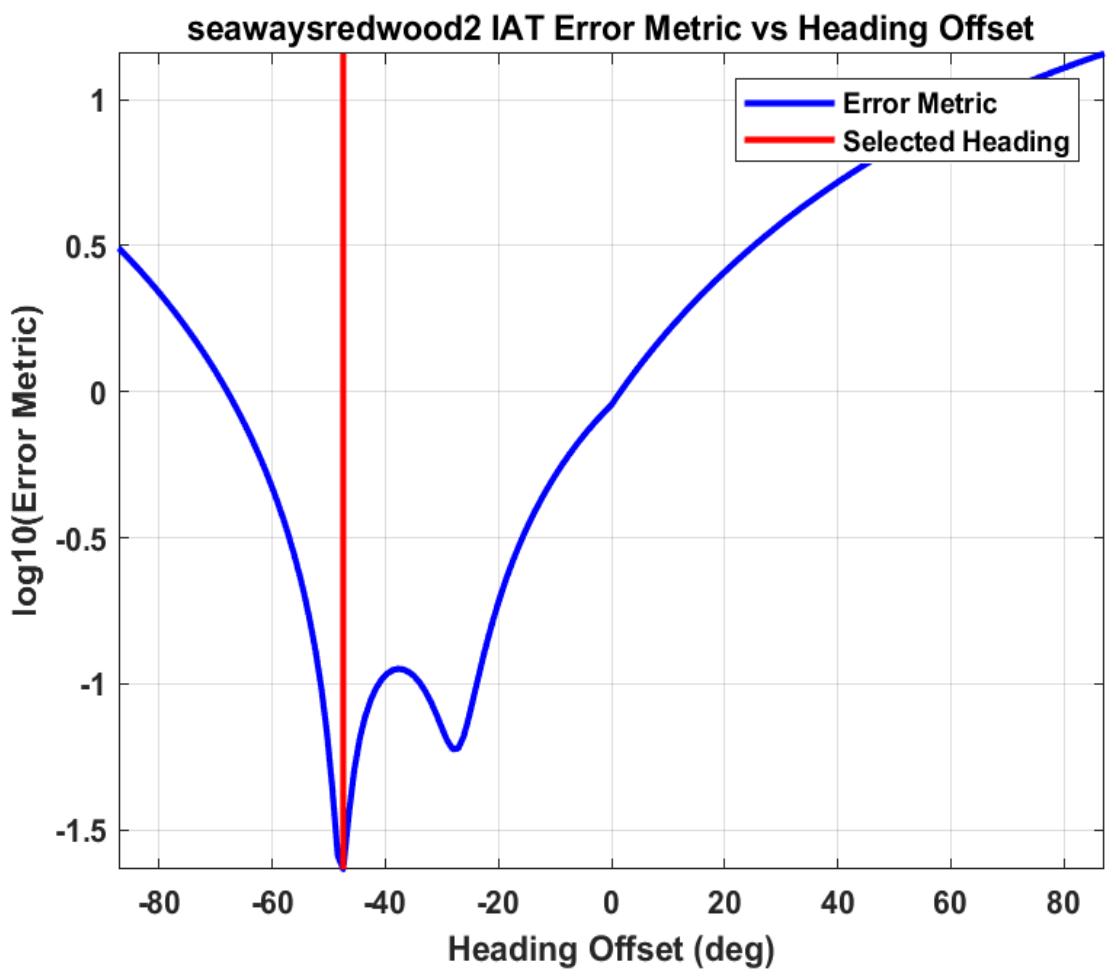

***Figure 13: Error metric versus heading offset for seawaysredwood2***

The IAT has an RMS heading error of 8.04 degrees versus 26.56 degrees for the Tracker. The ship speed error has also improved from 4 m/s to 1 m/s. The details are:

```
Tracker RMS Heading Error (deg):           26.56
Tracker Median Heading Error (deg):        15.24
Tracker RMS Aspect Error (deg):            26.27
Tracker Median Aspect Error (deg):         15.05
Tracker RMS Speed Error (m/s):              4.02
Tracker Median Speed Error (m/s):           0.77

IAT RMS Heading Error (deg):                8.04
IAT Median Heading Error (deg):             5.11
IAT RMS Aspect Error (deg):                 8.08
IAT Median Aspect Error (deg):              5.20
IAT RMS Speed Error (m/s):                  1.00
IAT Median Speed Error (m/s):               0.63
```

The improvement in the RMS error in course and speed for the IAT is very good but the errors have a long-tailed distribution – there are a few ship cases with large errors.

I noticed that there are sometimes close alternate solutions – these are handled by penalizing solutions that deviate too much from the tracker solution … this penalty is only about 3 dB at an angle offset of 90 degrees from the tracker heading.

Examination of the plots shows that the IAT can correct heading errors of over 80 degrees. But the residual errors after applying the IAT can still be large – up to 22 degrees for a case like seawaysredwood1, which has large ocean waves.

## VI. CONVERSION TO SHIP LENGTH

Focus3D has a robust estimator for ship length. The fundamental method is the Thin Ship Approximation. It also has a method for reducing the effects of multipath scattering and a correction for ship width that uses a standard naval architecture Rule of Thumb relating length to width for ships. It computes estimates of length for each frame of the ISAR movie using time-dependent estimates of RE and aspect angle and then forms the final estimate as a double median of these estimates (i.e., the median of those estimates in the top half of the raw estimates).

I converted the aspect angle results for the 22 cases to length errors using the Focus3D LOA algorithm:

| | | |
|---|---|---|
| Tracker LOA Error: | 88.4 m | (35.3%) |
| IAT LOA Error: | 40.1 m | (16.0%) |
| AIS LOA Error: | 34.5 m | (13.8%) |

The Appendix gives a spreadsheet that contains the important parameters of all 22 cases with the calculation of the above statistics. Since the AIS aspect angles are presumed to be correct, it appears that the IAT result is within a dB or so of the desired minimum error of 10% but above the 8% error for the 43-case data set of Section III. Note that the IAT error is a combination of the Inconsistency Error of Section IV and the error in the Range Extent, which is due to the normal physical issues (shadowing, multipath effects, target fading, weak scatterers at the ends of the ship). This would likely be lower if the algorithm for the deterministic motion compensation were known. As with the distribution of errors in aspect, the RMS error in length is dominated by a few large values – thus all these error estimates are sensitive to the choice of cases.

## VII. DISCUSSION OF THE WORST LOA CASES

There are two main purposes for examining the worst cases. First, we want to understand the errors to exclude some cases in the statistics used to evaluate the IAT. For example, I decided to ignore 3 cases that have aspect angles less than our lower limit of 5 degrees. However, studying these cases gives insight into this parameter region. The second purpose is to identify the best methods to improve IAT. Thus, I will discuss some cases that are included in the final evaluation statistics that still have large LOA errors. These cases illustrate the problems we need to solve to improve the algorithms.

As previously noted, the distributions of the errors in either angle or LOA are long-tailed – the RMS error is dominated by a few cases with large errors. For this data set the following table summarizes the errors for 4 cases that have large LOA errors (meters):

| Case Name | Tracker | AIS | IAT |
| --- | --- | --- | --- |
| TransibBridge2 | +113 | +100 | +102 |
| AtlanticSail1 | -68 | -81 | -80 |
| SeawaysRedwood1 | -92 | -12 | -93 |
| Shouchenshan1 | +44 | +64 | +56 |

To understand these large errors, we need to examine the details of the results.

Transib Bridge is an oiler with a normal deckhouse configuration at the rear (Figure 14) and it heads towards the radar. This case has an AIS aspect angle of 4.2 degrees.

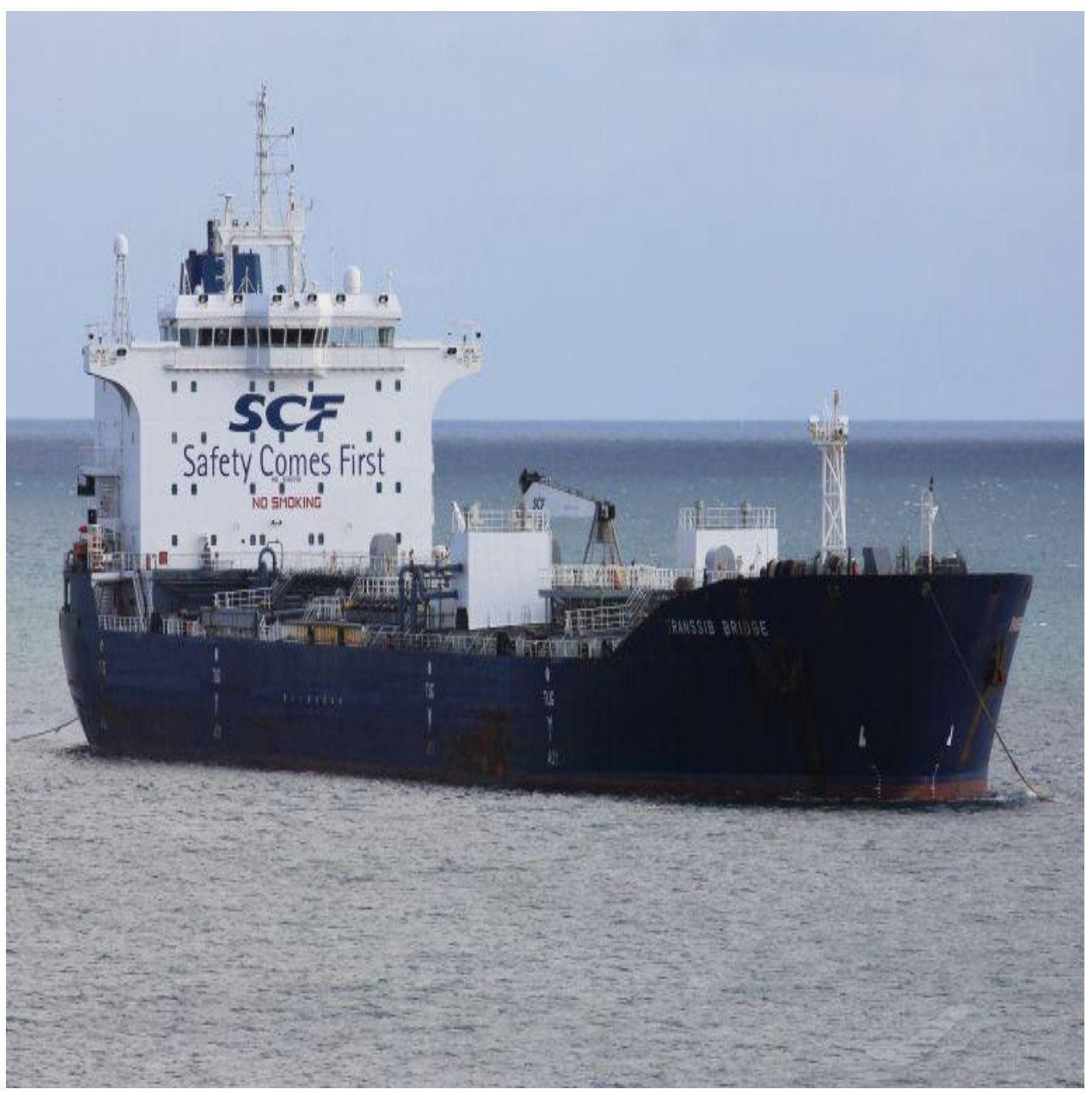

*Figure 14: Oiler Transib Bridge*

Atlantic Sail is an unusual ship that is a combination of a container ship and a RORO (Roll-On Roll-Off) with a strange deckhouse configuration midship (Figure 15). This case has an AIS aspect angle of 3.5 degrees.

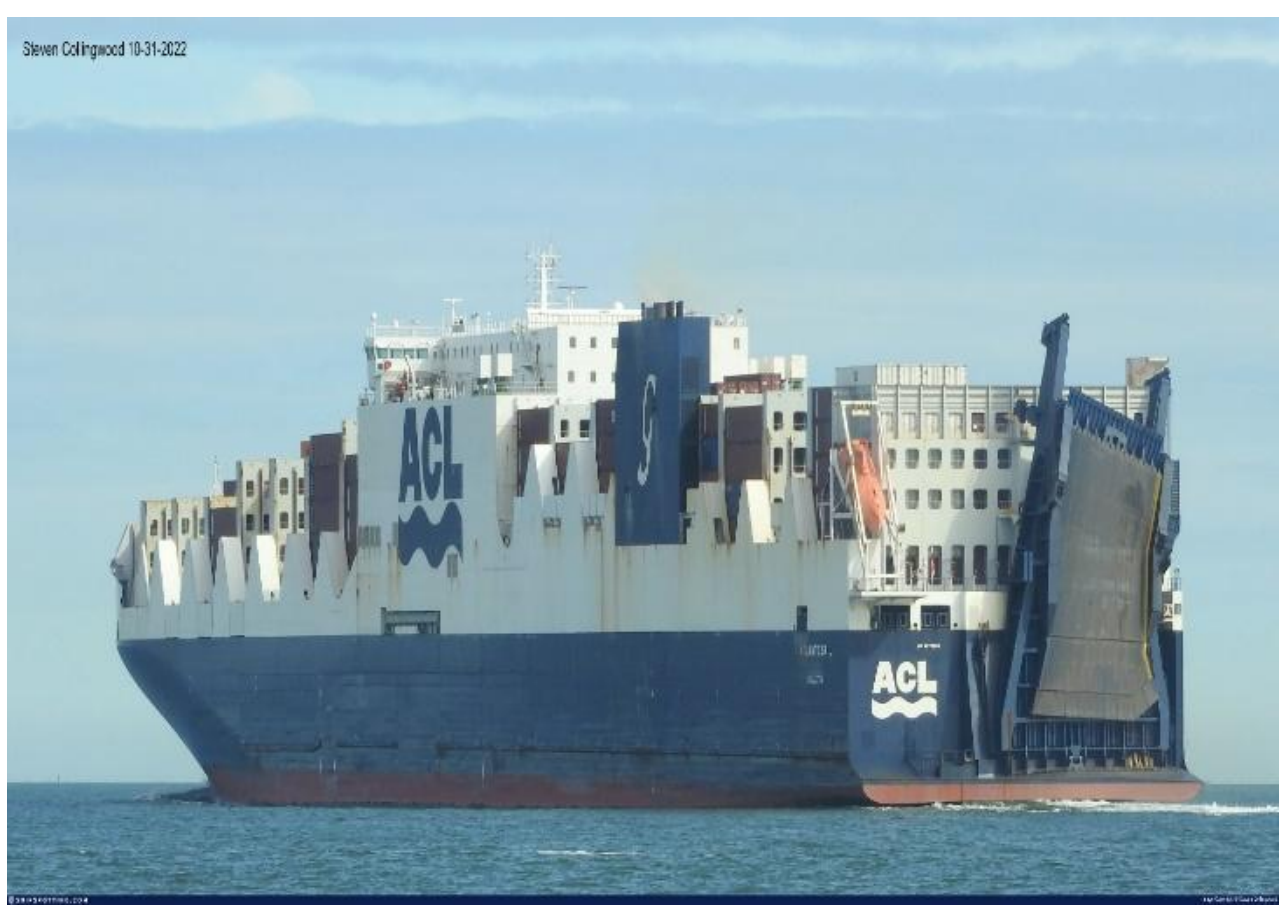

*Figure 15: Container/RORO Atlantic Sail*

Seaways Redwood is an oiler with a normal deckhouse configuration at the rear (Figure 16). This case has an AIS aspect angle of 60 degrees.

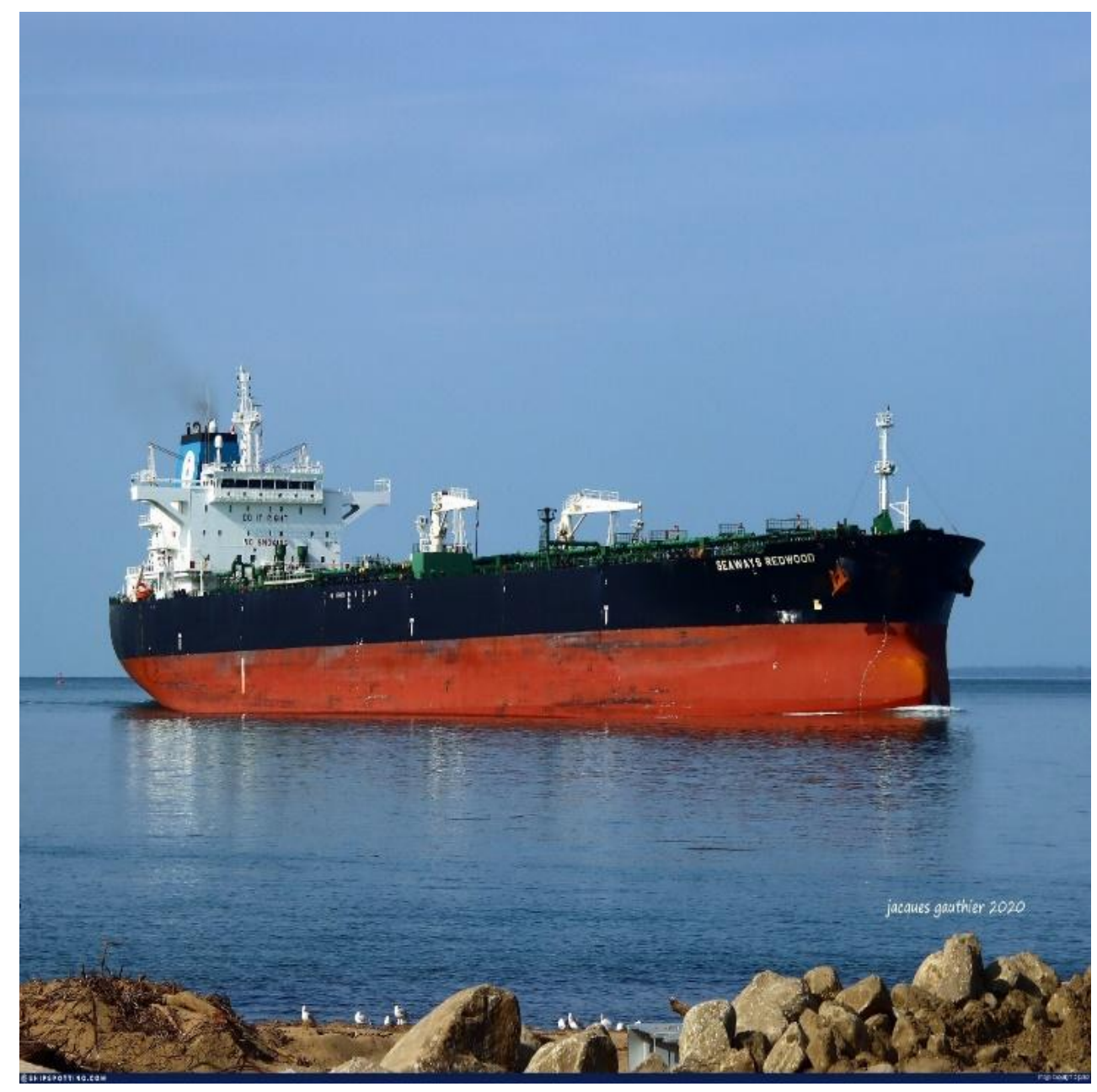

*Figure 16: Oiler Seaways Redwood (Alt: Overseas Redwood)*

Shou Chen Shan is a bulk carrier with 4 large cranes heading toward the radar. This case has an AIS aspect angle of 28 degrees.

First let's look at the two low aspect cases, Transib Bridge and Atlantic Sail. One has a large positive LOA error and the other has a large negative LOA error. The ISAR images for these cases (Figures 17 and 18) show very little Doppler extent, indicating that the motion compensation has worked well and that there is very little wave action. And at low aspect angle the variation in the cosine term in the Thin Ship Approximation is small. So, the main determinant of the LOA error is the Range Extent (RE).

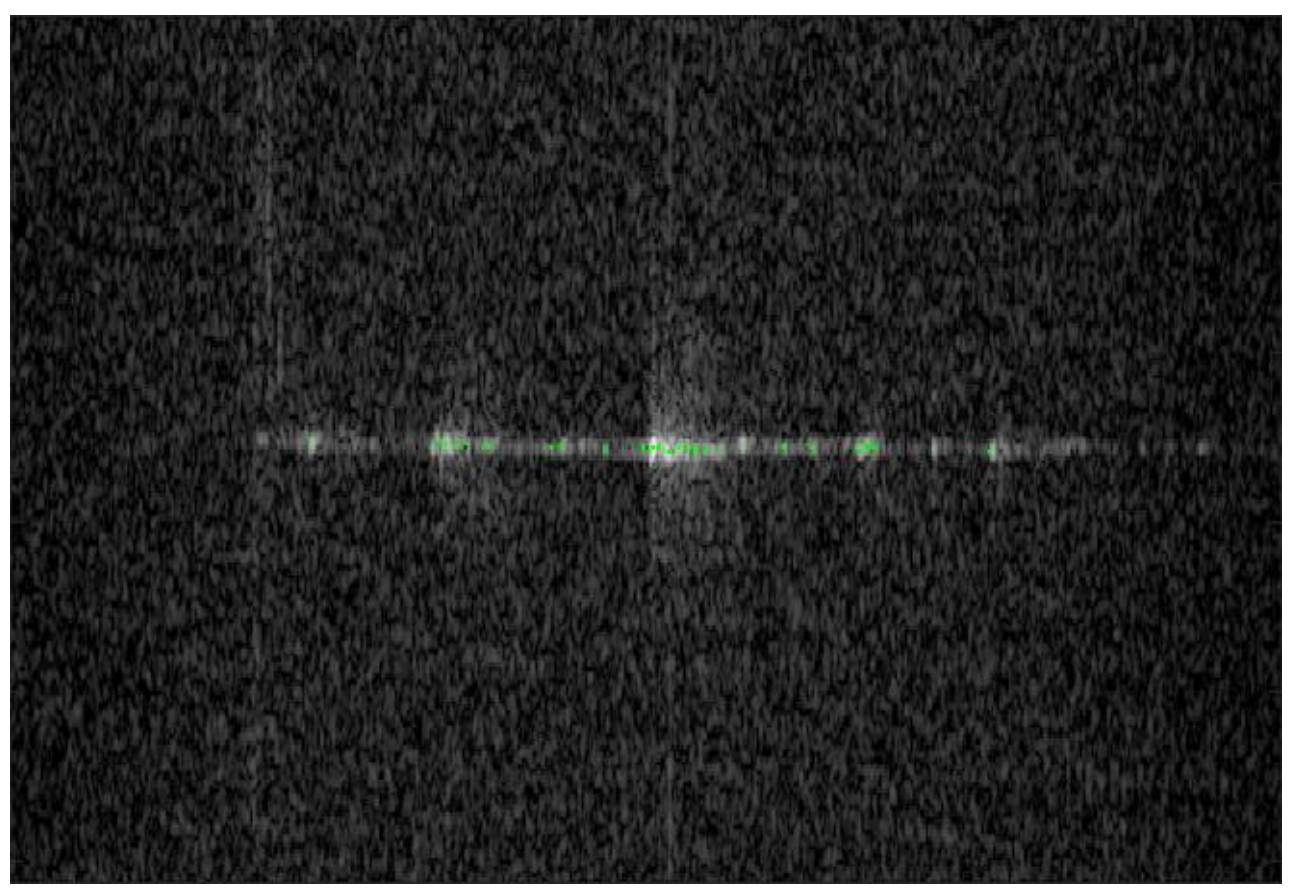

***Figure 17: ISAR image for transibbridge2***

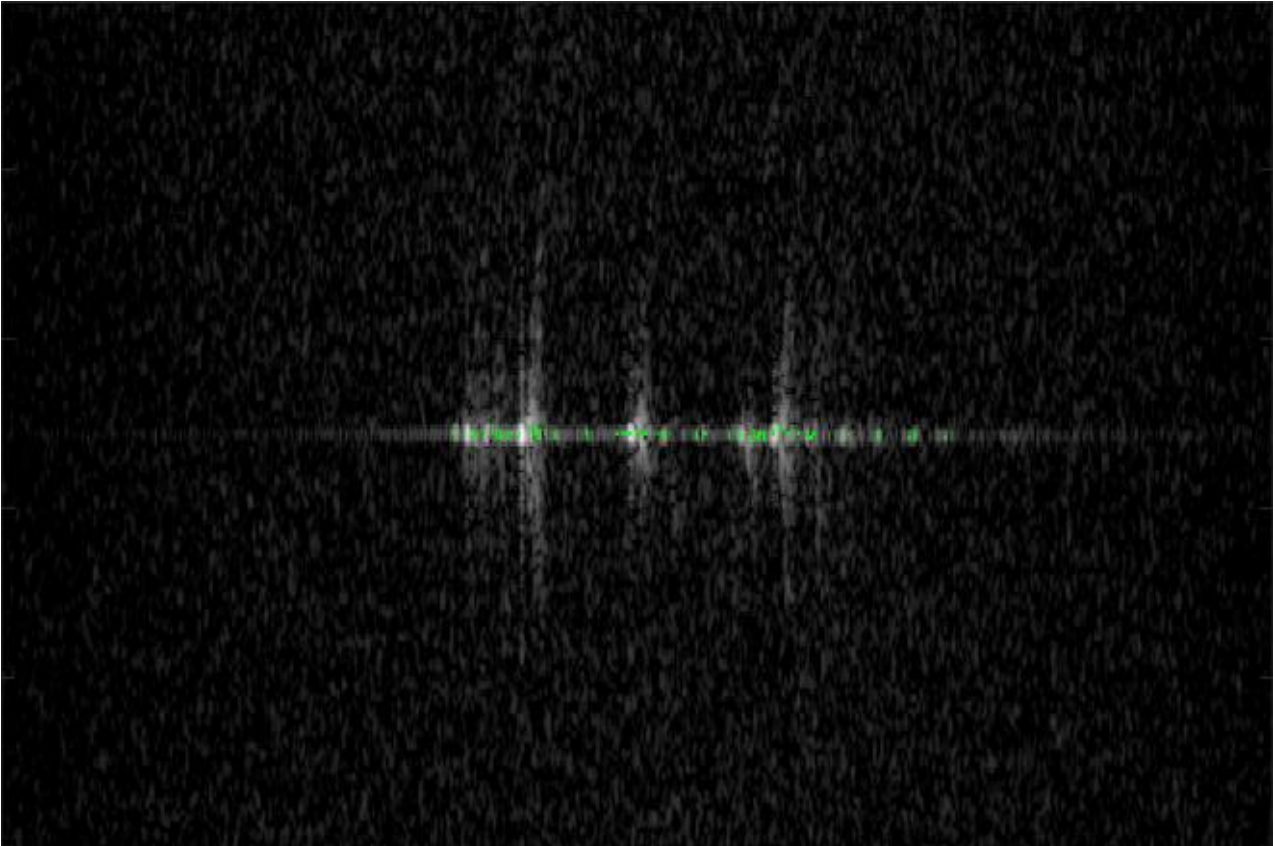

***Figure 18: ISAR image for atlanticsail1***

My analysis is that the Transib Bridge overestimation of the LOA is due to multipath scattering. This effect is commonly seen in ships heading towards the radar at low aspect angle. The physical scattering process is probably that the radar beam first bounces off the large 'deckhouse mirror', then off one of the pieces of machinery on the deck, then back to the mirror before returning to the radar. This creates an image of the machinery at an increased range equal to the distance the machinery is in front of the mirror. This hypothesis involves three bounces of the radar energy; however, two of the bounces are likely to have negligible spreading error since they are bounces off a mirror-like feature. In this case there are multiple instances of this process for various scattering centers in front of the mirror, yielding an LOA error of over 100 meters for all aspect methods. The small differences in LOA for the three methods are simply due to the small differences in the aspect angle estimates. It is difficult to assign the multipath pattern to the structures on the Transib Bridge since there are many potential large RCS structures visible in the photo. This may be possible for ships with very large regular structures like those on bulk carriers with cranes.

By contrast, the Atlantic Sail has a deckhouse midship and possibly, depending on the load, a lot of containers both in front of and behind the deckhouse. The deckhouse and containers effectively shadow any scattering from the parts of the ship behind the deckhouse, causing a significant underestimation of the length.

Interestingly, in the 22-case set there is a third case (colomboexpress2) with an even lower aspect angle, about 1 degree. The Colombo Express is a container ship, and most web photos show it fully loaded. Thus, the deckhouse mirror is largely blocked by containers and the main effect on LOA is the small shadowing of the far end of the ship, yielding a small underestimate of the LOA of about 32 meters (10 percent). To be consistent with the intent to apply the LOA algorithms to only ships with aspect angles greater than 5 degrees, I excluded Colombo Express from the final evaluation along with Transib Bridge and Atlantic Sail.

Turning now to seawaysredwood1, the third case of large LOA error. The AIS aspect angle for this case is 60 degrees so the problems at low aspect do not apply. Instead, this may be an example of the large ocean waves clearly visible in Figure 8 causing errors in the motion compensation that affect the accuracy of the IAT. The IAT has a narrow band filter to try to deal with this but the large difference in the values at the beginning and end of the time series make this task difficult. The IAT estimate of the aspect angle is 38 degrees, 22 degrees too small – enough to cause an error of the secant in the Thin Ship Approximation of 40 percent.

Another aspect of the Seaways Redwood results is that there is a lot more Doppler information than for the low aspect cases. The image frame in Figure 19 shows the Doppler of the superstructure plus some large scattering centers at the bow and stern that have the opposite Doppler excursion compared to the main superstructure response. These scattering features are probably due to scatterers below the average height of the ship. The one at the bow is the anchor that is secured at the port or starboard hawse pipe and the one at the stern is associated with the rudder or similar feature near the waterline. The anchor is clear in the photo of Figure 16. Other photos on the web show features on the waterline at the stern.

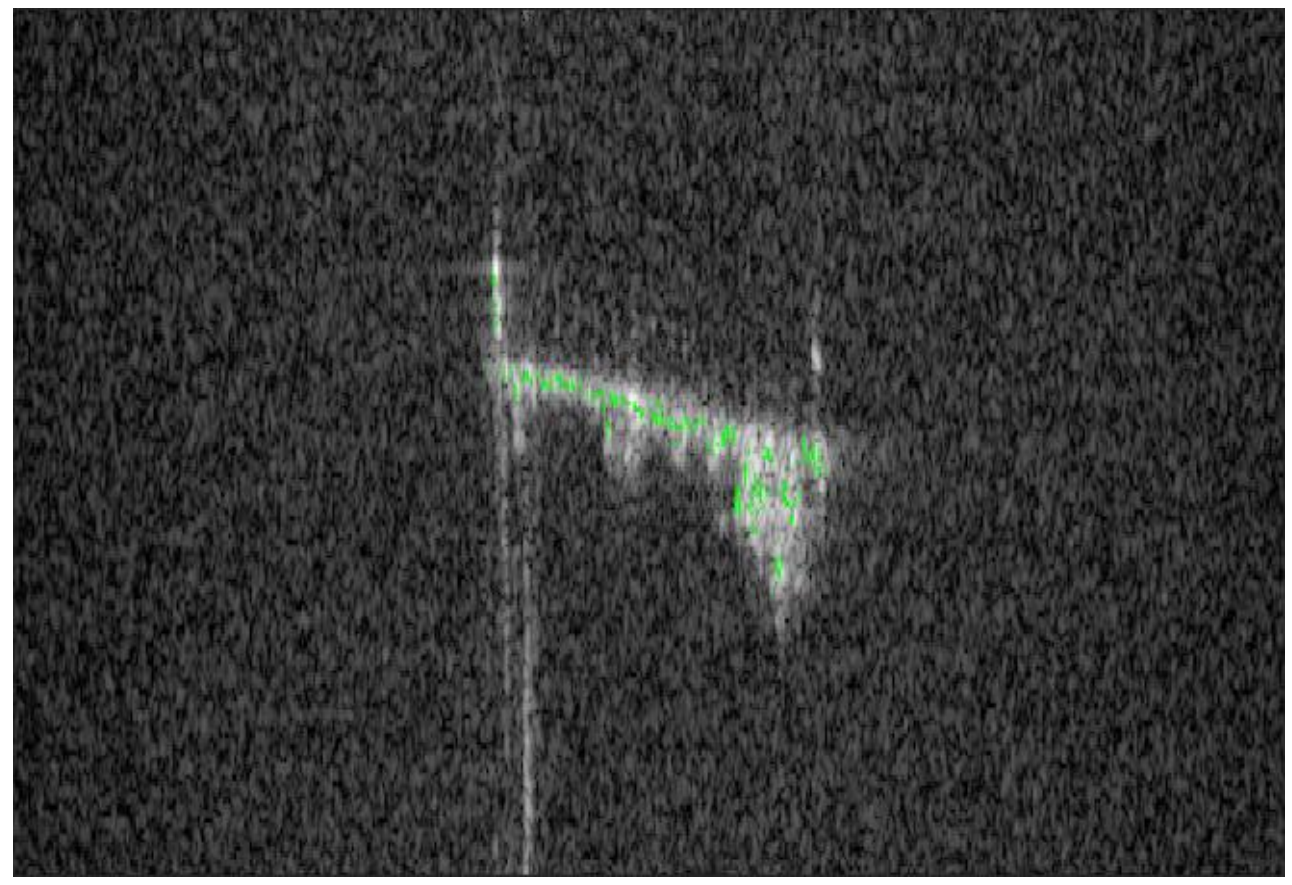

***Figure 19: ISAR image for seawaysredwood1***

The seawaysredwood1 case is only one example so it is tempting to ignore it. However, this case is the one with the largest ocean wave response and it thus represents an important part of the expected operational regime for maritime ISAR. We do not want to design an algorithm that just works for calm seas. Therefore, I followed up on the case by doing a deep dive into the data.

First, I checked to see if the ocean wave signature was responsible for the error in the IAT aspect. I isolated this signature and inserted it into a simulation that covers the aspect angles from 5 to 85 degrees. But the IAT program did not suffer any major error due to this moderate wave motion. The RMS course error was about 5 degrees – and about the same as if I used the noise measurement of any of the other 21 cases.

Then I noticed that the data seemed to indicate that the ship had a large cross-range offset that yielded a course that better agreed with the AIS report. Also, this case had one of the largest values of the mean adaptive motion compensation velocity. These effects are consistent with a large cross-range offset in the target cue.

Due to the problems of multipath and shadowing for the estimation LOA at very low aspect angles it is reasonable to avoid this region. This is difficult in a real radar system because this decision may have to be made based on the tracker estimate of aspect, which may not be very accurate. Therefore, I envision that the process would make any culling decisions for aspect angle based on the IAT answer.

The wave effects can possibly be reduced by using the results of the full Focus3D analysis, which makes estimates of the variations of aspect and tilt angles from the ISAR imaging process. Figure 20 shows the spectrum of the two angles estimated by Focus3D. The tilt angle spectrum has a clear peak at about 0.125 Hz (8-second period). This peak in the tilt angle spectrum is 18 dB above the peak for aspect angle – this means that the cause is likely to be a roll motion of the ship in response to an oblique angle to the wave direction. This is about the same as the period in the motion compensation plot, Figure 8. These two estimates of the wave period use different methods – one uses the bulk motion of the ship and the other uses the range-Doppler covariance functions of the target reports. Thus, there is a reasonable expectation that this error can be reduced in the full Focus3D algorithm.

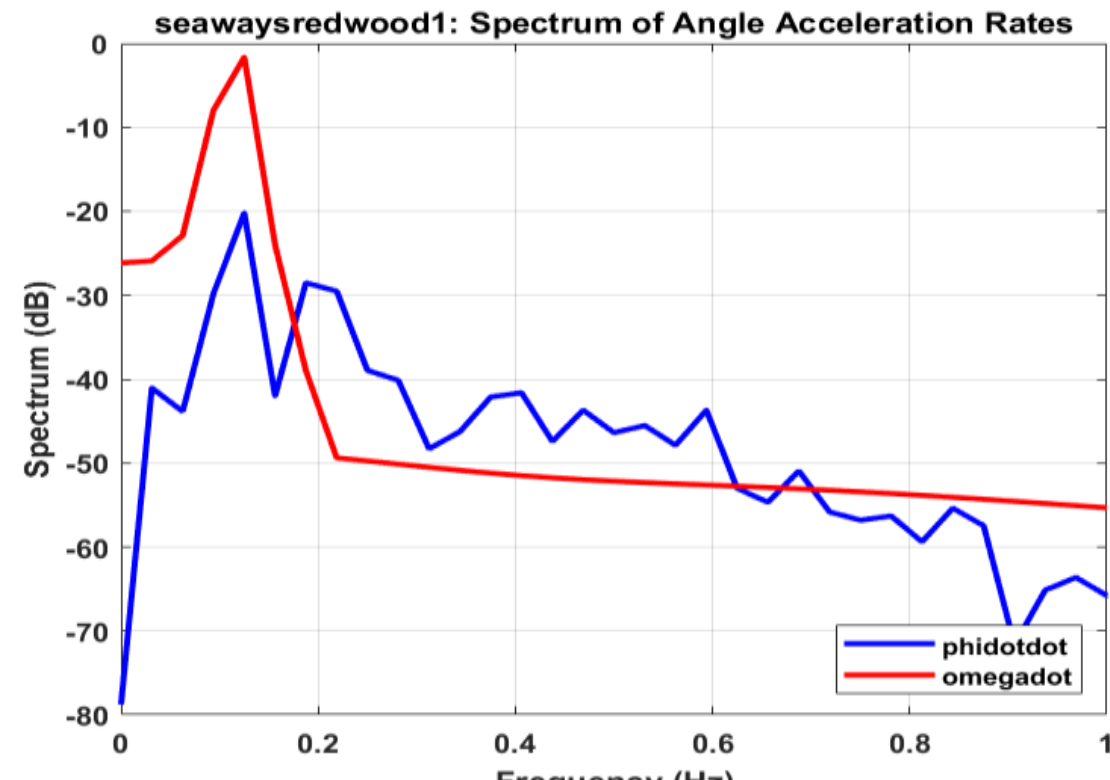

***Figure 20: Spectrum of angle acceleration for seawaysredwood1 from the Focus3D ISAR processor. The red line is the spectrum of the tilt rate estimate.***

The final case with large LOA error is shouchenshan1, a bulk carrier with a large 'deckhouse mirror' and 4 large cranes. Figure 21 is a photograph of the ship and Figure 22 is an ISAR image. Since this ship is heading towards the radar and has a flat 'deckhouse mirror' it should not be surprising to see multipath scattering – this is visible at the far range (beyond the deckhouse) in the ISAR image. In this image the green dots represent the range-Doppler points that were identified as point scatterers used in the autofocus and LOA algorithms and the red dot is a target that was rejected by the Focus3D multipath filter. This multipath scattering is a little surprising since the aspect angle of 28 degrees is not low enough to create a robust mirror of the deckhouse – the bounce off the deckhouse should be directed off the radar line of sight. My interpretation is that this effect is overwhelmed by the large radar cross section of the cranes – i.e., the multipath scattering may be due to only the bounces among the cranes. This case was included in the evaluation because I do not see how a radar system can select it for exclusion when the aspect angle is so large.

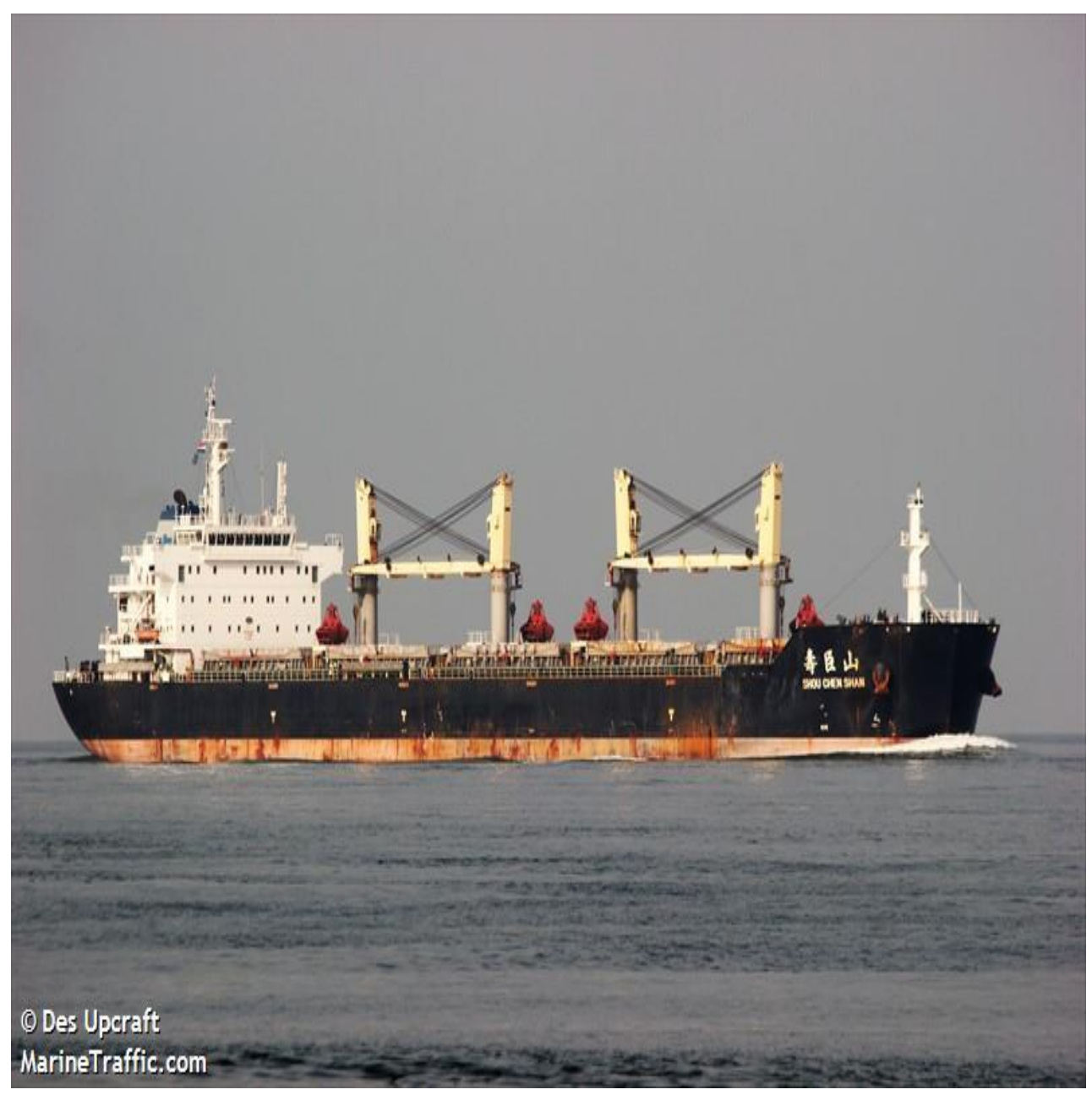

***Figure 21: Bulk carrier Shou Chen Shan***

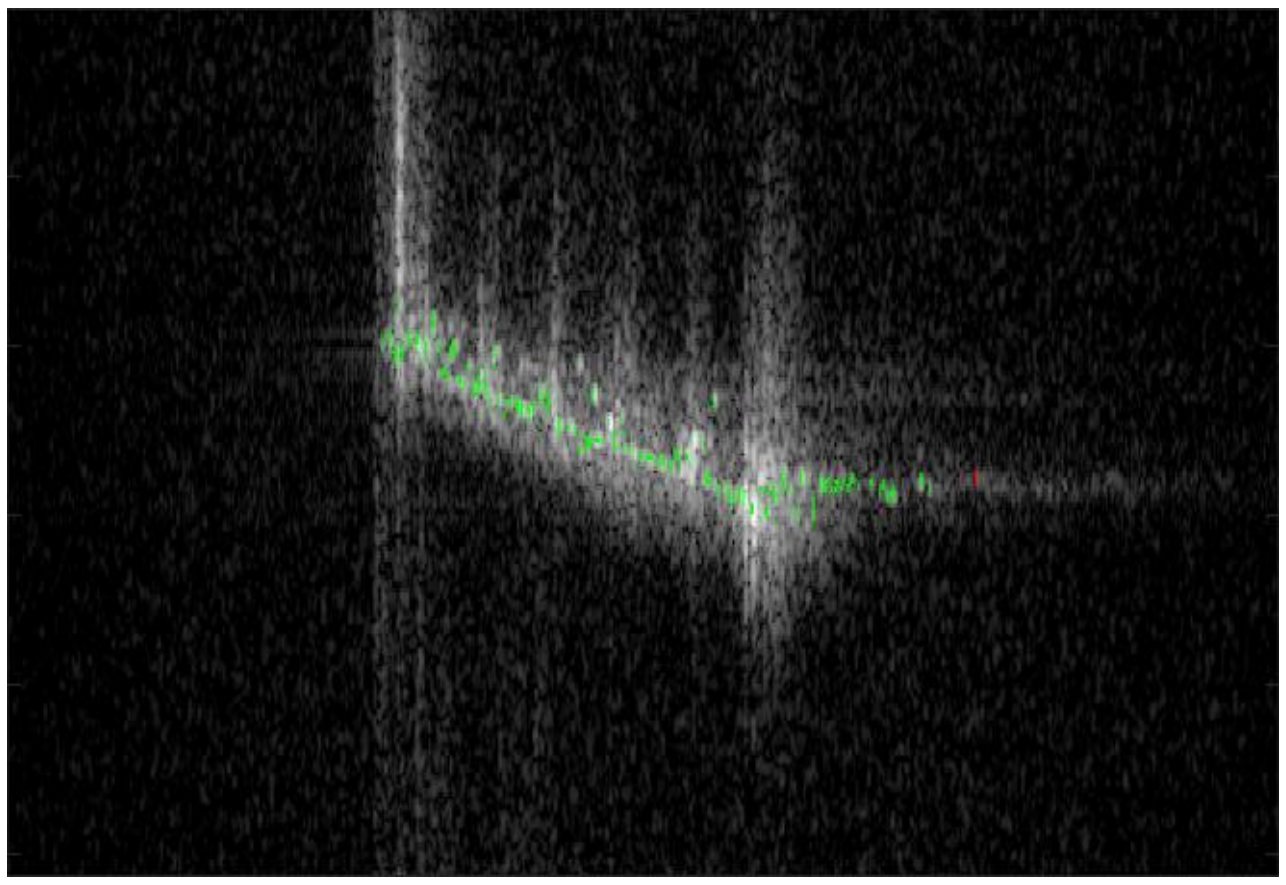

***Figure 22: ISAR image of the Shou Chen Shan. The green dots are scattering centers identified by the ISAR processor and the red dots are those eliminated by the algorithm for multipath removal.***

## VIII. THE EFFECT OF TRACKER LOCATION ERRORS

One potential source of error in the ship's course and speed is the error in the initial geographical location. We can evaluate this factor easily since the AIS reports also have the latitude and longitude of the ship measured by its GPS. The simplest way to illustrate this effect is to plot the magnitude of the Relative Vector Error. This is the ratio of the magnitude of the difference between the Course/Speed vectors for the tracker and the AIS, scaled by the AIS speed. Figure 23 plots this non-dimensional ratio when using the Tracker position estimates. Figure 24 repeats this calculation but replacing the tracker-estimated Lat/Lon positions with the AIS true positions. The mean value of this metric is significantly lower for the AIS positions. The RMS aspect angle error is reduced from 8 to 4.6 degrees and the mean vector error is reduced by the same factor. This indicates that the errors in target position have a major effect on the course and speed estimates.

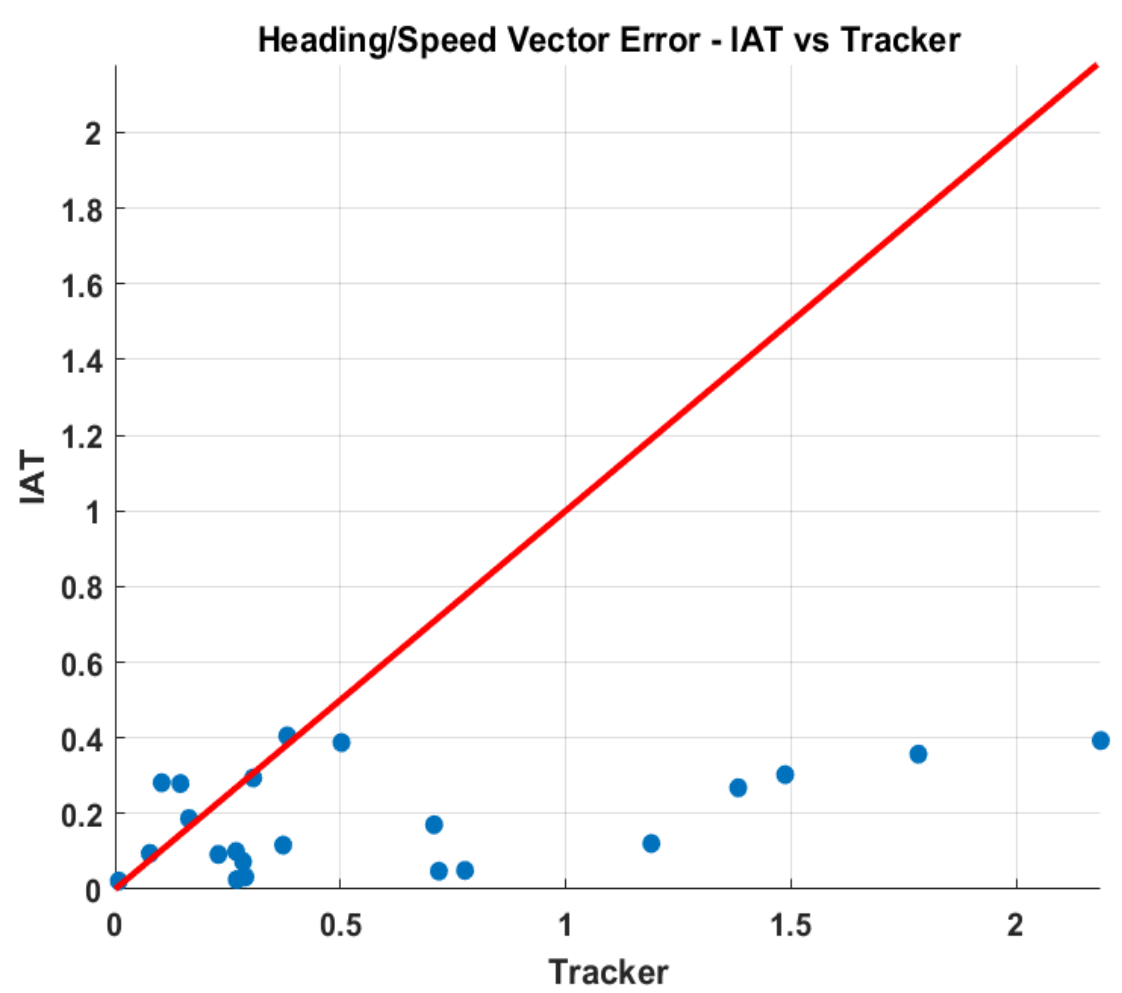

***Figure 23: Vector error metric assuming Tracker Lat/Lon Positions***

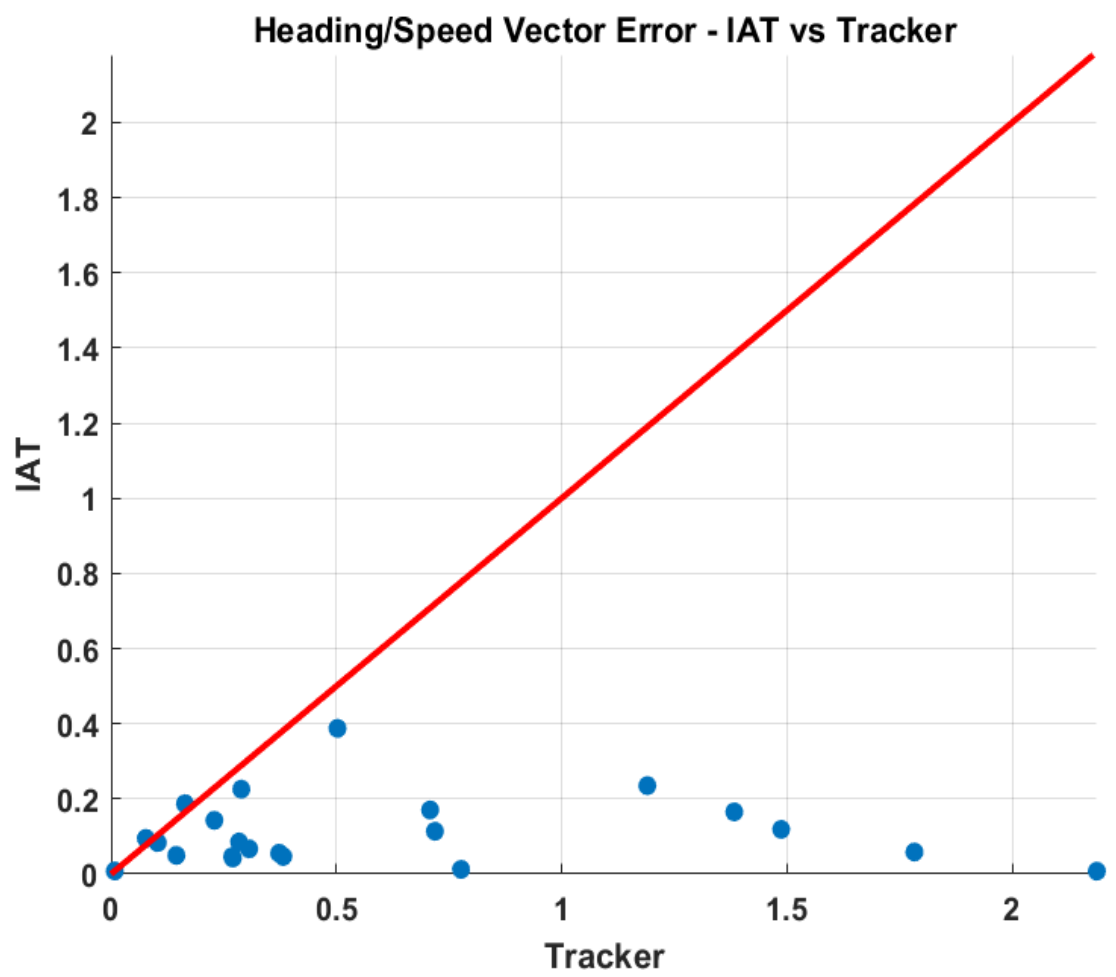

***Figure 24: Vector error metric assuming AIS Lat/Lon Positions***

This result is not surprising. Errors in cross-range position have a direct effect on the mean radial velocity – the product of the angle offset (radians) and the cross-range component of the *platform* velocity. For a broadside-look direction a 1 km error at a range of 100 km and a platform speed of 100 m/s yields a 1 m/s error in the *ship* radial velocity. At low aspect angles this is not too serious since it mainly affects the error in ship speed. But at large aspect angles the estimated in-range component of velocity is highly sensitive to this error. In other words, the implied error in the mean radial velocity due to the position error is in direct competition with the most important signal used by the IAT.

From a system perspective the critical measurement that determines the initial position is the initial target bearing – the range to the target is less controversial and is readily corrected in the ISAR processor acquisition algorithm. It makes sense to assume that the target bearing from the tracker can be improved by adding a method in which the ISAR mode could correct for this by varying the bearing to search for the bearing that maximizes the received power. Or an advanced radar with multiple receive apertures could compare the phase from adjacent beams to correct the bearing.

It is useful to consider the logical progression of the IAT algorithm as three levels:

Level-0 – Application of the IAT to existing systems with minimal changes to the system hardware or software. Implementation to the existing radars requires the storage of both the deterministic and adaptive motion compensation results and the documentation of the system method for the deterministic data following the logic of Section VI. All the calculations provided in this paper were done with this level of processing except for Figures 23 and 24.

Level-1 – Add a method to improve the estimate of the bearing to the ship, either through wiggling the beam to find the angle of maximum return or employing multiple apertures. This would be most effective near the broadside looks, where the velocity error due to the cross-range position error is the largest. Note that this need not be as elaborate as the interferometric methods used for fully 3D calculations – the IAT only needs one phase estimate calculated from the ship signature.

Level-2 – Modify the bearing estimation method of Level-1 to also estimate the time derivative of the bearing. This would enable the system to compute a direct estimate of the cross-range velocity of the ship and to merge this estimate with the IAT result. This would be especially important near the azimuth angles ahead of and behind the radar where the Analytical Method of the IAT is singular.

## IX. APPLICATION TO THE SCATR DATA

An earlier reviewer noted that all the data used here is from very large ships under relatively low sea states. To partially remedy this, I have analyzed some data from the ONR Small Craft ATR (SCATR) program. In 1997-8 I and my colleague Dr. Kenneth Melendez were participants in this program, responsible for doing the ISAR processing and passing this information to the ATR team lead by Scott Musman. In the literature review I discussed Musman's report and noted that its good LOA results were driven by the aspect angles estimated from GPS recorders on the target boats and thus were not a fair test of LOA accuracy for unknown ships and operational radars. However, the fact that this data set has been distributed widely and includes the full GPS records convinced me to explore what insight we could glean from it.

The purposes of this analysis are:

1. To validate the basic physical assumptions behind the IAT
2. To start a process that can build a model that estimates the IAT performance versus ship type, sea state, range and other parameters
3. To develop an algorithm for extrapolating data from a shore-based radar to a moving radar at a different range. This might be of wide interest since a lot of such data has been collected in many countries for the development of maritime ISAR systems.
4. To eventually use the data and methods to develop fully 3-D ISAR systems. But this use is not covered here.

The SCATR data consists of about 20 target boats, usually traversing an octagon shaped course between 2 and 10 kilometers from a shore-based X-band radar. I chose to analyze data from 8 sections of this data for two U.S. Coast Guard cutters, the Point Stuart (25 m) and the Tybee (33.5 m). Both these ships are fully ocean-going vessels while many of the other targets were small USN work boats that might be more difficult to analyze given the low range resolution of 0.37 m. Since the auxiliary and GPS data from the Tybee is much better than that of the Point Stuart the full analysis (Figure 29 and above) was limited to the Tybee.

Since the radar is on shore and does not move and does not have a tracker to provide a first guess of the ship course and speed, the difference between the cross-range platform velocity and the cross-range tracker velocity is effectively zero and the Map Drift equation (Eq, 3) reduces to:

$$A * R = u^2 \quad \text{Eq. 4}$$

Two implications of this formula are:

a. It implies that the acceleration must be positive. In other words, both left-right and right-to left travel yield a positive acceleration. This makes sense because in both cases the ship begins getting closer to the radar and then it gets further from the radar.

b. Since the cross-range velocity only appears as a squared quantity this radar cannot tell the sign of $u$.

c. Using a non-moving radar has a major limitation since for the MSR2 data set the linear term is about 30 times the value of the quadratic term. This means that the acceleration is 30 times the expected value for the shore-based case. Thus, it is in the linear range of the map-drift equation and, since the acceleration is larger it is easier to estimate versus noise and clutter effects.

For purposes of estimating the LOA of the ship the first two limitations do not matter since they do not affect the magnitude of the aspect angle. But the smaller expected acceleration needs to be considered in our interpretation of the results.

To simplify the analysis here I chose to simply compare the two sides of this equation using the GPS and radar data.

One unique feature of the SCATR data is that it includes GPS spreadsheets that give the time history of the latitude and longitude of the ships, sampled at a 1 Hz rate. This data was derived from a pair of GPS recorders provided by the Naval Research Laboratory; one recorder at the radar site and the other on the Tybee. This differential GPS appears to have a location accuracy better than 0.5 m for time periods of at least 10 seconds. This allows the estimation of the mean in-range and cross-range components of the ship velocity, and their time derivatives averaged over the ISAR dwell times but does not allow us to see the wave response when the encounter period is less than 10 seconds (see the discussion of Figures 33, 34 and 35 for more detail).

Figure 25 plots the GPS heading and aspect angles for the Tybee over the 42 minutes required for the 8 cases. During each leg of the octagonal pattern the heading is nearly constant, indicating that the helmsman was able to steer a steady course. However, the aspect angle changes over time due to the cross-range motion of the ship (recall that aspect is the difference between the bearing and the heading).

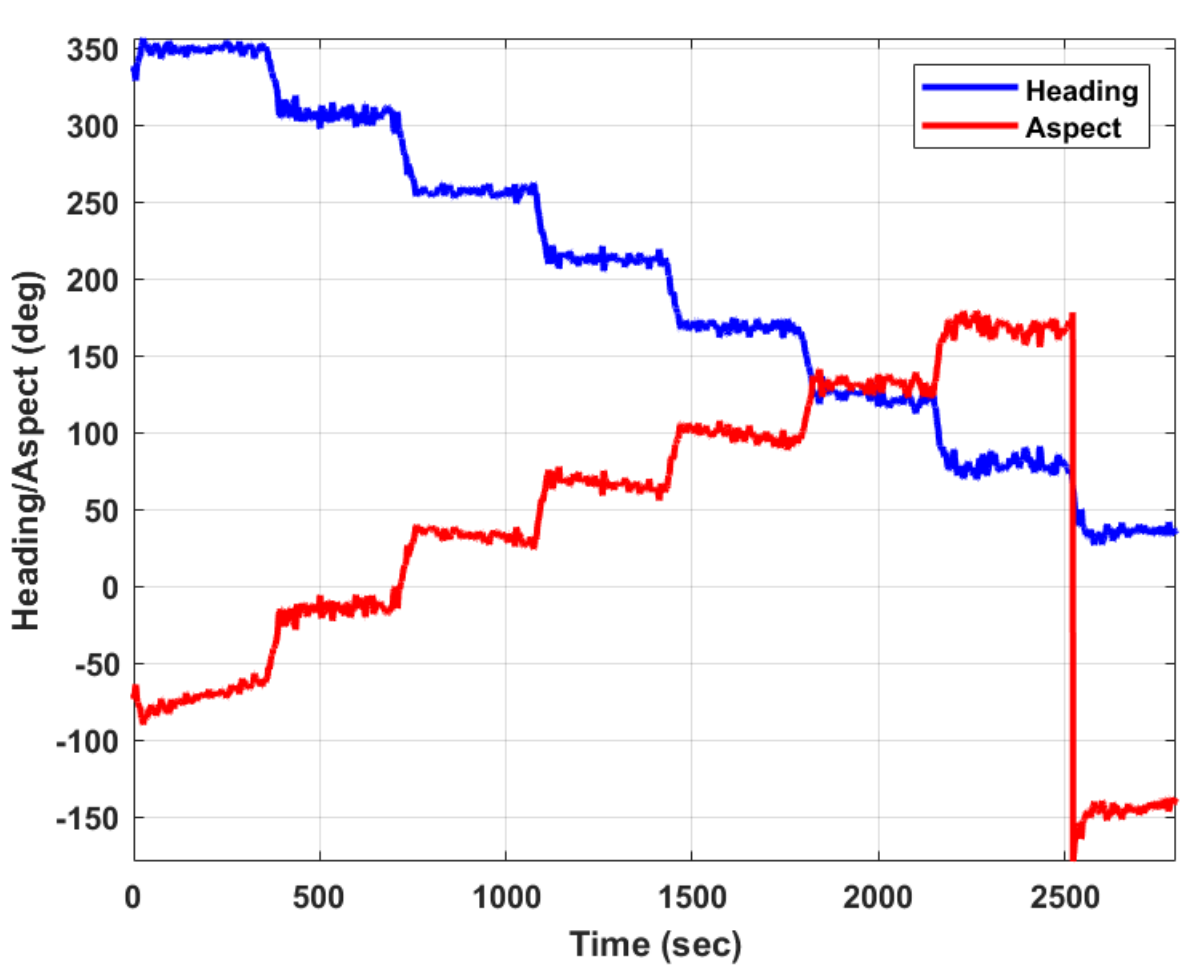

*Figure 25: Heading and aspect angle for the Tybee during the SCATR experiment – derived from GPS recording*

The recorded GPS data allows the estimation of the left- and right-hand sides of the equation from the mean motion, ignoring the radar measurements. Figure 26 compares the LHS and RHS for this model assumption for both the Point Stuart and Tybee cases. The fact that they agree is encouraging since it means that the equation is correct. However, Figure 27 gives this same relation but using the radar data to estimate the acceleration and cross-range velocity. This does not have as good a relation between the RHS and LHS. In fact, there are even 5 cases where the acceleration is negative, a result inconsistent with a real value of the cross-range velocity from Eq. 4.

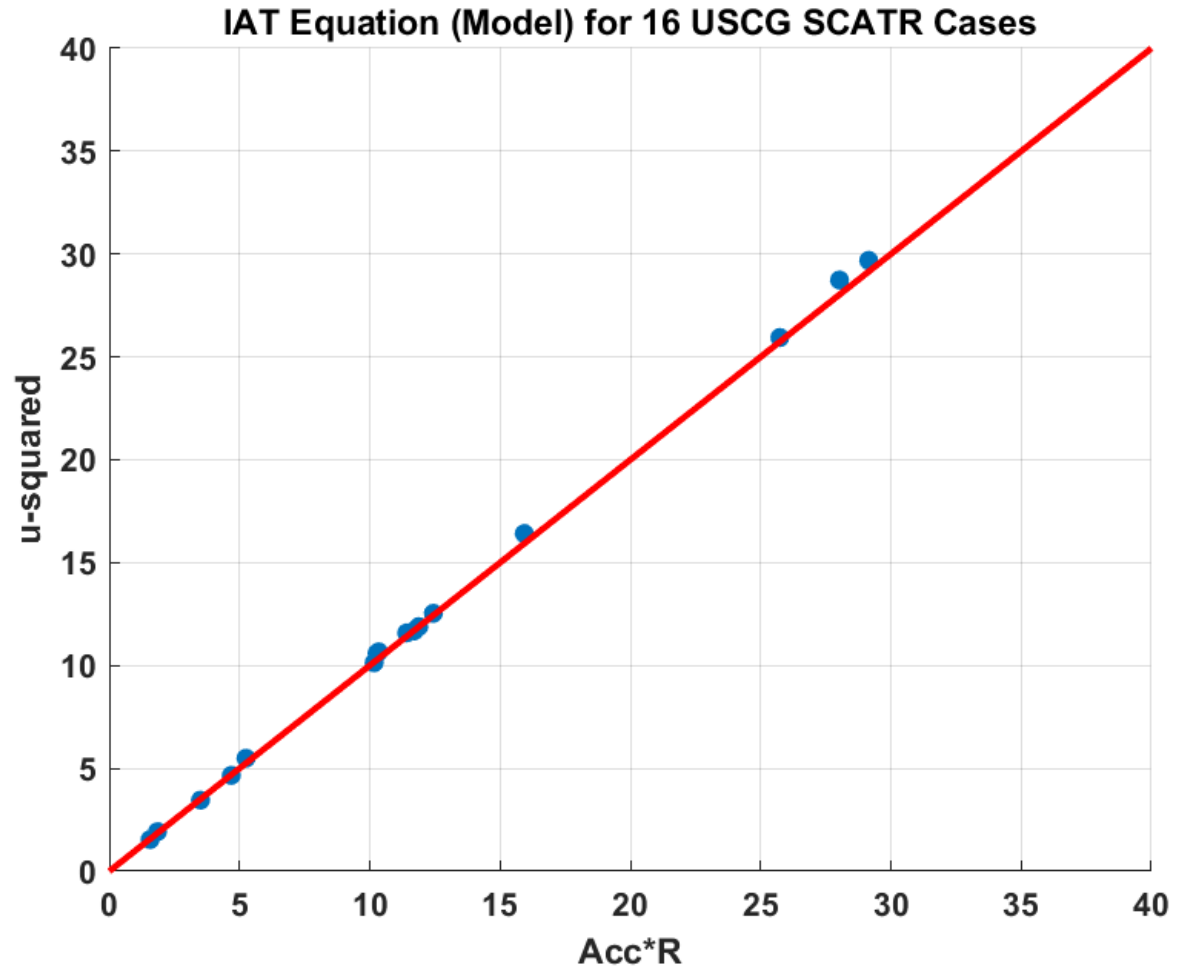

*Figure 26: IAT Equation (RHS vs LHS) for 16 SCATR Cases using a theoretical model based on the GPS data*

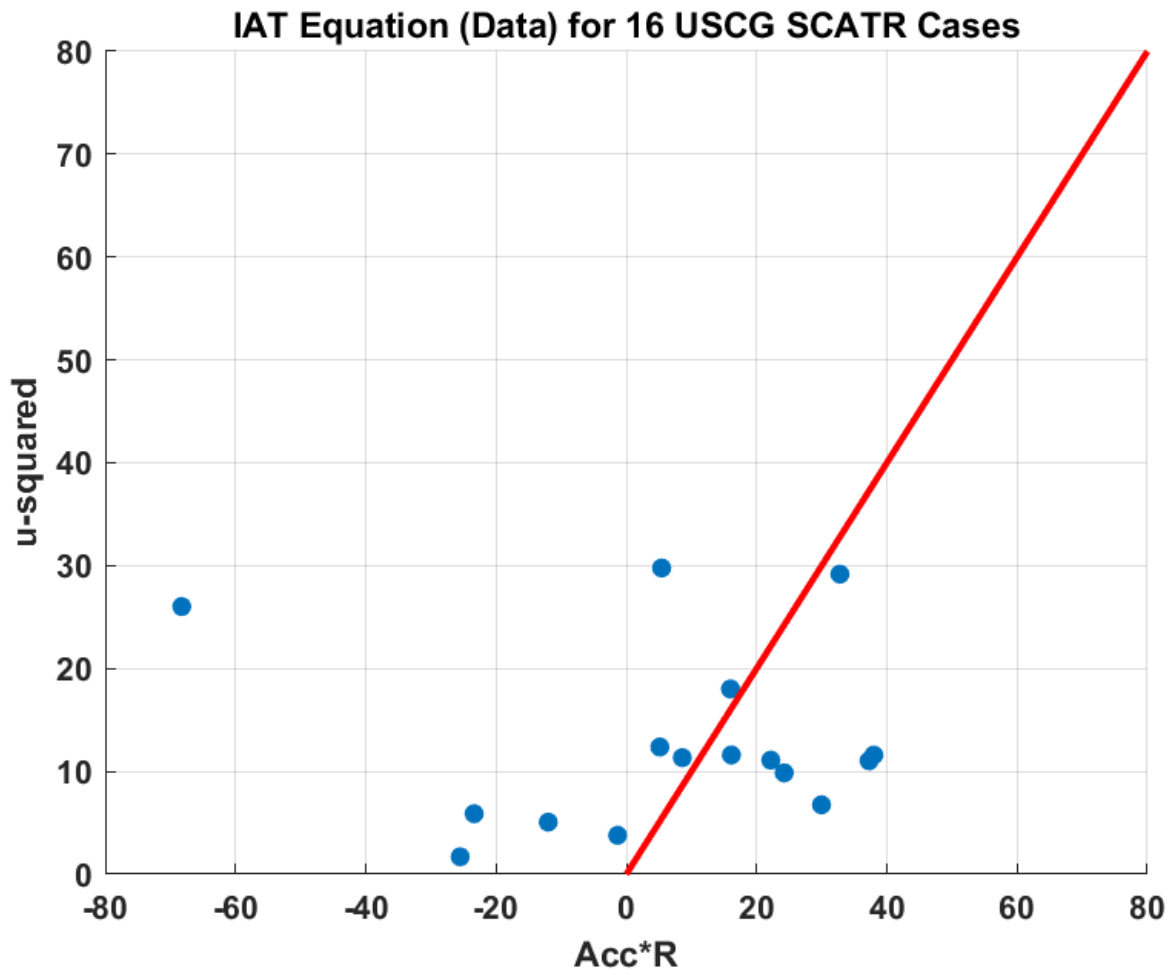

*Figure 27: IAT Equation (RHS vs LHS) for 16 SCATR Cases using data from the SCATR Radar*

To diagnose this poor performance of the radar data I calculated the integrated acceleration value for 60 seconds to get the difference in the motion compensation velocity over this period to compare with the wave velocities. Figure 28 plots these velocity differences for the model and radar cases. It is possible that the error in estimating the acceleration explains most of the errors in Figure 27. Note that this result was anticipated due to the use of a stationary radar. The simulation results in Section X show similar issues with the forward and rearward azimuths. This is why I refer to this region as a partial blind spot for the IAT.

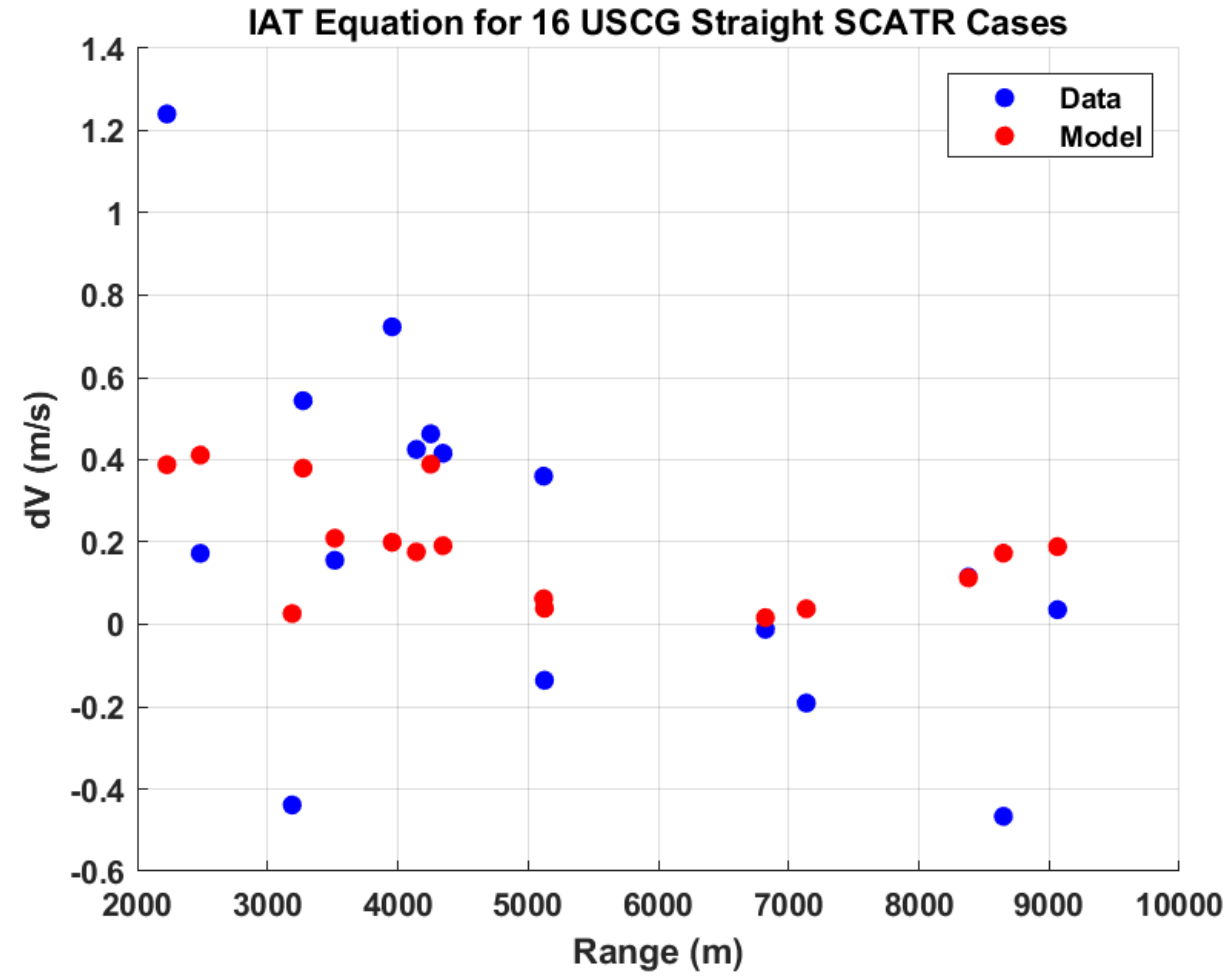

*Figure 28: Integrated acceleration for 16 SCATR cases, comparing the data calculation with the model*

I believe that the main reason that the SCATR results are not as good as for the MSR2 radar is that the shore site is not moving and thus the map-drift equation is in its quadratic mode. However, I cannot rule out the conclusion that the IAT may not work as well for smaller ships. Other factors are:

(1) The SCATR data has fewer range cells for each case – the range resolution is about 0.37 m and some of the cases have high aspect angles.

(2) The SCATR radar is not as high a quality radar as the MSR2, producing much lower values of the signal to clutter ratio.
(3) The PRF of the SCATR system is only 200 Hz versus 512 Hz for the MSR2, making it more difficult to produce good motion compensation data.
(4) The SCATR data have a different range of aspect angles – the MSR2 data has only one case over 45 degrees but the SCATR data has 5 cases over 75 degrees.

To resolve this issue and to validate the basic physical assumptions behind the IAT, I did a more detailed analysis of the GPS data recorded by the USCG Tybee. Following the logic of the IAT, this involved looking at the time histories for range, range velocity and range acceleration.

First, I calculated the average of the in-range velocity from the GPS data over the 60-second radar records and compared it to the same quantity estimated by the ISAR processor (Figure 29). As expected, these two quantities agree. Not shown is the initial calculation, which showed that there is a 60 second time difference between the two values. This time offset is also evident in the range comparison. The reason for this is lost to history. My guess is that the analyst in 1998 initially set the initial times for the ISAR modes based on the times of the turns in the octagon but then realized that the ship had not fully adjusted its motion after the turns and then just selected a later section of data for this distribution. At any rate, the time shift is very clear in all the legs of the octagon and for both range and velocity.

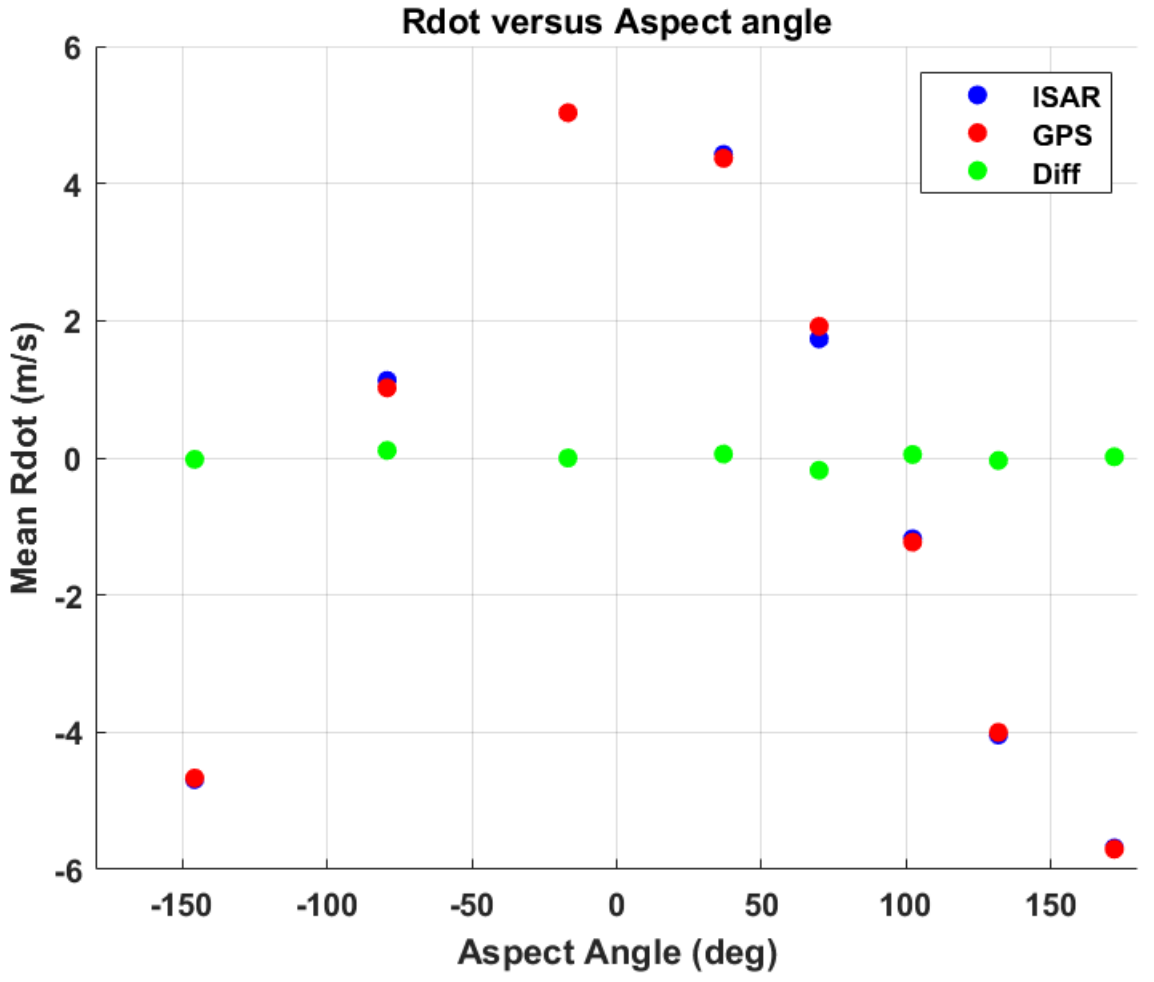

*Figure 29: Mean radial velocity, ISAR vs GPS, for the Tybee SCATR cases*

Next, I looked at the time variation of the range velocity to check the consistency of the narrow band filter used to correct for ocean waves. The chapeau method for implementing this filter estimates the mean wave period as observed by the radar. Since the radar does not 'see' the water because of the low grazing angle and wide beam, this period is that of the ship's motion in response to the waves. In other words, it is an encounter period. The apparent encounter periods for the 8 legs of the octagon range from 2.6 seconds to over 18 seconds, depending on both the intrinsic ocean wave period and the ship's velocity. I fit these values to a simple deep-water wave model using a search over the intrinsic wave period and wave direction and concluded that the ocean wave has a period of about 5 seconds (corresponding to a 39 m wavelength and a 7.8 m/s phase velocity) and is from due west. Figure 30 compares the ISAR estimates of the encounter period with the wave model. The oscillatory motion of the Tybee at all 8 legs of the octagon is explainable by this wave model.

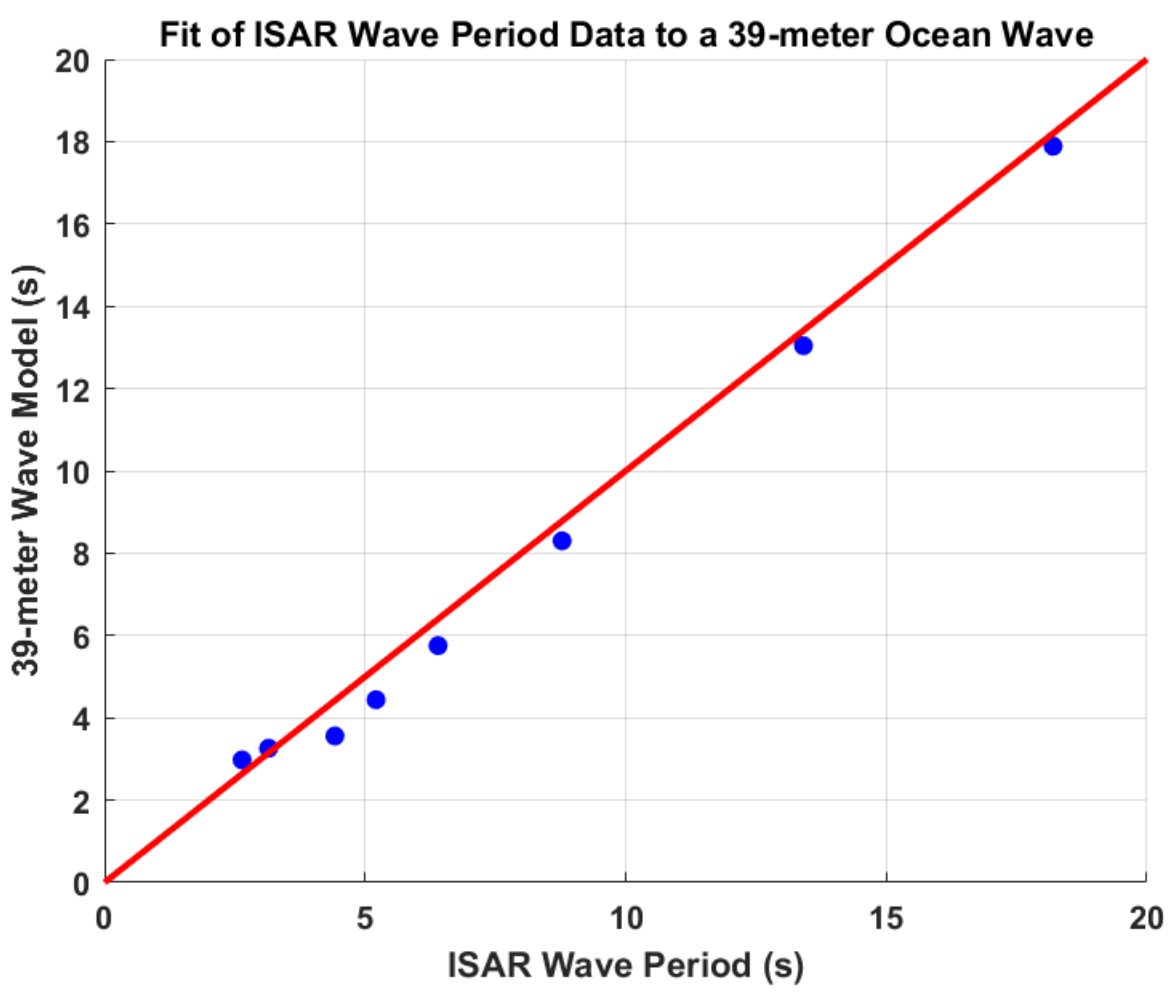

*Figure 30: ISAR estimated wave encounter period vs the best-fit (5 second, 39-meter, direction 270 deg.) ocean wave model*

To check the result of the encounter model I asked the Scripps Institution of Oceanography if they had any historical wave data during the experiment. They found two measurements from a high-quality directional wave buoy during the 50 minutes of the Tybee test. The buoy is only 6.7 km away to the northwest so it should be representative of the Point Loma coastal area. The buoy wave period is 5.3 seconds, and the wave direction is 279 degrees. That is a difference of 6% in wave period and 9 degrees in direction. This is close to exact agreement given the location difference and the fact that these are two different types of measurements. The wave amplitude from the radar data is only about half that for the wave buoy but that is expected since the buoy report of significant wave height is based on an integral over the whole wave spectrum, not just the peak period. Also, especially for the cases at close to right angle to the waves, the 33.5-meter ship tends to average out the velocity signature of the 39-meter waves.

The results for the IAT for the simple stationary radar used at SCATR will not be presented here since our goal is to understand the performance of flying systems. Instead, I have developed an extrapolation algorithm that allows one to use data from a stationary radar to estimate the performance of the IAT and other algorithms at a larger range and for a moving radar. The method computes the exact solution based on the desired range and radar velocity and the GPS mean estimate of the ship's motion and then adds a noise term based on the normal IAT data preparation routine plus an acceleration error measured from a blend of the values from GPS and ISAR data.

This method works because the added terms to the equation are due to the radar motion, which in a real system would be known very accurately from the aircraft GPS and its Inertial Navigation System.

It is important to understand the details of the extrapolation logic since this knowledge can give insight into the IAT algorithm and help one develop a better plan for implementing it in an operational radar. The method uses both the GPS and the ISAR data. The GPS data measures the motion of one part of the Tybee with a time scale of perhaps 5 seconds and thus it misses the wave motion for most of the legs of the octagon that have short encounter periods. But for long enough time scales it agrees with the ISAR data. When the encounter period is greater than 10 seconds even the wave motion agrees for the two measurement methods. However, there is no reason to doubt that the GPS data has the same accuracy for cross-range motion as it has for in-range motion – it has no knowledge of the radar. Therefore, it is reasonable for us to regard the GPS data as truth data for long term motion in any direction. Its use in the extrapolation method is to define the mean velocity vector of the Tybee.

The role of the ISAR data in the extrapolation method is to define the motion compensation noise. First, the mean value of the in-range motion is subtracted from the ISAR motion since this is a duplication of the mean for the GPS data. Then the cross-range velocity from the GPS data is used to correct the linear trend of the ISAR data since both methods should agree. The remainder of the ISAR data consists of wave motion and the steering variation so this is added to the GPS truth data to get the extrapolated signal plus noise.

The extrapolation to a moving platform at a different range is the simplest part of the process since we can assume that the radar platform measures its own motion very accurately using a GPS receiver assisted by an inertial navigation system. In the end the extrapolation process computes the hypothetical motion of a ship, including the important errors from the ISAR data, that is imaged by a moving radar at a different range than the Point Loma shore site.

The extrapolation algorithm then uses the Map-Drift equation of Eq. 4 with the square completed since there is no need for a deterministic motion compensation stage to remove the square of $V$:

$$A * R = (V - u)^2 \qquad \text{Eq. 5}$$

Here $A$ is the time derivative of the in-range component of velocity, $R$ is the extrapolated range, $V$ is the synthetic cross-range platform velocity and $u$ is the ship's cross-range velocity. There is no need to add an in-range component of the platform velocity since this would be removed at the deterministic motion compensation stage. Note that even in this general form of the equation the acceleration must be positive to yield a real value of $u$.

The extrapolation of the Tybee data used two major parameters. One parameter multiplies the range by 10, changing the maximum range from 9 km to 90 km. The second parameter changes the cross-range platform motion from 0 to 50 m/s. In other words, instead of a stationary radar we are simulating a radar moving in the cross-range direction at 50 m/s. This motion is equivalent to a radar traveling at 100 m/s shifting its heading by 30 degrees to identify a target dead ahead. This seems like a reasonable maneuver to ask an aircraft to do to identify a high-value target.

As expected, adding a platform motion shifts the mathematics of the Map-Drift equation from a balance of the acceleration term and the square of the cross-range ship velocity to a nearly linear equation. For this example, the acceleration increases by a factor of 80 and the term linear in $u$ is 18 times the term quadratic in $u$, making the errors in the acceleration much less important. The platform motion effectively acts as an amplifier to the acceleration signature used by the IAT, without increasing the acceleration noise. The larger conclusion to this logic is that the IAT works best when the ISAR system has some skill in SAR imaging. However, either a stationary radar or a radar looking ahead or behind has no angle diversity and thus no SAR capability.

With the extrapolated data the problem with the negative accelerations (forcing the square root of a negative number) goes away and all accelerations for the 8 cases are positive. Figure 31 plots the cross-range velocity estimated from the IAT algorithm against the value estimated from the GPS data. Figure 32 plots the values for the aspect angle. There is still one point that is a bit off the 'perfect' line, but all the estimates are sensible in that they follow from square roots of positive numbers, and they allow us to determine both the magnitude and sign of the cross-range velocity.

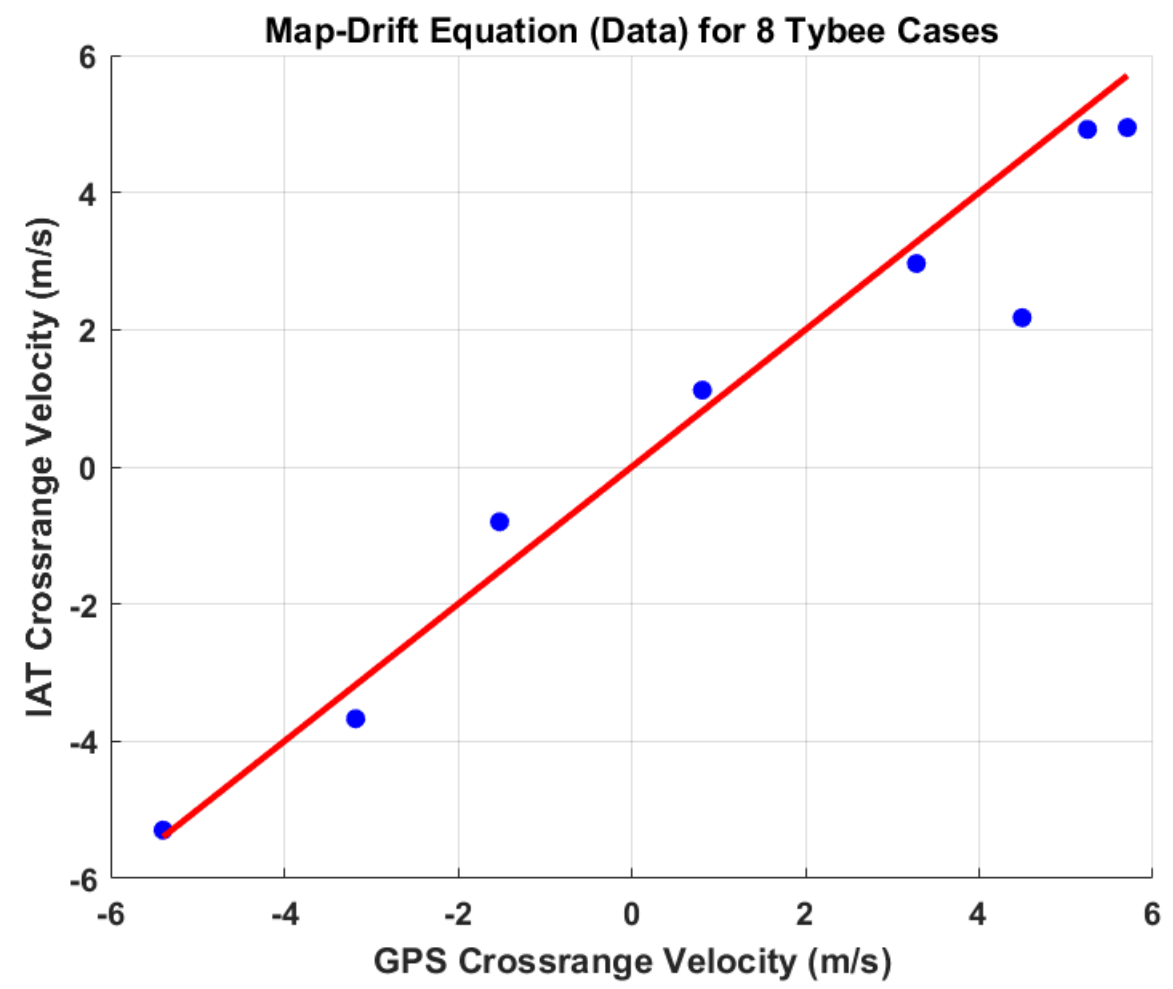

***Figure 31: Cross-range Velocity IAT vs GPS for the Tybee in SCATR, extrapolated to a radar moving cross-range at 50 m/s and ranges 10 times the true range.***

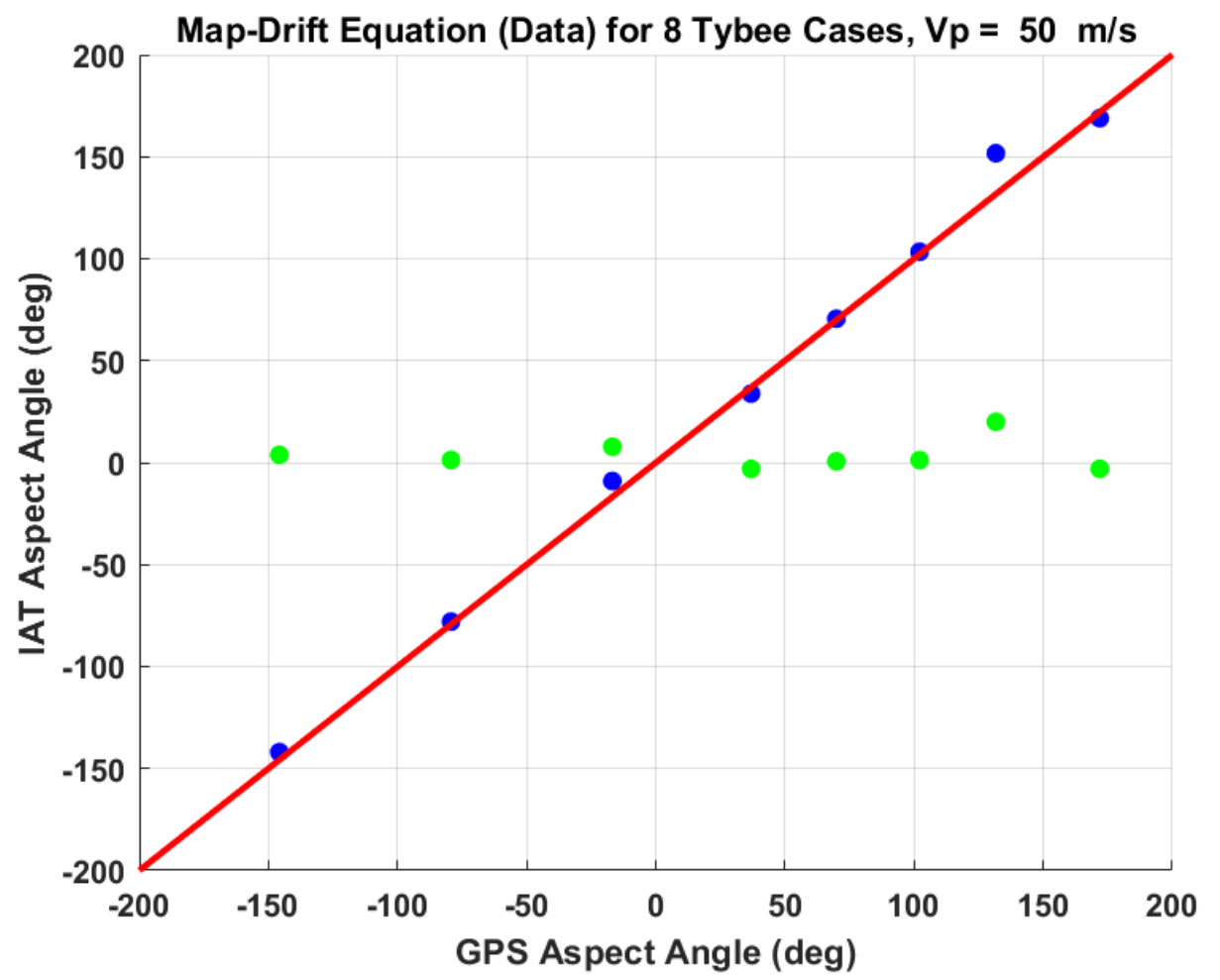

***Figure 32: Aspect Angle IAT vs GPS for the Tybee in SCATR for the extrapolated simulation***

The results so far have used the 60-second data segments defined by the NRL's official SCATR data set distributed widely to researchers. However, the negative accelerations observed for this data were of serious concern because they might be a first symptom of a major problem for the IAT. I looked for a way to explain them. One idea to run this to ground was to use the longer original data sets instead of the NRL 60-second segments.

To understand the difference between the two data types, it is useful to summarize the data collection method employed at SCATR. The experiment site was just offshore of Point Loma, California. The ships usually had an octagonal path formed by directing the ships to start at a given point and then to execute 8 nominally straight paths at a speed of about 5.6 m/s for about 6 minutes each. Thus, each leg of the octagon is about 2 km long and the entire octagon covers an area of about 5 km square. At each corner the heading is incremented by about 45 degrees. In executing this pattern, the Tybee helmsman did not have any visual cues. And the GPS receiver on its deck was not available to guide him since it had no readout – in any case the accuracy of the differential GPS was only available after the recordings of both systems were analyzed back at the NRL. Thus, the Tybee was simply operating in its normal open water mode, trying to steer a course with a constant heading and speed.

After the analysis of the standard NRL 60-second data the original data files from the radar were pulled from the archives. These files were each about 6 minutes long and the files were changed at each vertex of the octagon. Since the first section of each of these files consisted of a course change and an adjustment to it, we followed the NRL analyst and skipped the first minute of each record; and we skipped a piece at the end of each record when the Tybee was preparing for the next course change. This process provided 4 minutes of data for each leg where we were certain that the Tybee was trying to steer a steady course at constant speed. During these long dwells the Tybee traveled through several radar beamwidths of azimuth angle; but this effect was corrected during the experiment by a manual tracking process using a bore-sighted video camera.

These 4-minute data segments allow a choice of integration times – from 16 15-second pieces to a single 240-second piece. The results from this process can be summarized by Figures 33 and 34, which compare the acceleration values for the GPS and ISAR data for 60 seconds (32 examples) and 240 seconds (8 examples). For the 60-second data Figure 33 has two interesting properties. First, there is a high correlation, about 0.87, between the GPS and ISAR accelerations. Second, both measurements have about one third negatives. Since these estimates are from very different physical methods, I assumed that the acceleration estimates are due to real variations in the Tybee's motion. In other words, the Tybee was not perfectly following a constant course and speed. This Steering Variation could be very important to the evaluation and development of the IAT algorithm. Figure 34, for the 8 full 4-minute data segments, is also interesting because the negative accelerations have been greatly reduced. There is a single value that is slightly negative. But this point is close to zero and thus handled by the algorithm inserting zero to guarantee real values of the result of the map-drift equation. We cannot ensure that negative values never occur since zero is a legitimate acceleration value when the ship is heading exactly towards or away from the radar. It is reasonable to conclude that the large values of negative acceleration for the 60-second data were due to insufficient integration time, not a flaw in the map-drift logic.

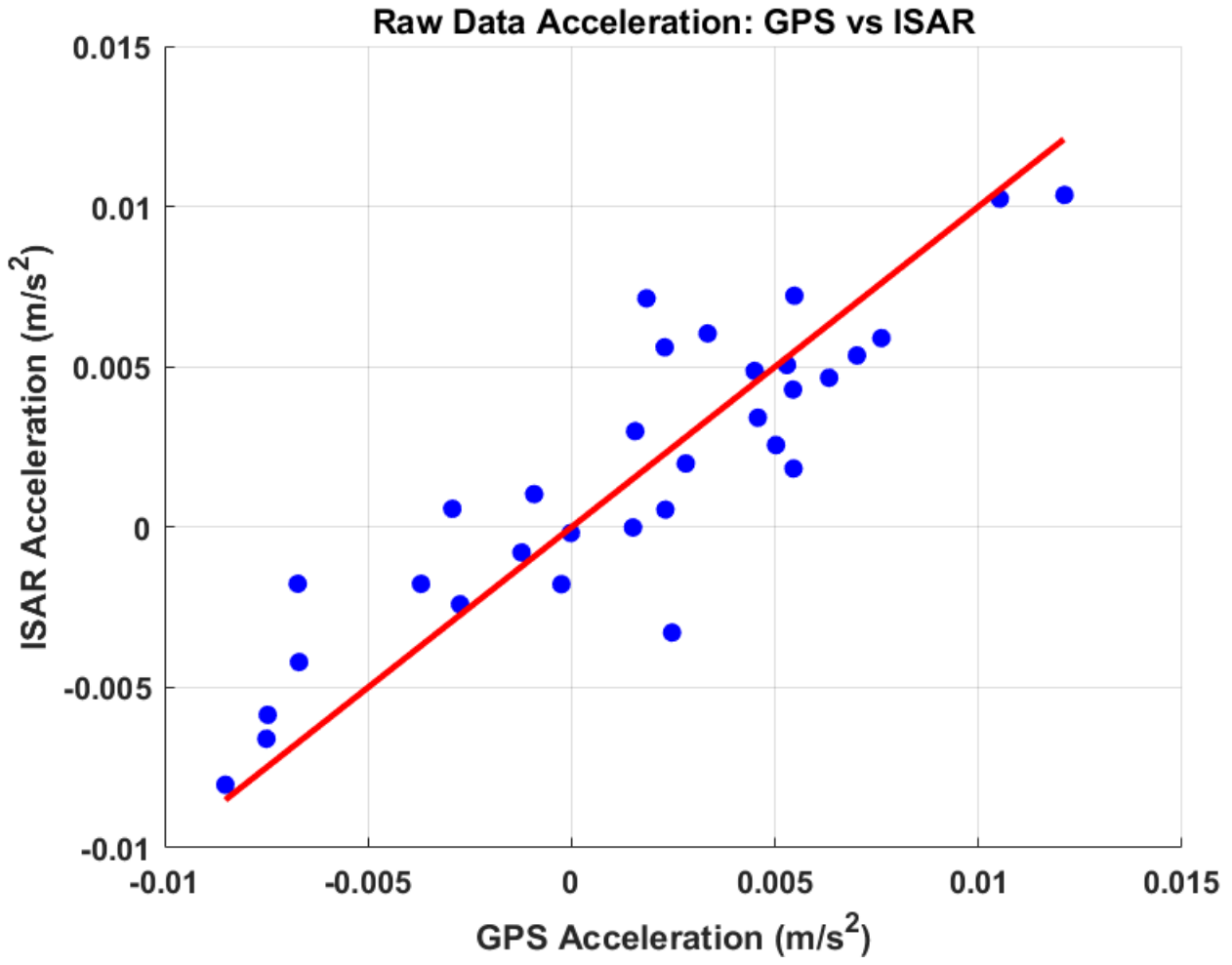

***Figure 33: Acceleration – GPS vs ISAR for 32 60-second dwells***

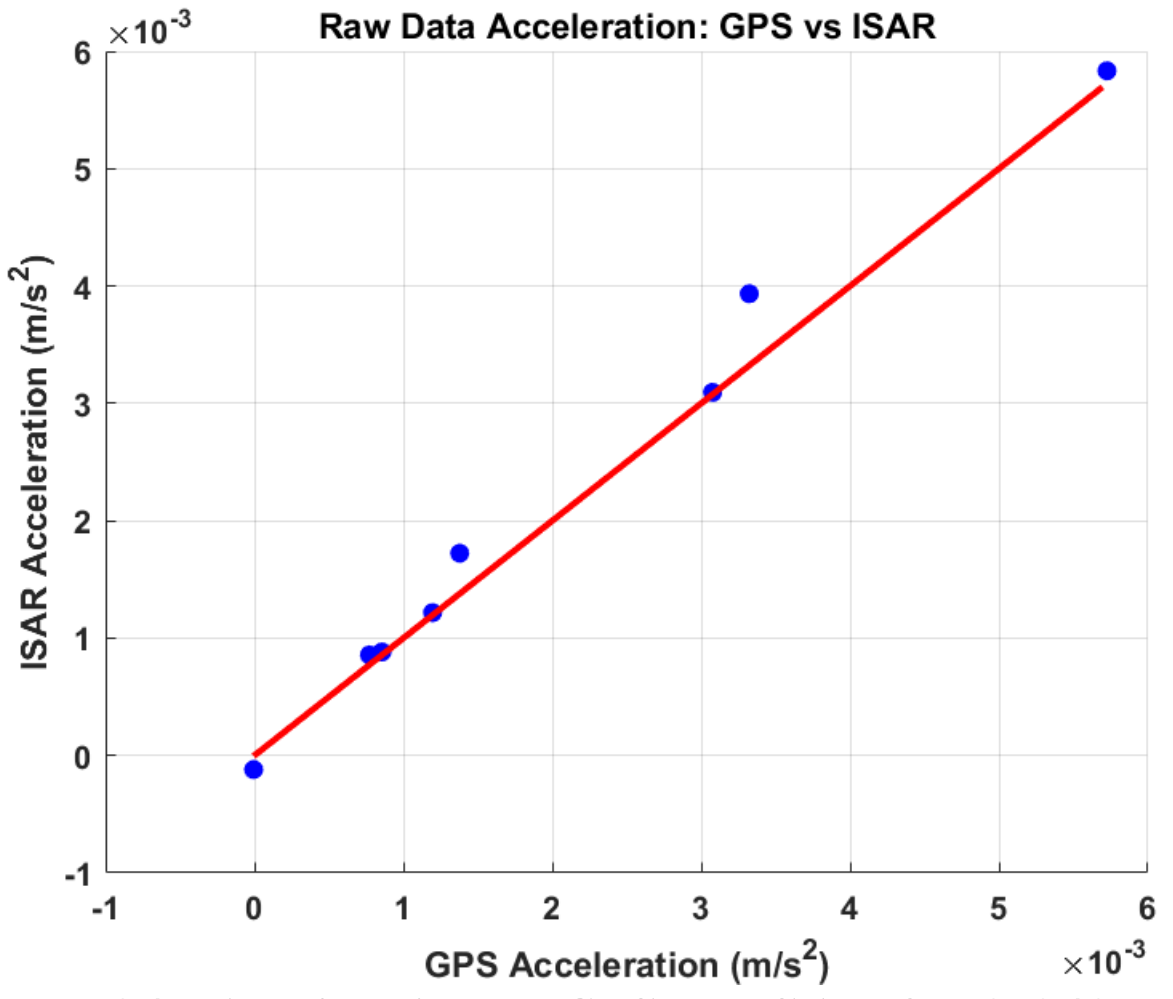

***Figure 34: Acceleration – GPS vs ISAR for 8 240-second dwells***

To confirm the Steering Variation hypothesis I analyzed the full plots of the comparison of the GPS and ISAR velocities for the 4-minute segments. Most of the octagon legs are dominated by the 6 shortest encounter periods shown in Figure 30. These plots are complicated to analyze since the GPS does not respond to the high frequency waves, probably due to the filtering of the GPS itself or the software in the box. However, there are 2 cases that have encounter periods of 13.4 and 18 seconds for legs when the Tybee and the waves have the same direction of travel. For these cases the GPS data should accurately measure the low frequency wave motion. Figure 35 plots the in-range velocity for the GPS and the ISAR data for the full 240 seconds for the 6th leg of the octagon, Sc15a12, which has an encounter period of 13.4 seconds. The plot for the case with the 18 second encounter period is similar. Clearly the two types of measurements track each other well over the entire time series, implying that the Steering Variation is real.

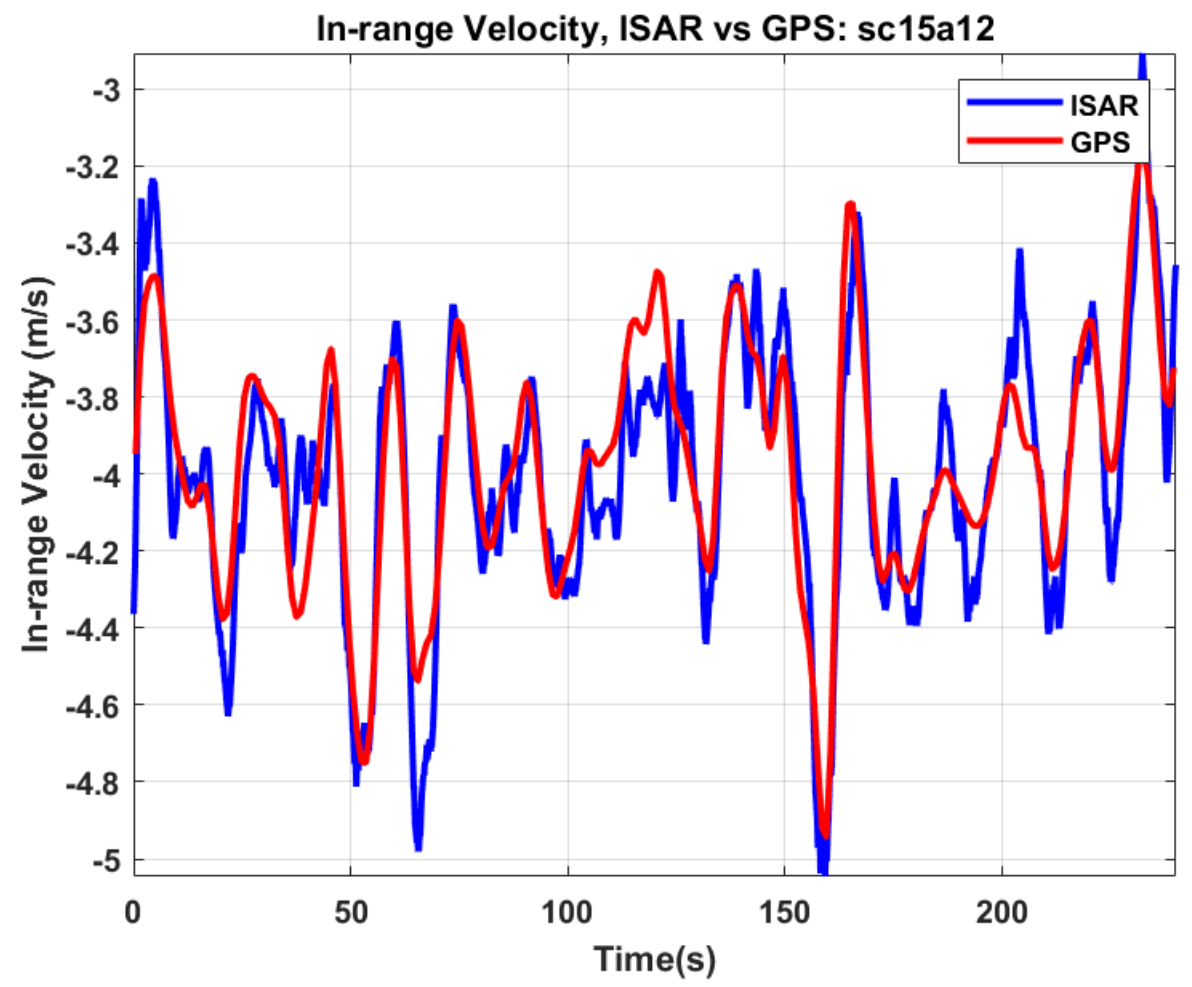

***Figure 35: Acceleration – ISAR vs GPS for 240 seconds when the wave and ship have nearly the same direction of travel***

Table 3 gives the results for the ship length estimates for the 8 Tybee cases using the 4-minute data and the extrapolation algorithm using a range of 10 times the actual range and a cross-range platform speed of 50 m/s. The RMS error using the IAT aspect angles is 9.5%, a little less than the goal of 10% here. If I had used the GPS values of aspect angle the error would have been 7.3%. The GPS value is mainly due to the inaccuracy of the range extent; the IAT is only responsible for the 2.2% difference. Note that this error covers a large distribution of aspect angles.

**Table 3: LOA for SCATR data of the USCG Tybee**

| Case | GPS Aspect | IAT Aspect | IAT LOA | LOA Error |
|---|---|---|---|---|
| Sc15a07 | -72.1 | -71.1 | 33.4 | -0.1 |
| Sc15a08 | -14.2 | -16.7 | 32.4 | -1.1 |
| Sc15a09 | 34.0 | 34.5 | 30.3 | -3.2 |
| Sc15a10 | 67.9 | 65.4 | 33.7 | 0.2 |
| Sc15a11 | 99.5 | 98.3 | 39.8 | 6.3 |
| Sc15a12 | 131.2 | 123.4 | 36.6 | 3.1 |
| Sc15a13 | 168.8 | 165.9 | 30.8 | -2.7 |
| Sc15a14 | -143.7 | -148.0 | 29.9 | -3.6 |
| | | | Mean | -0.14 |
| | | | RMS (m) | 3.18 |

My conclusion from this analysis of the Tybee ISAR and GPS data is that the SCATR radar has done a very good job of measuring the motion of the Tybee. And, given the improvement in the mathematics due to extrapolating the data to a moving platform, we now have a better idea for how the IAT should be used in an operational scenario.

Future experiments need to carefully plan the collection to ensure that the radar system used is properly matched to the requirements for the IAT. For example, shore experiments can be useful; but they need to be analyzed with software capable of extrapolating the results to moving platforms at different ranges. And airborne tests should concentrate on broadside looks to put the map-drift equation in a linear mode.

## X. IAT REQUIREMENTS AND SIMULATION RESULTS

Since we do not have data with an adequate range of operating conditions and ship examples, I have developed a simulator to fully explore the IAT algorithm over the full range of requirements.

The IAT needs to operate over a wide range of ship aspect angles and at all azimuth angles relative to the radar. However, for low altitude operations we may need to avoid the smallest aspect angles due to the problems with self-shadowing of the ship due to superstructure or stacks of containers. Shadowing can cause an underestimation of LOA. Also, at low aspect angles multipath scattering can cause an overestimation of LOA. Very high aspect angles (i.e., ship motion across the radar beam) can be difficult to interpret due to the intense reflected energy when the side of the ship is perpendicular to the beam. Thus, I have set the goal of operating at aspects between 5 and 85 degrees.

Since many countries operate circular-scan ISAR systems it is important to have the IAT algorithm work at any azimuth. The IAT is naturally more accurate at side looks and less accurate looking ahead or behind the radar. This is because the cross-range component of ship velocity is driven by the time derivative of the motion compensation velocity, the mocomp acceleration. The acceleration signature at side looks is proportional to twice the along track velocity of the radar, which is zero at the ahead and behind views. Since this only affects the cross-range component of ship velocity I call this a partial blind spot. To remedy this problem, I envision using a more direct method for cross-range velocity at these angles.

The simulator does not try to produce raw radar data from a model ship. Instead, it generates a time series of the adaptive motion compensation velocity as the sum of the exact solution plus a noise term. It was executed for a sample that uses a range to the target of 100 km and a platform speed of 100 m/s. The ship is assumed to be at broadside and has a speed of 6 m/s. To make the simulation more realistic a noise term was added to the theoretical motion compensation velocity. To generate realistic noise data for this purpose I simply removed a linear regression line from each of the 22 real cases and used the resulting data as a realization of noise. This allows me to swap the noise files between the cases or to use any one of them for simulations like this. The particular noise used here was from the seawaysredwood1 case, since it had the most severe wave signatures I have seen in data, with a peak velocity of 0.5 m/s. Ocean waves can of course have higher velocities but the ship tends to yield a filtered response due to averaging over the ship length and due to the response function of the ship hull.

The MocompIAT routine was executed for aspect angles from 5 to 85 degrees and for a variety of assumed errors in the tracker estimates. The tracker estimates of aspect angle (or ship course) error was +/- 25 degrees; for ship speed +/- 2 m/s and for the initial cross-range position of the ship +/- 1 km. The IAT Analytical Method was used for all calculations here. And all calculations were done with Level 0 processing option, which is readily applicable to all existing ISAR systems with only modifications to the ISAR processor code. An exception is the final calculation in Figures 31 and 32, where Level 1 correction to the ship bearing was used to show the performance gain for improved pointing. I did not consider errors in the in-range position since we can assume that the ISAR processor would easily correct this in its centering algorithm.

First, we will analyze the errors in aspect angle to gain some physical insight into the algorithm. Then we will analyze the errors in LOA since that is the main goal here.

The results for aspect angle errors due to errors in ship course, ship speed and ship position are given in Figures 36, 37 and 38. The x-axis for these plots is the true aspect angle for the ship; the y-axis is the fractional error in the aspect angle in degrees. The variations in the tracker cues for aspect angle and ship speed only yield a few percent error in LOA over the entire required range of 5-85 degrees. However, errors in the cross-range position cause slightly larger errors in aspect angle, especially at large aspect angles.

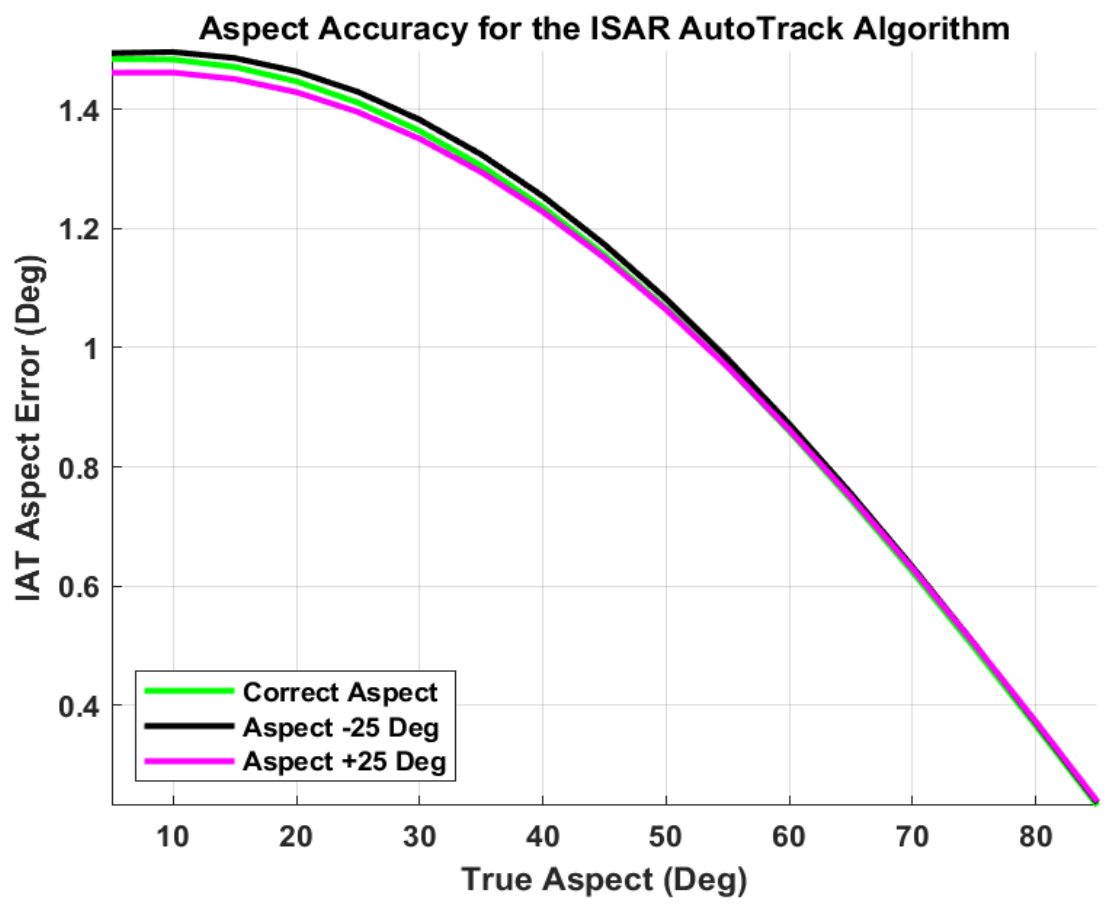

***Figure 36: Aspect Angle Error vs True Aspect - Varying Tracker Aspect Error***

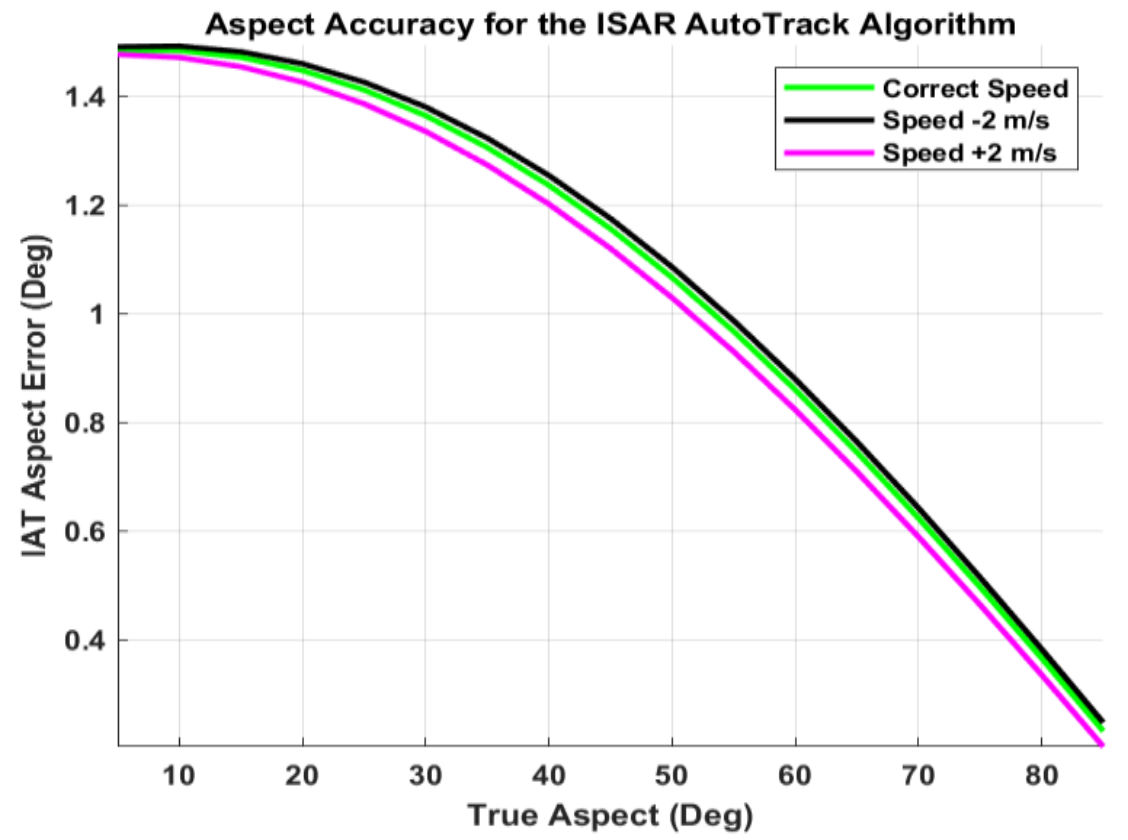

***Figure 37: Aspect Angle Error vs True Aspect - Varying Tracker Speed Error***

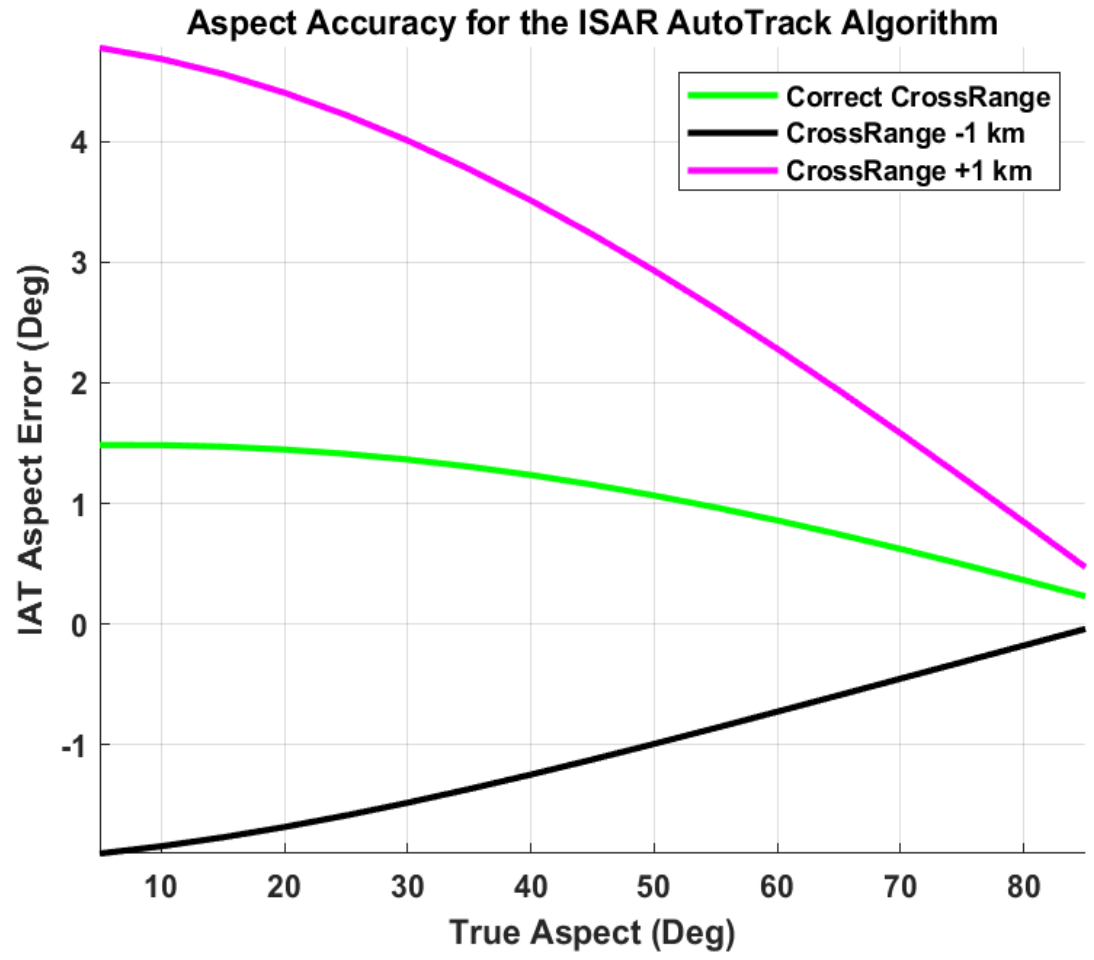

***Figure 38: Aspect Angle Error vs True Aspect - Varying Cross-range Position Error***

The results for LOA errors due to errors in ship course, ship speed and ship position are given in Figures 39, 40 and 41. The x-axis for these plots is the true aspect angle for the ship; the y-axis is the fractional error in the LOA. The variations in the tracker cues for aspect angle and ship speed only yield a few percent error in LOA over the entire required range of 5-85 degrees. However, errors in the cross-range position cause large errors in LOA, especially at large aspect angles.

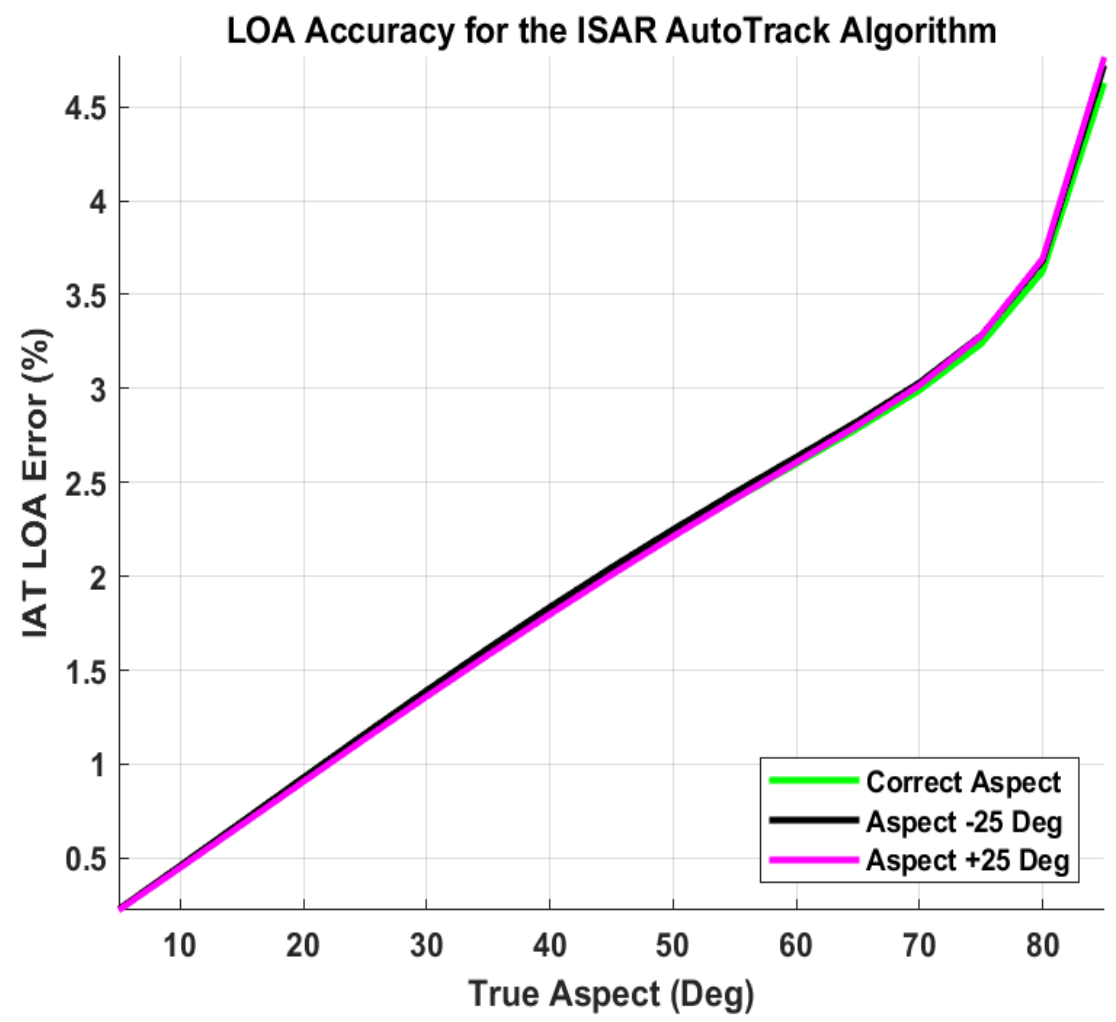

***Figure 39: Fractional Length Ship Error – Varying Tracker Aspect Error***

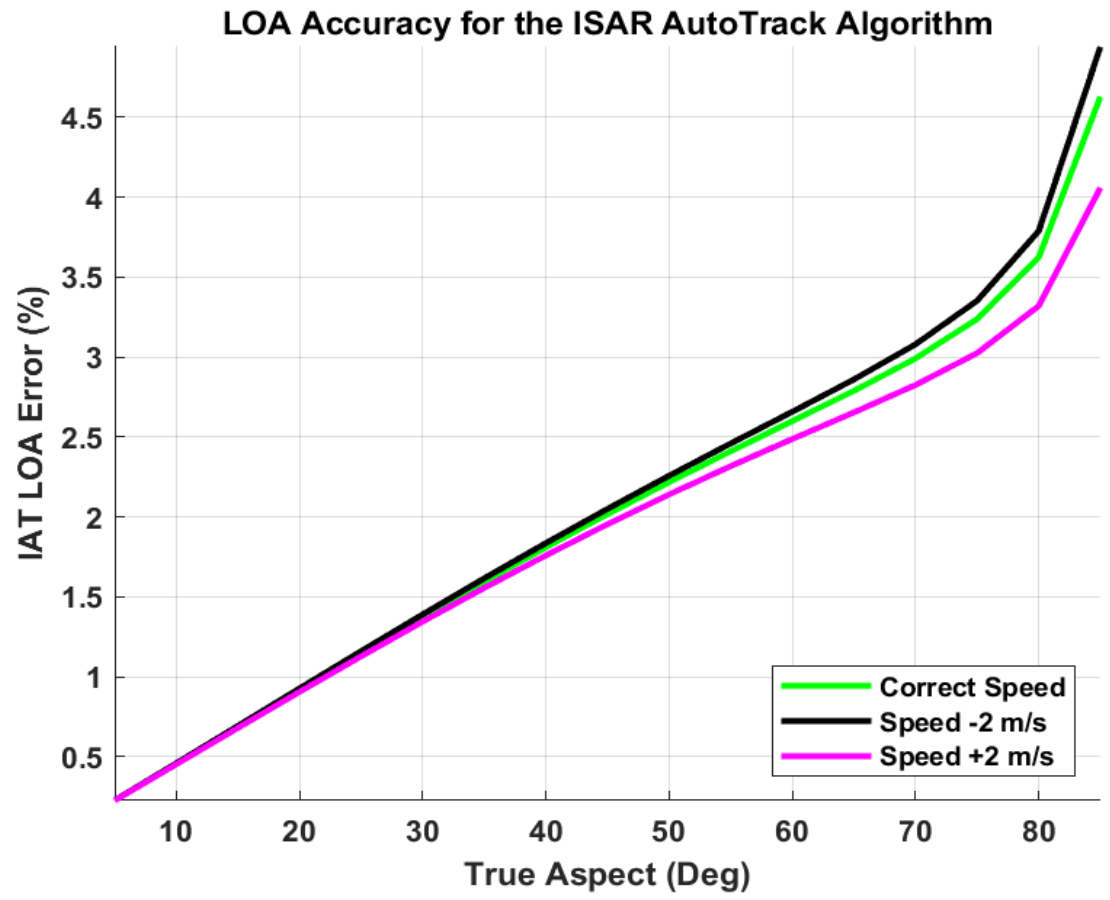

***Figure 40: Fractional Ship Length Error – Varying Tracker Speed Error***

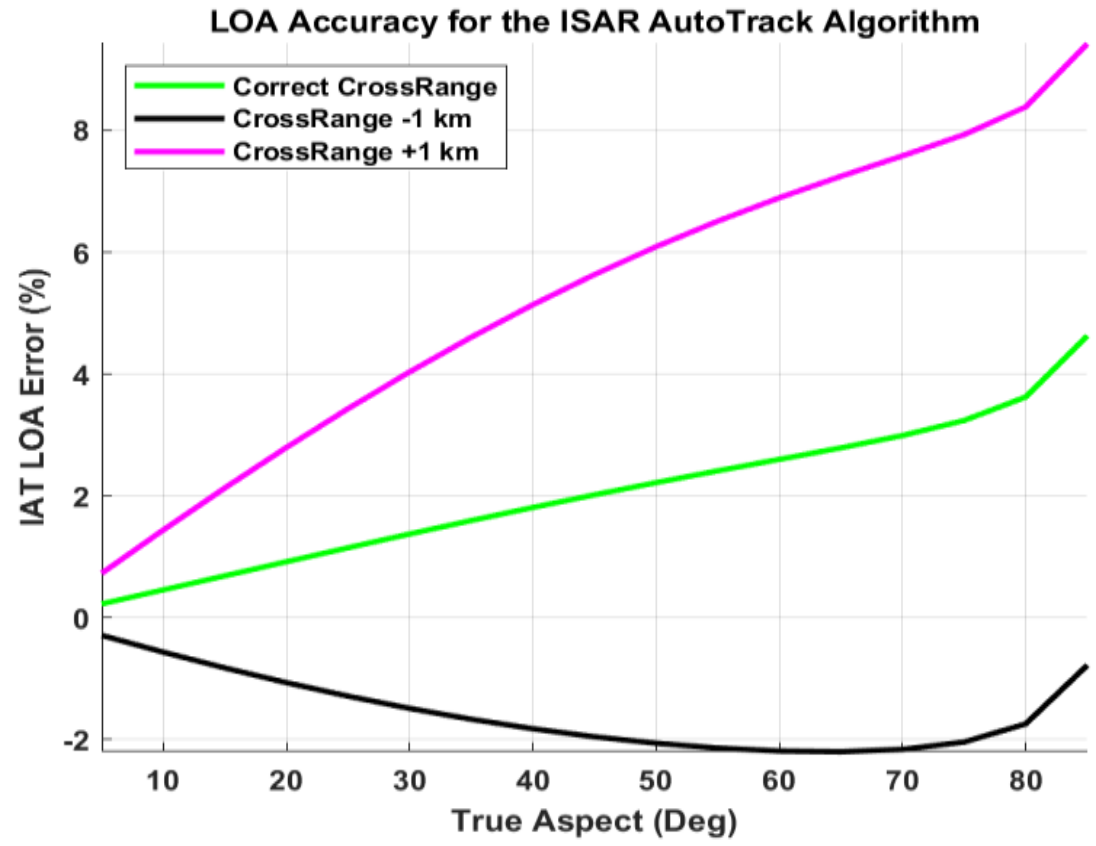

***Figure 41: Fractional Ship Length Error – Varying Cross-range Position Error***

To investigate the potential performance when all three types of tracker error are present Figures 42 and 43 are the aspect angle error and the LOA error with a 25-degree aspect angle, a 2 m/s speed error and a 1 km cross-range position error. The curves are plotted for three azimuth angles relative to the aircraft nose – 85, 45 and 5 degrees.

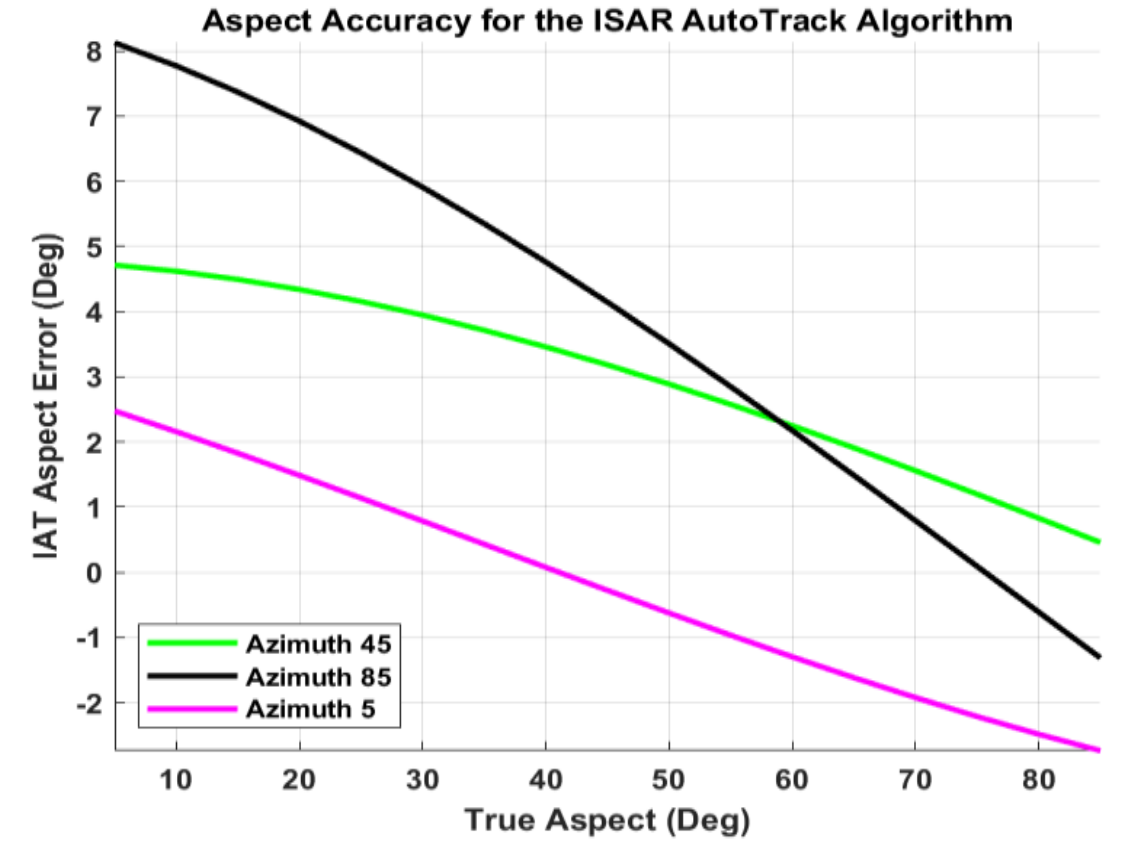

***Figure 42: Aspect Angle Error vs True Aspect - Varying Aircraft Azimuth***

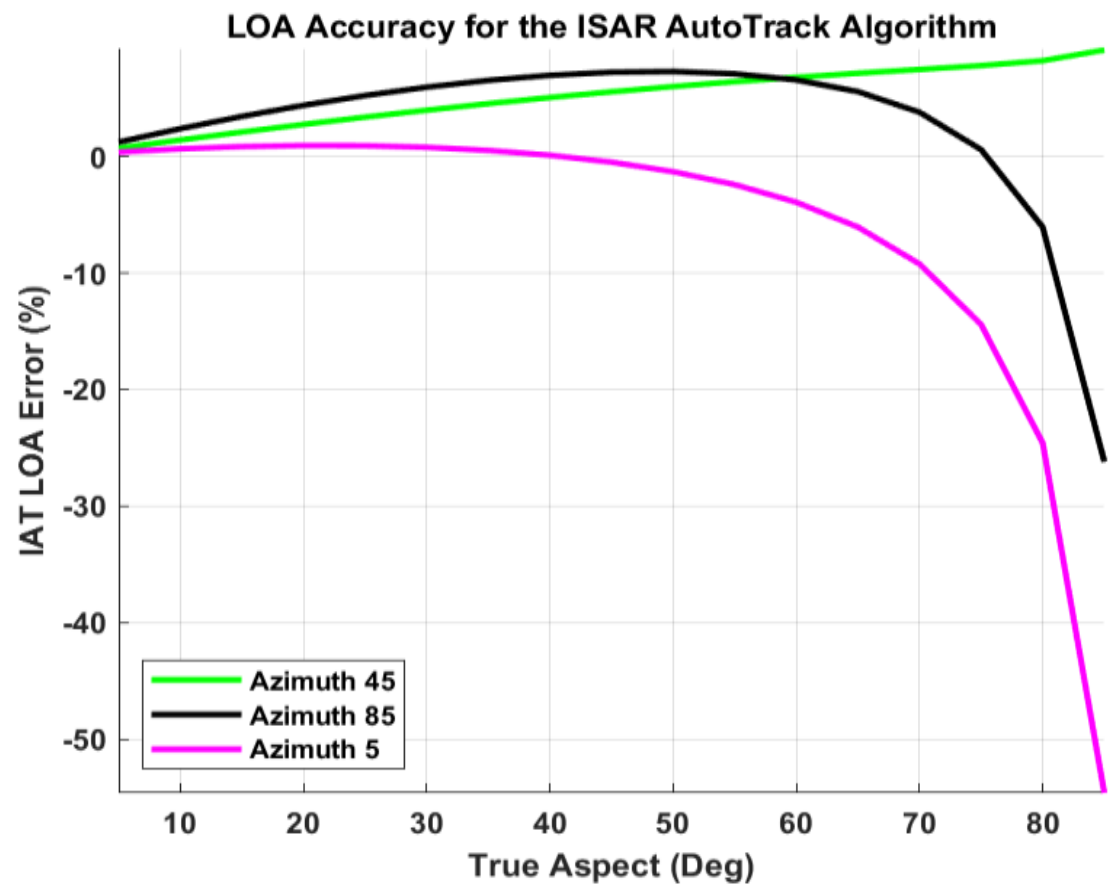

***Figure 43: Fractional Ship Length Error – Varying Aircraft Azimuth***

I have also simulated Level 1 processing by assuming that the cross-range position error is zero – Figures 44 and 45. This confirms that the enhanced bearing option in software or hardware would improve the IAT performance for aspect angle. But the goal of 10 percent LOA error is still violated at the extreme values of aspect angle. However, note that the ocean noise component used here is at the high end of observations, so this is not a forecast of the performance for large ships in *typical* sea conditions. And note that we did not evaluate the use of Level 2 processing since the simulator does not yet support this option.

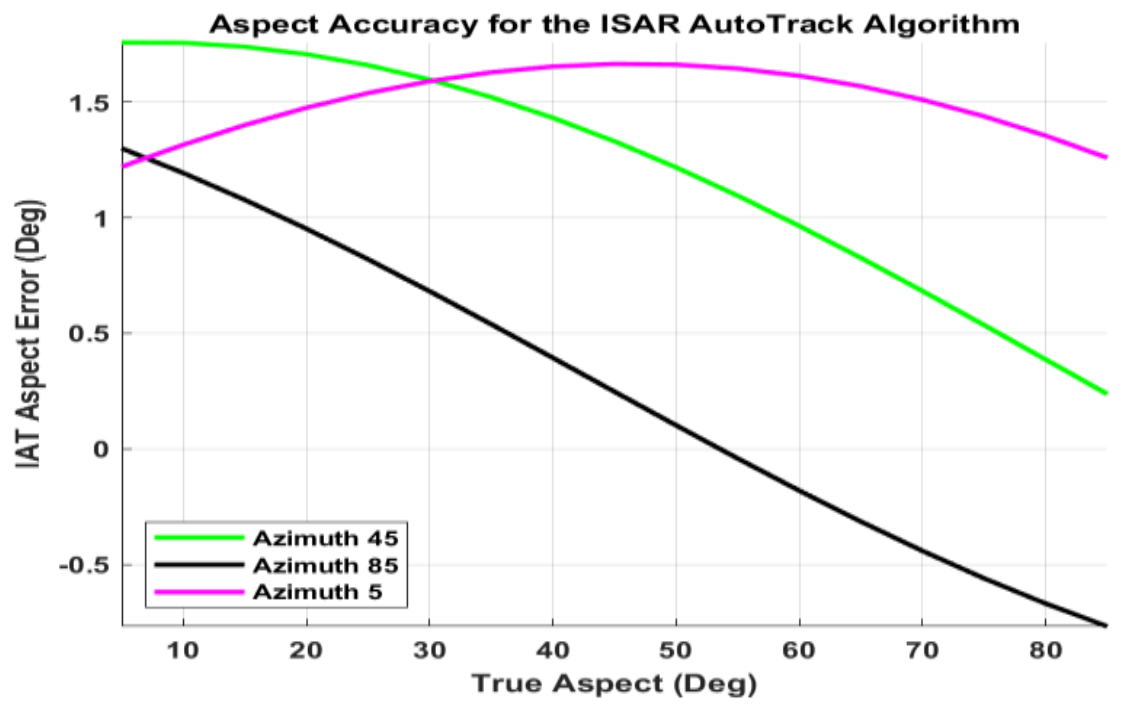

***Figure 44: Aspect Angle Error vs True Aspect - Varying Aircraft Azimuth, with Correction for Ship Lat/Lon Position***

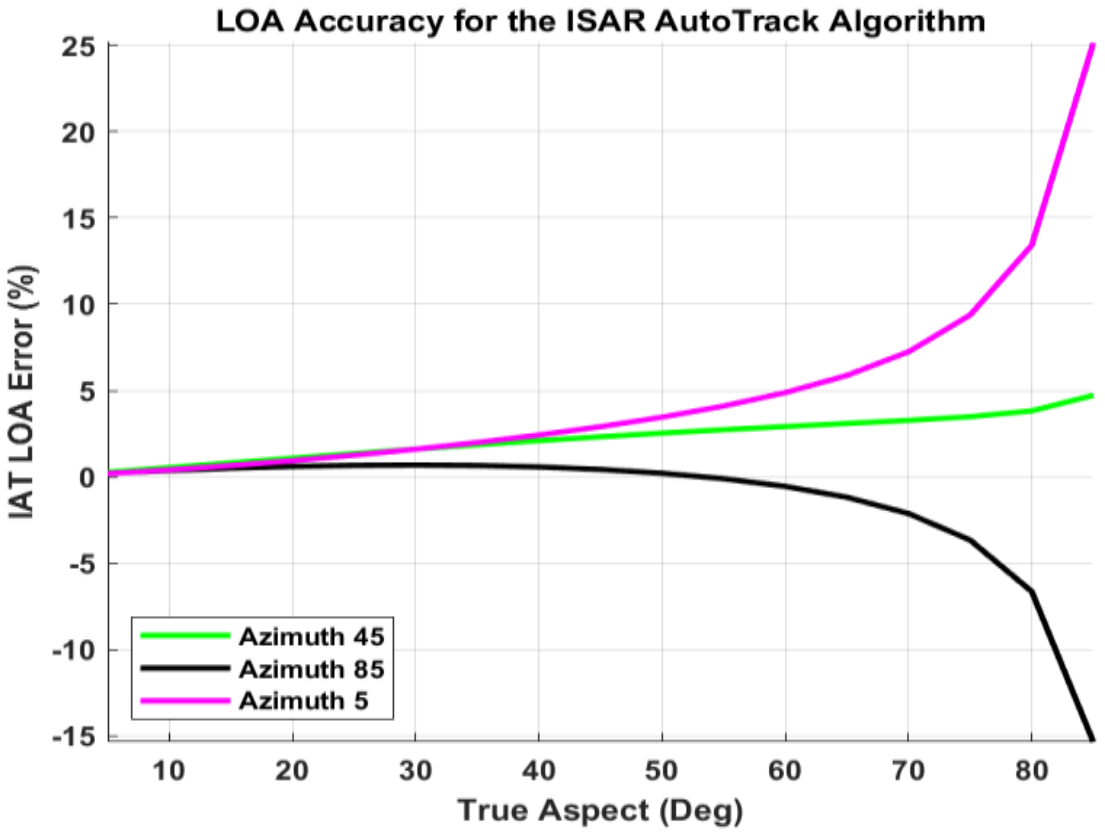

***Figure 45: Fractional Ship Length Error vs True Aspect - Varying Aircraft Azimuth, with Correction for Ship Lat/Lon Position***

## XI. SUMMARY: HOW WELL CAN AN ISAR MEAURE SHIP LENGTH?

This question from the Introduction can now be addressed with more technical insight. First, the large increase in aspect angle error at larger aspect is less than predicted by simple application of trigonometry rules. This is because the cross-range component of the ship velocity is easier to estimate at higher aspect angle where it is larger. I have not seen any evidence for such an increase in the available data and the IAT simulations show very limited effects. But, most importantly, the aspect angle can be improved by exploiting the adaptive motion compensation data using the IAT. This improvement from the tracker to Level-1 IAT processing is about 15 dB in variance (RMS from 26.6 to 4.6 degrees). An examination of the detailed data in the Appendix shows that this gain is not sensitive to the range of the target ships or the signal-to-clutter ratio.

I believe that this level of accuracy can be improved even within the constraint of using software-only changes to existing operational ISAR systems. One factor that limits the results here is the lack of knowledge of the actual algorithms for the deterministic motion compensation process. Also, adaptive beam pointing algorithms can greatly reduce the effect of errors in the target bearing. Finally, as we achieve better aspect angle accuracy, we may be able to make additional progress by addressing the errors in the range extent of the ship.

However, for LOA many of the largest errors are for cases with either significant multipath scattering, shadowing by superstructure, or large ocean waves. The multipath problem can probably be addressed with improvements in the 3D LOA algorithm. Shadowing can be reduced at short ranges by using a higher aircraft altitude or at any range by avoiding the very low aspect angles – these methods would ensure that both ends of the ship are visible to the radar. The issue with ocean waves may be addressed by the narrow band filter used in the IAT or by utilizing the Focus3D algorithm for estimating the time dependence of the aspect and tilt angles of the ship. These estimates are produced by the range-Doppler covariance functions of the detected scatterers over the entire ISAR images and are thus independent of the IAT estimates from the motion compensation data.

What range of aspect angles should the IAT algorithm be applied to? For the currently available data, we do not have adequate coverage of the higher aspect angle region except for the SCATR data. However, there is no reason to think that the higher aspect region would yield larger errors. In fact, the tracker simulation discussed in the Introduction (Figure 2) indicates that the error in aspect angle should decrease with aspect angle; and the IAT simulation agrees with this result.

Based on these calculations and on my experience with data I have set the goals for LOA accuracy of 10 percent for an aspect range of 5-85 degrees. We do not yet have solid evidence of this for a wide range of ship types, but we are probably in the ballpark. So, these may be 'stretch' goals – but they seem within range to me, especially if we implement the improvements identified here.

Is it realistic to expect good ATR performance from a system that only estimates the ship length from the aspect angles available from a large-scale tracker? I doubt it. The data from the MSR2 system clearly shows that the tracker errors in aspect angle are large. Thus, the goal here is to improve the IAT so that the LOA error is close to the error due to the residual effects in determining the range extent (shadowing, multipath, target fading, weak scatterers at the ends of the ship). This analysis shows that the IAT can do this. And we have identified some clear routes to improving the algorithm.

The IAT algorithm was designed to work with existing single-aperture ISAR systems without requiring excessive code modifications. It is applied at the end of the image movie formation and uses only information that should be readily available to the real-time computer. The algorithm just does a pre-processing stage where the adaptive motion compensation velocity is screened for anomalous pulses that may be due to interference from another radar. Then the time series is smoothed, and an adaptive narrow-band filter is used to reduce the wave response of the ship. Finally, the corrected time series is used to calculate the mean and the linear trend via linear regression. The two regression parameters are used to estimate the error in the tracker's velocity vector with either the Search or the Analytical method.

It is important to note that the IAT has an additional major advantage over using a large-scale tracker to estimate the aspect angle – the IAT produces an aspect estimate *during* the ISAR time window. Therefore, the IAT should work better for maneuvering ships; and it does not depend on associating over 100 independent detections to the correct ship. A further advantage of using the IAT is that it can greatly reduce the radar resources required for the tracker – the IAT can accurately measure the aspect angle after only a coarse estimate of the bearing and location of the ship. For one of the data cases here the IAT corrected an 85-degree error in ship heading.

Improving the estimation of the aspect angle has a direct effect on the accuracy of the ship length, a critical value for classifying ships. In addition, improved aspect angles can enable a better estimation of the rotation rates of the ship in the vertical and horizontal planes, leading to identification of Profile and Plan frames for input to a full ATR algorithm.

To build on this work, I recommend that we:

- Collect additional test data, using Level-1 IAT processing (Standard IAT with adaptive beam pointing), and concentrating on cases where the cross-range component of the platform motion is 50 m/s or greater. The need for adaptive beam pointing and for a cross-range component of platform motion are based on the theory, simulations and data analysis presented in this paper.

- Transfer the code to engineers for an operational system – this is a straightforward task since the IAT uses only the information normally recorded in most ISAR systems … and the main routine is only about one page of Matlab code.

- Add an algorithm to the ISAR processor for automatic correction of the bearing to the target by adjusting the bearing to maximize the returned power. This stage could be implemented by a simple process of varying the bearing during the ISAR data collection to find the bearing of maximum radar return. Or, alternatively, an advanced multiple-aperture radar could compare the phase between apertures to improve the bearing estimate.

- To address the problem of the partial blind spot for circular-scan radar at azimuth angles ahead and behind the radar, I would extend the above bearing improvement algorithm to also estimate the time derivative of the bearing. Then the estimate of the cross-range velocity is the product of the bearing derivative and the range to the target. Note however, that this might not be necessary if the system chooses to simply turn the aircraft when implementing the ISAR mode to move the target from the blind spot.

- Develop a strategy for combining IAT with the scan mode of the radar to improve the aspect and ship length estimates and beam pointing.

- Test the full Tracker/IAT system with shorter tracking times.

- Test the algorithm on targets with larger aspect angles – my analytical work shows that the IAT based system should work well for large aspect angles.

- Possibly integrate the IAT with other 3D algorithms – this may be useful to remove the errors due to ocean waves since the 3D code explicitly estimates these effects. At present the IAT code does not do this – it uses a simpler method that uses a narrow band filter to remove the wave effects on the motion compensation velocity.

Ideally this process should be done in an airborne radar with the capability of recording AIS data from the ships. However, experiments with known ship motions and a shore-based radar could contribute to the solution if we develop algorithms for extrapolating the data to longer range and moving platforms. Shore-based radars could also aid in improving the estimation of the range extent of ships.

In conclusion, note that this process clearly calls for a collaborative program between analysts and the major radar contractors. There is no current data collection that can test all the IAT requirements. Also, implementation of the IAT requires access to the system software for calculating the deterministic motion compensation. And, finally, a full implementation of the IAT that is valid over the required range of aspect and azimuth angles would require software modifications to the radar system to improve the estimates of the bearing to the target and its time derivative.

## REFERENCES

[1] John R. Bennett. 2025. “Focus3D: A Practical Method to Adaptively Focus ISAR Data and Provide 3-D Information for Automatic Target Recognition”. http//arxiv.org/abs/2504.13321.

[2] Murali M. Menon, Eric R. Boudreau, and Paul J. Kolodzy. “An Automatic Ship Classification System for ISAR Imagery”. Volume 6, Number 2,1993, THE LINCOLN LABORATORY JOURNAL

[3] Hongxin Yang,Fulin Su, Jianjun Gao. (2015). “The length estimation of ship targets in ISAR images”. 2015. 1910-1913. 10.1109/ICOSP.2014.7015325.

[4] Hartmut Schimpf. “Radar ATR of Maritime Targets”. NATO Paper SSTO-EN-SET-172-2013 3 – 1.pdf. 28 pp.

[5] Kenneth A. Melendez and John R. Bennett. 1998. “ISAR Target Parameter Estimation with Application for Automatic Target Recognition”. *Proceedings of SPIE*, W. Miceli Ed., **Vol. 3162**, pp. 2-13, San Diego.

[6] S. Musman, D. Kerr, C. Bachmann. “Automatic recognition of ISAR ship images”, IEEE Transactions on Aerospace and Electronic Systems, Year: 1996 | Volume: 32, Issue: 4.

[7] F. E. McFadden and S. A. Musman. "Optimizing ship length estimates from ISAR images," *Proceedings of the IEEE-INNS-ENNS International Joint Conference on Neural Networks. IJCNN 2000. Neural Computing: New Challenges and Perspectives for the New Millennium*, Como, Italy, 2000, pp. 163-168 vol.1.

[8] C. Y. Pui, S. Ghio,, B. Ng, E. Giusti and L. Rosenberg. “Robust 3D ISAR ship classification”, IEEE Radar Conference 2023.

[9] K. Suwa, T. Wakayama, and M. Iwamoto,. “Three-Dimensional Target Geometry and Target Motion Estimation Method Using Multistatic ISAR Movies and Its Performance,” IEEE Transactions on Geoscience and Remote Sensing, vol. 49, no. 6, pp. 2361‑2373, 2011.

[10] C. Y. Pui, B. Ng, L. Rosenberg, and T.-T. Cao. “3D-ISAR Using a Single Along Track Baseline,” in 2021 IEEE Radar Conference (RadarConf21), 2021, pp. 1‑6.

[11] C. Y. Pui, B. Ng, L. Rosenberg, and T. Cao, “3D ISAR for an Along-Track Airborne Radar,” IEEE Transactions on Aerospace and Electronic Systems, vol. 58, no. 4, pp. 2673‑2686, 2022.

[12] Gang Xu, Meng-Dao Xing, Xiang-Gen, Xia, Qian-Qian Chen, Lei Zhang, Zheng Bao. “High-Resolution Inverse Synthetic Aperture Radar Imaging and Scaling with Sparse Aperture”. IEEE Journal of Selected Topics in Applied Earth Observations and Remote Sensing, Year: 2015 | Volume: 8, Issue: 8.

[13] https://www.telephonics.com/uploads/standard/46045-TC-Maritime-Classification-Aid-Brochure.pdf

[14] Marco Martorella. “Novel approach for ISAR image cross-range scaling”. IEEE Transactions on Aerospace and Electronic Systems, Year: 2008 | Volume: 44, Issue: 1.

[15] J. Dall (1992). “A fast autofocus algorithm for synthetic aperture radar processing. IEEE International Conference on Acoustics, Speech, and Signal Processing. Proceedings, 3, 5 - 8.

**[16]** en.wikipedia.org/wiki/Automatic_identification_system

## DATA AVAILABILITY

The software and data for the Tybee data is in github.com/mulebennett5-source/IAT_Published_Tybee. This site includes all the code and ISAR results for the Tybee case in Section IX, the code for the simulation discussed in the Introduction, and the simulations in Section X. The raw data for Maritime Surveillance Radar #2 is proprietary; I have been given permission to use it to publish my algorithms but not to redistribute it. However, the main IAT code used in the section on MSR2 is IAT_Est.m, which is included in the directory since it is used in the simulation of Section X.

The results in Section IX for the Tybee can be viewed by downloading the software and data and then executing three Matlab scripts: TybeeGPS, TybeeIAT and TybeeLOA. The first code reads the differential GPS data provided by NRL and outputs a file that contains the true heading and speed, the range rate and the aspect angle. This file is then read by TybeeIAT, which executes the IAT algorithm under several sets of assumptions. Finally, TybeeLOA executes the algorithm for estimating the ship length from the IAT aspect angles.

## ACKNOWLEDGEMENTS

Dr. Kenneth A. Melendez of Leidos and Dr. Raghu G. Raj of the Naval Research Laboratory provided constructive reviews of this work. James A. (Jay) Trischman of Leidos, who collected the SCATR data in his previous job at the USN Naval Information Warfare Pacific, generously provided analysis of the GPS data from SCATR and important details of the experiment from his memory and notes. The directional wave buoy data during the SCATR experiment were furnished by the Coastal Data Information Program (CDIP), Integrative Oceanography Division, operated by the Scripps Institution of Oceanography, under the sponsorship of the U.S. Army Corps of Engineers and the California Department of Parks and Recreation.

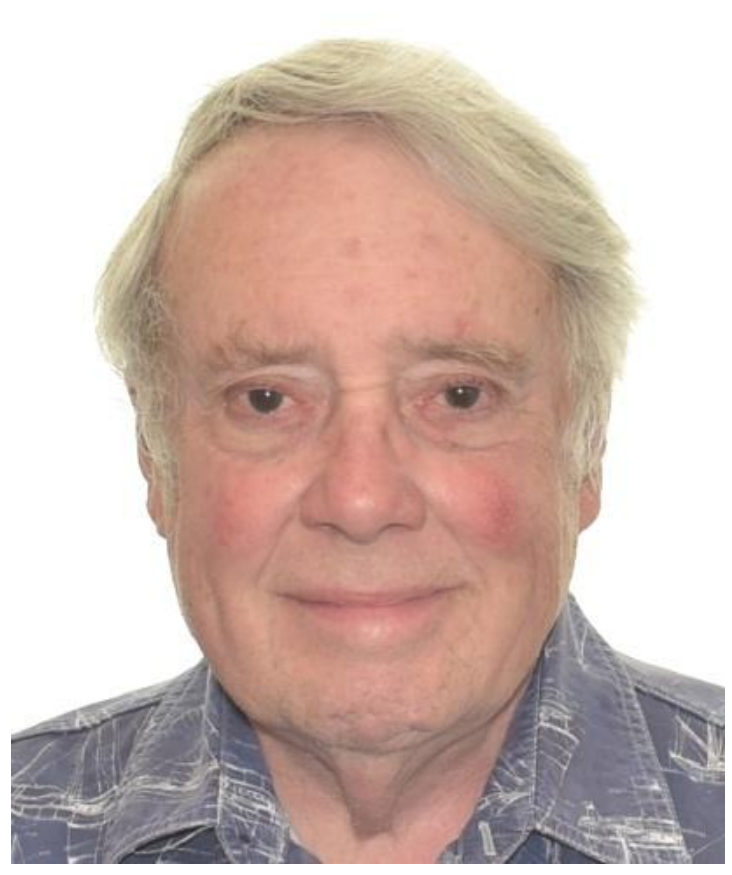

**John R. Bennett** received a PhD in meteorology from the University of Wisconsin, Madison Wisconsin USA in 1972. For the next 13 years he did research at NOAA and taught oceanography at MIT. In 1986 he joined the Environmental Research Institute of Michigan to work on radar measurements of ocean waves and water depth, with some experience on ocean surface targets. For 24 years starting in 1991 he was Senior Scientist at SAIC/Leidos in San Diego, concentrating on radar and lidar signal processing. In this position he wrote and tested the initial ISAR processors for several important radars. From January 2015 to June 2022, he was Chief Scientist at RDRTec where he worked on ISAR processing. At present he is semi-retired but is working to see 3-D ISAR algorithms applied to real-world radars and reviewing papers for IEEE Transactions on Aerospace and Electronic Systems.

## APPENDIX – Results for the ISAR AutoTrack Algorithm for the MSR2 Radar

### Table 1: Basic Parameters for 22 MSR2 Cases

| RangeNmi | TrackCourse | TrackSpeed | TrackAspect | AisMMSI | AisRangeNmi | AisSpeedKts | AisAspectDeg | AisCourse | AisSpeed m/s |
|---|---|---|---|---|---|---|---|---|---|
| 51.7 | -47.7 | 7.9 | 34.8 | 538006719 | 51.7 | 10.7 | 42.3 | 304.3 | 5.50 |
| 67.4 | -73.9 | 14.5 | -19.4 | 636013748 | 67.5 | 13.8 | -3.7 | 270.1 | 7.11 |
| 88.3 | -72.0 | 16.9 | -22.4 | 235114667 | 88.4 | 15.2 | -3.1 | 268.6 | 7.82 |
| 88.9 | -171.9 | 13.0 | 46.6 | 338418000 | 89.0 | 14.4 | 43.8 | 190.5 | 7.41 |
| 45.8 | -27.9 | 16.6 | 149.1 | 563006800 | 45.9 | 16.2 | 144.8 | 336.2 | 8.33 |
| 64.7 | -14.0 | 13.4 | 13.7 | 211433000 | 64.6 | 12.6 | 1.1 | 358.3 | 6.48 |
| 69.5 | -168.5 | 8.9 | 139.0 | 319498000 | 69.5 | 7.6 | 149.7 | 180.5 | 3.91 |
| 91.9 | -135.7 | 13.9 | 121.9 | 356330000 | 91.9 | 8.6 | 136.5 | 209.1 | 4.42 |
| 11.6 | -137.1 | 12.4 | -6.6 | 477767300 | 11.6 | 13.5 | -22.2 | 238.2 | 6.96 |
| 67.9 | -123.2 | 9.6 | -15.9 | 636091480 | 67.8 | 11.0 | 28.7 | 192.3 | 5.66 |
| 47.0 | 179.9 | 13.9 | -22.6 | 366495000 | 47.2 | 10.7 | -26.2 | 184.1 | 5.51 |
| 28.4 | -145.6 | 9.3 | 148.0 | 229769000 | 28.4 | 11.0 | 151.3 | 211.2 | 5.66 |
| 38.8 | 76.9 | 8.7 | -36.2 | 538004934 | 38.8 | 9.2 | -58.1 | 99.3 | 4.73 |
| 44.0 | 136.6 | 9.1 | 145.3 | 538004934 | 44.1 | 9.9 | -167.4 | 89.1 | 5.09 |
| 38.9 | 2.5 | 14.1 | -21.4 | 477183900 | 38.8 | 12.5 | -17.2 | 358.9 | 6.43 |
| 85.8 | -139.0 | 22.4 | 114.7 | 372008000 | 85.7 | 10.1 | 139.4 | 195.4 | 5.20 |
| 90.0 | 168.4 | 11.3 | 161.9 | 477076500 | 90.1 | 11.2 | 161.6 | 168.6 | 5.77 |
| 62.4 | -150.7 | 33.4 | 110.0 | 636019033 | 62.5 | 11.0 | 135.0 | 182.6 | 5.66 |
| 76.3 | -152.2 | 33.2 | 116.6 | 636018151 | 76.3 | 13.0 | 146.9 | 176.0 | 6.69 |
| 90.8 | -160.7 | 25.3 | 123.1 | 477076500 | 90.8 | 11.2 | 152.3 | 168.7 | 5.76 |
| 94.7 | 91.9 | 10.3 | -142.7 | 477076500 | 94.8 | 11.2 | 141.3 | 168.4 | 5.78 |
| 48.1 | -167.2 | 14.3 | 53.8 | 220416000 | 48.0 | 9.9 | 43.1 | 203.8 | 5.09 |

### Table 2: LOA Results for 22 MSR2 Cases

| Focus3D Ship Case | Ship LOA (m) | Pulse Cor. | SCR (dB) | Track LOA | Error | AIS LOA | Error | IAT LOA | Error |
|---|---|---|---|---|---|---|---|---|---|
| elandraspruce1 | 183 | 0.82 | 6.6 | 153 | -30 | 168 | -15 | 168 | -15 |
| transsibbridge2 | 182.32 | 0.64 | 2.5 | 295 | 113 | 282 | 99.68 | 284 | 101.68 |
| atlanticsail1 | 296 | 0.85 | 7.5 | 228 | -68 | 215 | -81 | 216 | -80 |
| maerskdenver1 | 299.54 | 0.58 | 1.4 | 264 | -36 | 254 | -45.54 | 291 | -8.54 |
| maerskshams2 | 299.92 | 0.97 | 15.1 | 272 | -28 | 284 | -15.92 | 314 | 14.08 |
| colomboexpress2 | 335.47 | 0.95 | 12.8 | 308 | -27 | 319 | -16.47 | 303 | -32.47 |
| stoltcreativity2 | 176.75 | 0.57 | 1.2 | 178 | 1 | 157 | -19.75 | 153 | -23.75 |
| msckrystal4 | 277.3 | 0.70 | 3.7 | 366 | 89 | 269 | -8.3 | 264 | -13.3 |
| shouchenshan1 | 189.99 | 0.97 | 15.1 | 234 | 44 | 254 | 64.01 | 246 | 56.01 |
| independenthorizor | 220.44 | 0.84 | 7.2 | 198 | -22 | 212 | -8.44 | 208 | -12.44 |
| overseaschinook3 | 183 | 0.90 | 9.5 | 187 | 4 | 192 | 9 | 181 | -2 |
| seamelody4 | 274.98 | 0.96 | 13.8 | 279 | 4 | 271 | -3.98 | 282 | 7.02 |
| seawaysredwood1 | 250 | 0.90 | 9.5 | 158 | -92 | 238 | -12 | 157 | -93 |
| seawaysredwood2 | 250 | 0.88 | 8.7 | 288 | 38 | 243 | -7 | 243 | -7 |
| coscojasmine1 | 366 | 0.94 | 11.9 | 369 | 3 | 362 | -4 | 347 | -19 |
| mscheidi3 | 331.99 | 0.85 | 7.5 | 587 | 255 | 330 | -1.99 | 349 | 17.01 |
| sagaadventure3 | 199.2 | 0.66 | 2.9 | 183 | -16 | 183 | -16.2 | 181 | -18.2 |
| mediatlantico6 | 199.98 | 0.82 | 6.6 | 385 | 185 | 190 | -9.98 | 167 | -32.98 |
| angeles4 | 225 | 0.79 | 5.8 | 373 | 148 | 205 | -20 | 219 | -6 |
| sagaadventure4 | 199.2 | 0.73 | 4.3 | 285 | 86 | 181 | -18.2 | 175 | -24.2 |
| sagaadventure5 | 199.2 | 0.96 | 13.8 | 190 | -9 | 197 | -2.2 | 234 | 34.8 |
| georgmaersk1 | 366.93 | 0.90 | 9.5 | 410 | 43 | 338 | -28.93 | 385 | 18.07 |
| **Mean** | **250.28** | **0.83** | **8.05** | | **31.08** | | **-7.37** | | **-6.33** |
| **Std Dev.** | **61.74** | **0.12** | **4.28** | | **82.78** | | **33.71** | | **39.55** |
| **RMS** | **257.78** | **0.84** | **9.11** | | **88.42** | | **34.50** | | **40.06** |